\documentclass[fleqn,usenatbib]{mnras}

\usepackage{newtxtext,newtxmath}
\usepackage[T1]{fontenc}

\DeclareRobustCommand{\VAN}[3]{#2}
\let\VANthebibliography\thebibliography
\def\thebibliography{\DeclareRobustCommand{\VAN}[3]{##3}\VANthebibliography}

\usepackage{graphicx}	% Including figure files
\usepackage{amsmath}	% Advanced maths commands
\title[Numerical simulations of macrospicules]{Numerical simulations of macrospicule jets under energy imbalance conditions in the solar atmosphere}

\author[Gonz\'alez-Avil\'es et al.]{
J. J. Gonz\'alez-Avil\'es,$^{1}$\thanks{E-mail: jjgonzalez@igeofisica.unam.mx}
K. Murawski,$^{3}$
A. K. Srivastava, $^{4}$
T. V. Zaqarashvili $^{5,6,7}$
and J. A. Gonz\'alez-Esparza $^{2}$
\\
$^{1}$C\'atedras CONACYT, SCIESMEX-LANCE, Instituto de Geof\'isica, Unidad Michoac\'an, Universidad Nacional Aut\'onoma de M\'exico, Morelia, Michoac\'an, M\'exico\\
$^{2}$ SCIESMEX-LANCE, Instituto de Geof\'isica, Unidad Michoac\'an, Universidad Nacional Aut\'onoma de M\'exico, Morelia, Michoac\'an, M\'exico \\
$^{3}$ Institute of Physics, University of Maria Curie-Sk{\l}odowska, 5 M. Curie-Sk{\l}odowskiej Sq., 20-031 Lublin, Poland \\
$^{4}$ Department of Physics, Indian Institute of Technology (BHU), Varanasi 221005, India \\
$^{5}$ IGAM, Institute f\"ur Physik, University of Graz, Universit\"atsplatz 5, A-8010 Graz, Austria \\
$^{6}$ Ilia State University, Cholokashvili ave 5/3, Tbilisi, Georgia \\
$^{7}$ Abastumani Astrophysical Observatory, Mount Kanobili, Abastumani, Georgia \\
}

\date{Accepted 2021 April 28. Received 2021 April 12; in original form 2021 February 17}

\pubyear{2021}

\begin{document}
\label{firstpage}
\pagerange{\pageref{firstpage}--\pageref{lastpage}}
\maketitle

\begin{abstract}

Using numerical simulations, we study the effects of thermal conduction and radiative cooling on the formation and evolution of solar jets with some macrospicules features. We initially assume that the solar atmosphere is rarely in equilibrium through energy imbalance. Therefore, we test whether the background flows resulting from an imbalance between thermal conduction and radiative cooling influence the jets' behavior. In this particular scenario we trigger the formation of the jets by launching a vertical velocity pulse localized at the upper chromosphere for the following test cases: i) adiabatic case; ii) thermal conduction case; iii) radiative cooling case; iv) thermal conduction + radiative cooling case. According to the test results, the addition of the thermal conduction results in smaller and hotter jets than in the adiabatic case. On the other hand the radiative cooling dissipates the jet after reaching the maximum height ($\approx$ 5.5 Mm), making it shorter and colder than in the adiabatic and thermal conduction cases. Besides, the flow generated by the radiative cooling is more substantial than the caused by the thermal conduction. Despite the energy imbalance of the solar atmosphere background, the simulated jet shows morphological features of macrospicules. Furthermore, the velocity pulse steepens into a shock that propagates upward into a solar corona that maintains its initial temperature. The shocks generate the jets with a quasi-periodical behavior that follows a parabolic path on time-distance plots consistent with macrospicule jets' observed dynamics.

\end{abstract}


%%%%%%%%%%%%%%%%%%%%%%%%%%%%%%%%%%%%%%%%%%%%%%%%%%

%%%%%%%%%%%%%%%%% BODY OF PAPER %%%%%%%%%%%%%%%%%%

\section{Introduction}
\label{sec:Introduction}

Understanding the different layers and structures of the solar atmosphere has improved greatly in the last decades. However, the fine structures still pose many open questions to answer \citep{Judge_2006,Lipartito_et_al_2014,Kiss_et_al_2017}. Among these fine structures are spicules, which are spiky-like gas jets mainly observed at the chromosphere \citep{Secchi_1877,Sterling_2000,De_Pontieu_et_al_2004}. Depending on their size and lifetime, spicules are classified in two types: i) type I spicules with 7-11 Mm length, 5-15 minute lifetime, and 25 km s$^{-1}$ upward speed \citep{Beckers_1968,Zaqarashvili&Erdelyi_2009}; ii) type II spicules, with 5 Mm average height, 50-100 km s$^{-1}$ propagating speed and 10-150 s lifetime \citep{De_Pontieu_et_al_2007,De_Pontieu_et_al_2012,Sekse_et_al_2012,Kuridze_et_al_2015}. Additionally, type II spicules have been proposed that may play an important role in energy and material supply to upper layers in the solar atmosphere \citep{De_Pontieu_et_al_2011, Samanta_et_al_2019}. There are, however, very long spicules that are also abundant in the solar atmosphere, called macrospicules. These giant spicules were introduced by \citet{Bohlin_et_al_1975}, concerning the jets observed in Skylab's extreme ultraviolet (EUV) spectroheliograms at the solar limb. Besides, they are observed in polar coronal holes, where they reach heights from 7 to 70 Mm above the solar limb, maximum velocities from 10 to 150 km s$^{-1}$ and lifetimes ranging from 3 to 45 minutes \citep{Bohlin_et_al_1975,Withbroe_et_al_1976,Dere_et_al_1989,Karovska&Habbal_1994,Parenti_et_al_2002,Bennet&Erdelyi_2015,Kiss_et_al_2017,Sterling_2000,Wilhelm_2000,Loboda&Bogachev_2019}. In this context, a number of mechanisms have been proposed for the formation of macrospicules, for instance, by a pressure pulse \citep{Hollweg_1982}, through a velocity pulse \citep{Suematsu_et_al_1982,Murawski_et_al_2011}, magnetic recconection \citep{Shibata_1982}; among others \citep{Moore_et_al_1977,De_Pontieu_et_al_2004,Kamio_et_al_2010}.

Regarding the classical numerical models developed to explain the formation of spicules, we can find, for instance, \citet{Sterling_2000} simulated the formation of spicules with an energy input located at the photospheric level. A pressure pulse or Alfv\'en waves generate the energy input, which sharpens into shock waves. Also, \citet{Muraswki&Zaqarashvilli_2010} showed that a single initial localized velocity pulse could lead to the formation of consecutive shocks due to the nonlinear wake in the stratified atmosphere, which in turn is consistent with the classical 2D rebound shock model of \citet{Hollweg_1982}. Besides, \citet{Heggland_et_al_2007} performed 1D simulations of shock wave-driven jets in the solar atmosphere. There are also more sophisticated models, for example, \citet{Hansteen_et_al_2006} and \citet{De_Pontieu_et_al_2007} performed radiative MHD simulations to study the formation of dynamic fibrils due to slow magneto-acoustic shocks. Also, \citet{Martinez-Sykora_et_al_2009} used three-dimensional MHD numerical simulations to show that type I spicules can be driven by various mechanisms that include p-modes and convecting buffeting of flux concentrations. Another process to excite spicules is found in \citet{Smirnova_et_al_2016}, where authors carried out numerical simulations of a localized pressure pulse at the null point of a magnetic arcade. They found the formation of jets with some characteristics of type I and II spicules. 

Aside from \citet{Murawski_et_al_2011} that simulated the formation of plasma structures with many macrospicules features in the context of the ideal MHD equations, this paper includes non-adiabatic ingredients that make the model more realistic since the dissipation processes appear more naturally over the system. It is also different from \citet{Kuzma_et_al_2017a} since our paper does not consider a system with an initial energy balance between thermal conduction and radiative cooling compensated by an external heating term. Moreover, our simulation setup is more related to macrospicules' parameters than type I or type II spicules. Specifically, in this paper, we test whether background flows resulting from an imbalance between thermal conduction and radiative cooling influence the jets' behavior. In this regard, we can support this approximation by reviewing some investigations about numerical simulations and analytical models in the context of the solar atmosphere phenomena. In such articles, the authors do not explicitly consider energy balance and study the impact of non-adiabatic terms such as thermal conduction, magnetic resistivity, and radiative cooling separately. For instance, in \citet{Botha_et_al_2011} the authors performed numerical simulations and found that thermal conduction plays an essential role in the kink instability of coronal loops. Also, in \citet{Silva_et_al_2018}, the authors found that thermal energy is transported more efficiently to the upper chromosphere and lower transition region and led to earlier heating of the lower atmosphere.
Additionally, in \citet{Gonzalez-Aviles_et_al_2017, Gonzalez-Aviles_et_al_2018}, the authors investigated the importance of magnetic resistivity itself and its relation to the magnetic reconnection process on the formation and evolution of jet with Type II spicules' characteristics. Moreover, the radiative cooling of coronal plasmas is related to catastrophic cooling, characterized by a rapid fall in coronal temperature, mainly identified in flares, active regions, and coronal loops \citep{Cargill&Bradshaw_2013}. Besides, radiative cooling appears to be the energy source for the oppositely directed fluxes of electromagnetic energy on MHD models of the quiet Sun from the convection zone to the corona \citep[see, e.g.][]{Abbett&Fisher_2012}.

We organize the paper as follows. In Section \ref{Model&methods} we describe magnetohydrodynamic (MHD) equations, the numerical methods we use, the model of the solar atmosphere, the magnetic field configuration, and the perturbation. In Section \ref{sec:results_numerical_simulations}, we describe the results of the numerical simulations for the different test cases: i) adiabatic, ii) thermal conduction, iii) radiative cooling, and iv) thermal conduction + radiative cooling. We also compare the four cases' results, present the results' discussions, and some observational facts of macrospicules that support some of our simulation results. Section \ref{sec:conclusions} contains the conclusions and final comments.

% ------------------------------------------
% ----->.    Section    <----------
% ------------------------------------------
\section{Model and methods}
\label{Model&methods}

% --------------------------------    
% ------> Subsection <------
% --------------------------------
\subsection{The system of MHD equations}

We consider a gravitationally stratified solar atmosphere described by a plasma obeying the MHD equations with thermal conduction and radiative cooling. In particular, we write the equations in a conservative dimensionless form that is optimal for the adopted numerical methods:

\begin{eqnarray}
\frac{\partial\varrho}{\partial t} + \nabla\cdot(\varrho{\bf v}) &=& 0, \label{density}\\
\frac{\partial(\varrho{\bf v})}{\partial t} + \nabla\cdot(\varrho{\bf v}{\bf v}-{\bf B}{\bf B} + p_{t}{\bf I}) &=& \varrho {\bf g},  \label{momentum} \\
\frac{\partial E}{\partial t} +\nabla\cdot((E+p_{t}){\bf v}-{\bf B}({\bf v}\cdot{\bf B})) &=& {\varrho}{\bf v}\cdot{\bf g}\nonumber\\+Q_{rad}+\nabla\cdot{\bf q}, \label{energy} \\
\frac{\partial{\bf B}}{\partial t} +\nabla\cdot({\bf v}{\bf B} -{\bf B}{\bf v}) &=& {\bf 0}, \label{evolB} \\
\nabla\cdot{\bf B} &=& 0, \label{divergenceB} 
\end{eqnarray}

\noindent where $\varrho$ is the mass density, ${\bf v}$ represents the velocity, ${\bf B}$ is the magnetic field, $T$ is the temperature, $Q_{rad}$ represents the thin cooling radiation, and ${\bf q}$ is the anisotropic thermal conduction flux. Besides, $p_{t}=p+{\bf B}^{2}/2\mu_{0}$ is the total (thermal + magnetic) pressure, ${\bf I}$ is the unit matrix, and $E$ is the total energy density, that is, the sum of the internal, kinetic, and magnetic energy densities, viz. 

\begin{equation}
E = \frac{p}{\gamma-1} + \frac{\varrho{\bf v}^{2}}{2} + \frac{{\bf B}^{2}}{2\mu_{0}}. \label{total_energy}
\end{equation}
\noindent For the fluid we consider the adiabatic index $\gamma=5/3$. The system of equations (\ref{density})-(\ref{divergenceB}) is closed with the ideal gas law
\begin{equation}
p = \frac{k_{B}}{m}\varrho T, \label{eos}  
\end{equation}
\noindent where $T$ is the temperature of the plasma, $m=\mu m_{H}$ is the particle mass specified by a mean molecular weight value $\mu=$0.6, for a fully ionized gas and $m_{H}$ is hydrogen's mass, and $k_{B}$ is Boltzmann's constant. The gravitational source term on the right hand side of equations (\ref{momentum}) and (\ref{energy}) is given by ${\bf g}=[0,-g]$ with magnitude $g=274$ m s$^{-2}$, which represents an average over the solar surface. 

In this paper, we consider the term $Q_{rad}=-n_{H}^{2}\Lambda(T)$ that represents the optically thin radiative losses computed according to the version 7 of CHIANTI code \citep{Landi_et_al_2012}, assuming a hydrogen number density of $10^{9}$ cm$^{-3}$ and ionization equilibrium according to \citet{Dere_2009}. The radiative losses are physically consistent to an optically thin medium such as the solar corona and upper chromosphere. Therefore, we avoid the cooling term at the photosphere and lower chromosphere by setting $\Lambda(T)=0$, for $T<1\times10^{4}$ K. 

Additionally, the contribution from thermal conduction is through the heat flux vector that allows the heat propagation along the magnetic field lines, given as follows: 

\begin{equation}
{\bf q} = \kappa_{\parallel}\hat{\bf b}(\hat{\bf b}\cdot\nabla T), \label{thermal_cond}
\end{equation}
\noindent where $\kappa_{\parallel}$ is the thermal conduction coefficient along the magnetic field lines and $\hat{\bf b}={\bf B}/|{\bf B}|$ is the unit vector in the direction of the magnetic field. In this paper, we use the thermal conduction coefficient for a fully ionized plasma \citep{Spitzer_1962}, defined as follows:

\begin{equation}
\kappa_{\parallel} = \kappa_{0}T^{5/2}, \label{kappa} 
\end{equation}
with $\kappa_{0}=1\times{10}^{-6}$ erg s$^{-1}$ cm$^{-1}$ K$^{-7/2}$, which is a characteristic value in the solar atmosphere \citep{Priest_2014}.

% ---------------------------------------------------
% ---------->    SUB-SECTION    <----------
% ---------------------------------------------------

\subsection{Model of the solar atmosphere and magnetic field configuration}

\begin{figure*}
\centering
\includegraphics[width=8.0cm,height=6.0cm]{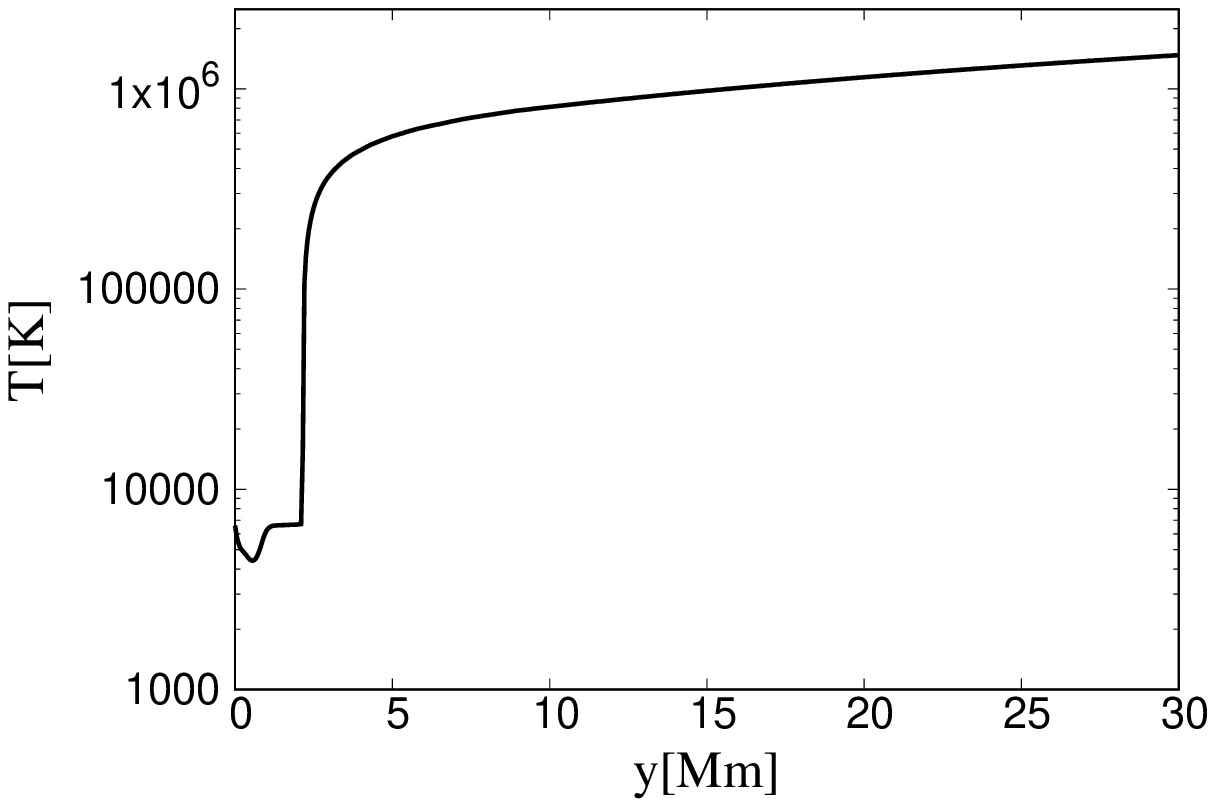}
\includegraphics[width=8.0cm,height=6.0cm]{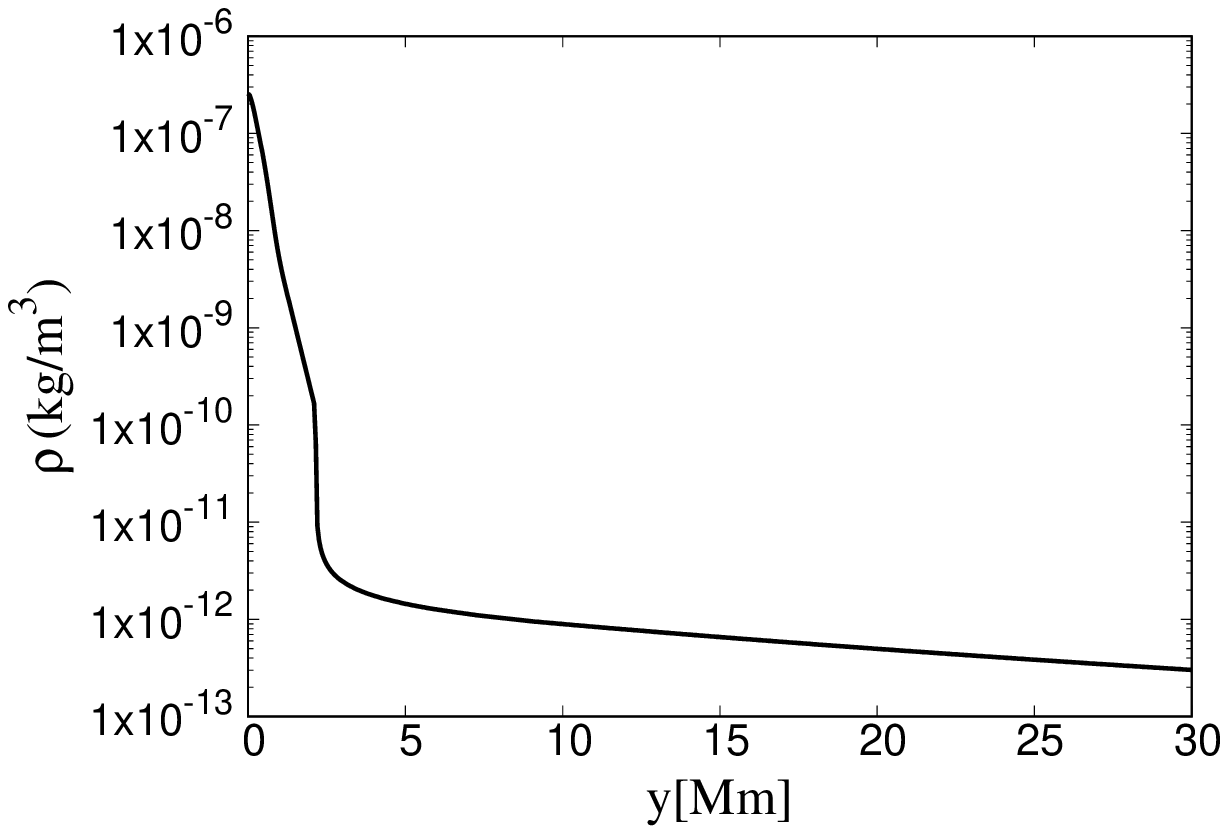}\\
\includegraphics[width=7.5cm,height=6.0cm]{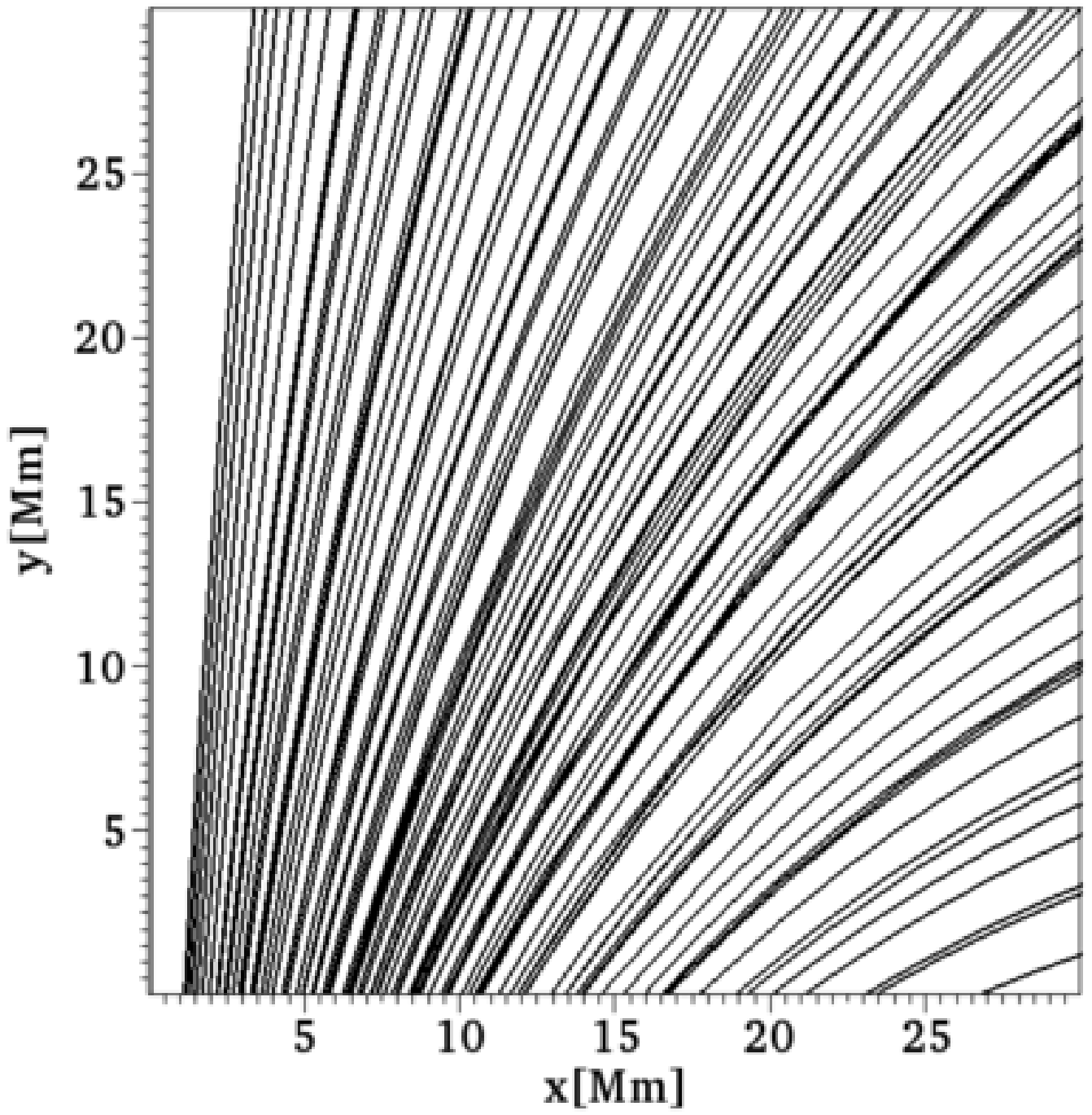}
\includegraphics[width=8.0cm,height=6.1cm]{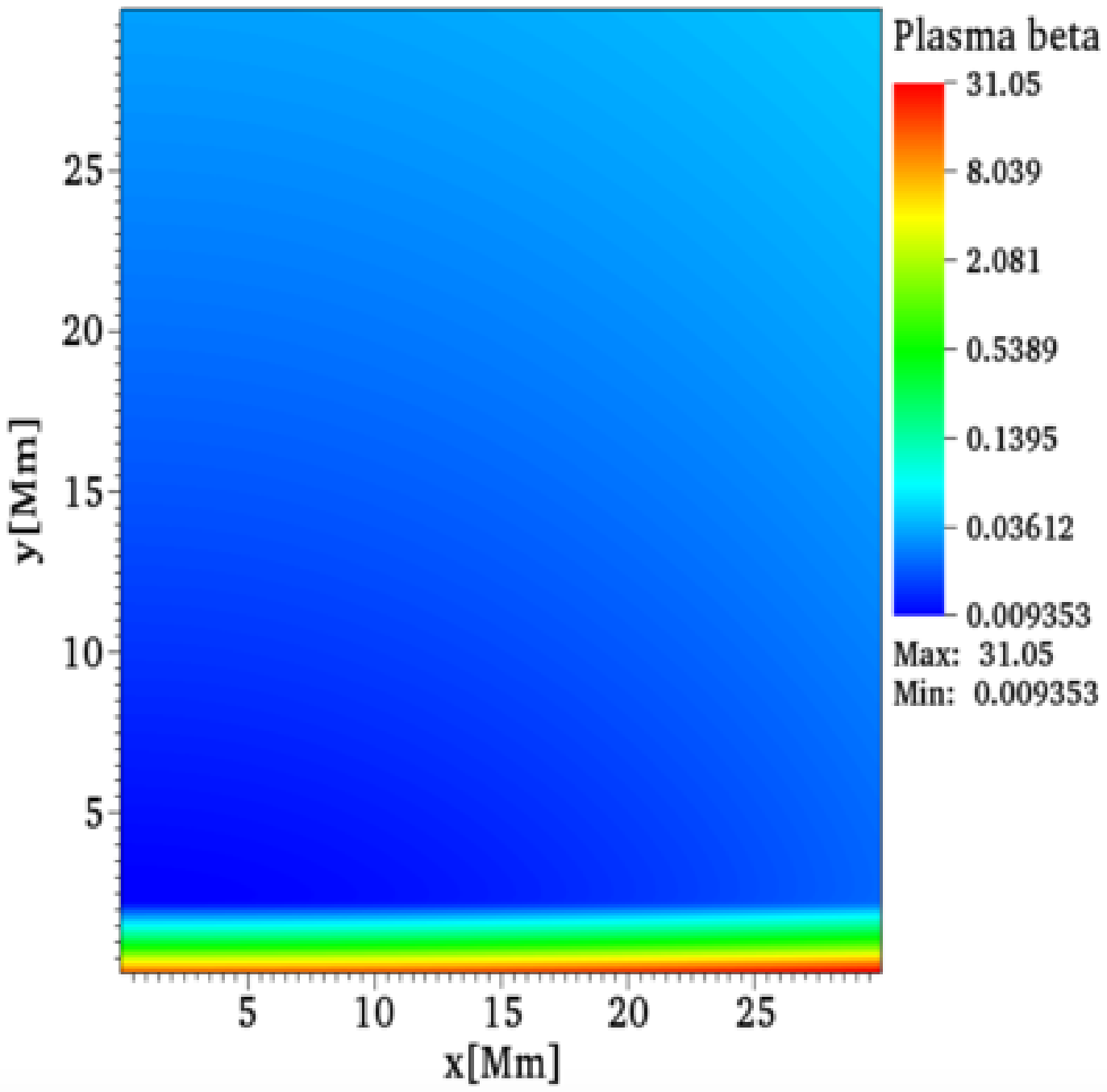}
\caption{(Top) The logarithm of temperature (left) and the logarithm of mass density (right) vs. height, $y$, for the C7 equilibrium solar atmosphere model. (Bottom) The magnetic field lines (left) and the plasma beta (right) in the 2D domain at $t=0$ s.}
\label{fig:atmosphere}
\end{figure*}

At the initial time ($t=0$ s) of the simulations, we assume that the solar atmosphere is in hydrostatic equilibrium but in energy imbalance. The C7 semiempirical model describes the temperature field in the chromosphere and transition region \citep{AvretLoeser2008}. For instance, at the top of Figure \ref{fig:atmosphere}, we show the initial temperature and mass density profiles. The magnetic field configuration ${\bf B}$ is force-free and current-free, with the following components \citep{Low_1985}:

\begin{eqnarray}
B_{x} = \frac{-2S(x-a)(x-b)}{((x-a)^{2}+(y-b)^{2})^{2}}, \\ \label{Bx}
B_{y} = \frac{S(x-a)^{2}-(x-b)^{2}}{((x-a)^{2}+(y-b)^{2})^{2}}, \label{By}
\end{eqnarray}

\noindent where $S$ corresponds to the pole's magnetic field strength given in units of G Mm$^{2}$, $a$ and $b$ are parameters that define the magnetic pole's location. In our simulations, we set $a=0$ Mm and $b=-40$ Mm. We choose $S$ in a way that at the reference point $(x=0,y=10)$ Mm, the magnitude of the Alfv\'en speed is about ten times greater than the speed of sound. For these settings, the magnetic field lines are discernibly diverged at about $(x=5,y=1.75)$ Mm, and the plasma beta $\beta$ decreases with height, as shown at the bottom of Figure \ref{fig:atmosphere}.

% ---------------------------------------------------
% ---------->    SUB-SECTION    <----------
% ---------------------------------------------------

\subsection{Numerical methods}

Equations (\ref{density})-(\ref{divergenceB}) are solved numerically using the PLUTO code \citep{Mignone_et_al_2007}. In all simulations, we set the Courant-Friedrichs-Levy (CFL) number equal to 0.3 and choose the second-order total variation diminishing (TVD) Runge-Kutta time integrator. Additionally, we use the Harten-Lax-van Leer discontinuities (HLLD) approximate Riemann solver \citep{Miyoshi&Kusano_2005} in combination with the minmod limiter.

\begin{figure*}
\centering
\includegraphics[width=9.0cm,height=6.0cm]{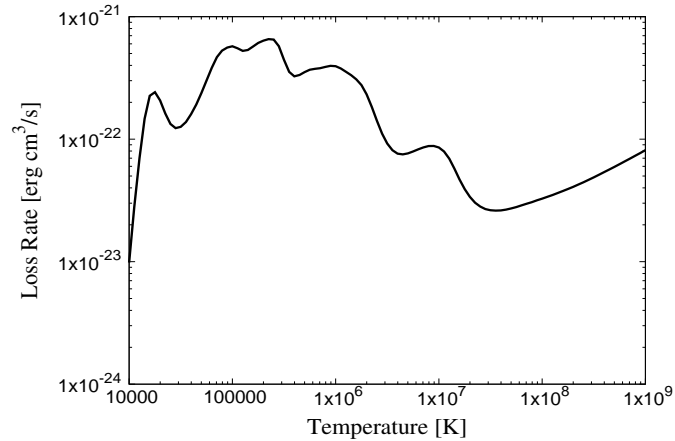}
\caption{Radiative loss rate $\Lambda(T)$ based on the CHIANTI coronal abundances.}
\label{fig:chianti}
\end{figure*}

The numerical evolution of the MHD equations can lead to the violation of the divergence-free constraint equation given by (\ref{divergenceB}), developing as a consequence unphysical results like the presence of a net magnetic charge. To control the constraint violation's growth, we use the extended generalized Lagrange multiplier method \citep{Dedner_et_al_2002}. This method is robust in problems involving strongly magnetized regions or low plasma beta scenarios, as implied in this paper. Regarding the optically thin radiative losses, PLUTO treats it in a fractional step fashion \citep{Mignone_et_al_2007}. We obtained the radiative cooling function $\Lambda(T)$ using a table lookup/interpolation method. For instance, in Figure \ref{fig:chianti} we show the density averaged radiative loss rate function $\Lambda(T)$ based on CHIANTI coronal abundances used in the numerical simulations of this paper. Finally, PLUTO evolves the thermal conduction separately from advection through operator splitting, using the super-time-stepping technique \citep{Alexiades_et_al_1996}. This method is robust enough to handle the highly parabolic (diffusion) behavior of thermal conduction in the MHD equations.    

Our simulations are carried out in a domain  $x\in[0,30]$, $y\in[0,30]$, in units of Mm, covered by 1200$\times$1200 grid cells, leading to the effective resolution of 25 km in each direction.  Here, $y=0$ Mm represents the bottom of the photosphere. At the domain edges specified by $x=0$ Mm, $y=0$ Mm, and $y=30$ Mm, we impose fixed in time boundary conditions, which means that we set all the plasma quantities to their equilibrium values. Whereas at the side given by $x=30$ Mm, we use outflow boundary conditions to avoid numerical reflections. 

% ---------------------------------------------------
% ---------->    SUB-SECTION    <----------
% ---------------------------------------------------

\subsection{Perturbation}

We perturb the initial background state by a localized Gaussian pulse set in the vertical component of velocity \citep[see, e.g.][]{Muraswki&Zaqarashvilli_2010,Murawski_et_al_2011, Kuzma_et_al_2017a}. We define the velocity pulse as follows:

\begin{equation}
v_{y}(x,y) = A_{v}\exp\left(-\frac{(x-x_{0})^{2}+(y-y_{0})^{2}}{w^{2}}\right), \label{perturbation}
\end{equation}
where $A_{v}$ denotes the amplitude of the pulse, $(x_{0},y_{0})$ is its position, and $w$ its width. We set and hold fixes the values of $x_{0}$ and $y_{0}$ to 5.0 and 1.75 Mm, respectively. Therefore, we launch the perturbation in the solar chromosphere below the transition region. In our simulations, we set $w=0.2$ Mm and $A_{v}=100$ km s$^{-1}$ which triggers similar jets to those obtained in \citet{Murawski_et_al_2011}. Note that this amplitude is higher than in \citet{Murawski_et_al_2011}; however, we do not use a high spatial resolution or static mesh refinement. These features would help avoid dissipation at the transition region and numerical reflections from the top of the domain. We use a higher velocity pulse to excite a faster jet and not let dissipation or numerical reflections dominate the system's evolution on timescales of interest and minimize numerical issues. Also, the initial amplitude of $100$ km s$^{-1}$ is the range of the average initial velocities from 70 to 140 km s$^{-1}$ estimated by \citet{Loboda&Bogachev_2019} through a statistical analysis of macrospicule jets observed in coronal-holes and the quiet Sun.

% -------------------------------------------
% ----->     SECTION     <-----
% -------------------------------------------
\section{Results of numerical simulations and discussion}
\label{sec:results_numerical_simulations}

We let the system evolve according to Equations (\ref{density})-(\ref{divergenceB}); for that, we apply a perturbation in the vertical component of velocity given by Equation (\ref{perturbation}) to trigger the collimated jet structures. Supposing an energy imbalance background between thermal conduction and radiative cooling we perform four different tests for the following scenarios: i) adiabatic case, ii) thermal conduction case (TC), iii) radiative cooling case (RC), and iv) thermal conduction + radiative cooling case (TC+RC). 

% -------------------------------------------
% ----->    SUB-SECTION     <-----
% -------------------------------------------
\subsection{Adiabatic case}
\label{subsec:results_adiabatic_case}

In particular, \cite{Murawski_et_al_2011} already studied this case; however, we decided to present this simulation again to highlight the difference with the other cases that include thermal conduction and radiative cooling more clearly. Therefore, in this subsection, we present some of the most representative results of the numerical simulations obtained in the adiabatic case. For instance, in the first column (a) of Figure \ref{fig:temp_evolution_comparison}, we show temporal snapshots of the plasma temperature and the velocity vector field. In particular, at the time given by $t=100$ s, we see the formation of a collimated jet that reaches a height of about 7.5 Mm. At $t=400$ s, this jet is getting smaller and hotter; it also shows a double structure at the top. This double-thread structure on spicule jets have been already observed and reproduced by numerical simulations \citep{Tanaka_1974,Dara_et_al_1998,Suematsu_et_al_2008,Murawski_et_al_2011,Kayshap_et_al_2013}. At $t=800$ s, the jet with the double-thread structure remains at the height of about 5 Mm and the velocity vector field shows a bi-directional flow, which is also consistent with the observed dynamics of spicule jets \citep{Tsiropoula_et_al_1994,Tziotziou_et_al_2003,Tziotziou_et_al_2004,Pasachoff_et_al_2009}. Note that the jets obtained in this case are smaller than the jets of Figure 5 of \cite{Murawski_et_al_2011}. However, the double-structure and bi-directional flows are consistent with previous results. Also, the maximum height of 7.5 Mm, a lifetime of around 10 min, and the speed of $\approx$ 50 km s$^{-1}$ are consistent with the observational data \citep{Georgakilas_et_al_1999, Sterling_2000}.  
     
% -------------------------------------------
% ----->    SUB-SECTION     <-----
% -------------------------------------------
\subsection{Thermal conduction case}
\label{subsec:results_thermal_conduction}

In this case, we show the results obtained from the numerical simulations that consider thermal conduction's effect along the magnetic field lines acting separately to radiative cooling. This effect has been found to slightly affect the temperature and morphology of type II spicule \citep[see, e.g.][]{Gonzalez-Aviles_et_al_2020}, and increase the energy and mass fluxes on solar tadpole-like jets \citep[see, e.g.][]{Navarro_et_al_2020}. According to our results, for example, in the second column (b) of Figure \ref{fig:temp_evolution_comparison}, we show snapshots of the plasma temperature and the velocity vector field at the same three times as for the adiabatic case. At $t=100$ s, we see a collimated jet structure reaching a height of about 6 Mm, which is smaller compared to the adiabatic case for the same time. At $t=400$ s, the jet reveals a small double-thread structure, and it is also evident that the velocity is higher than in the adiabatic case. This flow is generated by energy dissipation due to the thermal conduction, since it is highly efficient at redistributing the excess of energy throughout the solar corona \citep{Antolin_2020}. At $t=800$ s, the jet is smaller than in the adiabatic case; this jet neither shows double structure nor bi-directional flows. Instead, plasma flow moves upwards in the solar corona. Again, we could also relate the flow to the dissipation of the heat flux's energy produced. This effect is more visible in the regions surrounding the jet and especially in the shock front, as shown in Figure \ref{fig:mag_q_thermal_conduction_case}, where we display the magnitude of the heat flux $|{\bf q}|$ in W s$^{-2}$ at three times. Another feature observed is that heat flux distributes at the jet's boundaries. This feature could occur because the magnetic field and temperature gradient ${\nabla T}$ vector field are mainly perpendicular inside the jet structure and predominantly parallel to the surroundings, as shown in Figure \ref{fig:mag_q_thermal_conduction_case}.   

% -------------------------------------------
% ----->    SUB-SECTION     <-----
% -------------------------------------------
\subsection{Radiative cooling case}
\label{subsec:results_optically_thin_radiative_losses}

In this case, the radiative cooling function $\Lambda(T)$ describes an averaged radiative loss rate based on CHIANTI coronal abundances, as shown in Figure \ref{fig:chianti}. In this context, there are some numerical simulations of jets and other phenomena in the solar corona, where the radiation is described by an optically thin radiative loss term balanced with an external heating term \citep[see, e.g.,][]{Guarrasi_et_al_2014, Fan_2017, Petralia_et_al_2014, Reale&Landi_2012, Rempel_2017}. However, in the thorough analysis of this subsection, we do not consider any external heating term that balances the radiative losses, which let the possibility that cooling exerts an effect on the jets' behavior. Even though, in this paper, we excite the jet in the chromosphere. It results interesting to see how in a scenario where the background atmosphere is not in energy balance the radiative cooling can affect the jet when it propagates over the transition region and reaches coronal heights ($\approx$ 3 Mm).

In the third column (c) of Figure \ref{fig:temp_evolution_comparison}, we show snapshots of the plasma temperature and the velocity vectors at the same three moments time for the adiabatic and thermal conduction test cases. For example, at $t=100$ s, we see the formation of a broader and smaller jet compared to the adiabatic and thermal conduction cases. This jet reaches a height of about 5.5 Mm. At $t=400$ s, we note that the jet almost disappears, showing small amplitude oscillations compared to the adiabatic and thermal conduction cases. We also see that neither double structure is observed nor bi-directional flows, but we note a strong downflow that directly affects the jet and produces dissipation. In general terms, we can argue that radiative cooling naturally induces a cooling of the coronal plasma, which in turn decreases the gravitational scale height in the corona, leads to a mass flux towards the chromosphere. At $t=800$ s, the jet practically disappears, and we only see a tiny structure that remains just over the transition region. An important feature that we should analyze is how strong radiative cooling behaves around the jet. At the top of Figure \ref{fig:Qrad_mass_density_rad_cooling_case}, we show snapshots of the $-Q_{rad}=n_{H}\Lambda(T)$ given in W m$^{-3}$, at three moment of time. In this case, a positive value means that there is cooling. At $t=100$ s, plasma cooling is substantial over the transition region and at the jet's boundaries, but not inside on it. At $t=300$ s, the cooling grows and coincides with the time when the jet dissipates. At $t=800$ s, the cooling remains high near the transition region. This cooling behavior near the transition region has already been observed in advanced radiation MHD numerical simulations of jet-like structures in the solar atmosphere \citep{Leenaarts_2020, Rempel_2017}. Besides the radiative losses in the transition region (TR) and corona tend to have a sharp peak in the TR owing to the quadratic density-dependence \citep{Leenaarts_2020}. At the bottom of Figure \ref{fig:Qrad_mass_density_rad_cooling_case}, we show the mass density $\varrho$, in kg m$^{-3}$, at the same three moments ($t=100,200,800$ s) as the snapshots of $-Q_{rad}$. In these three snapshots, we see a denser jet structure that behaves accordingly as the cooling term does. This behavior is physically accepted since the cooling mainly depends on mass density; therefore, it is strong where density is high. Note that even the cooling is high and act as downflow on the jet, the temperature at the coronal level remains practically unchanged with values of the order $10^{6}$ K as shown in the set of sub-figures of column (c) of Figure \ref{fig:temp_evolution_comparison}.

% -------------------------------------------
% ----->    SUB-SECTION     <-----
% -------------------------------------------
\subsection{Thermal conduction with radiative cooling case}
\label{subsec:results_thermal_cond_rad_cooling_case}

\begin{figure*}
\centering
\centerline{\Large \bf   
      \hspace{0.07\textwidth}  \color{black}{\normalsize{(a) Adiabatic}}
      \hspace{0.1\textwidth}  \color{black}{\normalsize{(b) Thermal conduction}}
      \hspace{0.06\textwidth}  \color{black}{\normalsize{(c) Radiative cooling}}
      \hspace{0.095\textwidth}  \color{black}{\normalsize{(d) TC + RC}}
         \hfill}
\includegraphics[width=4.3cm,height=5.5cm]{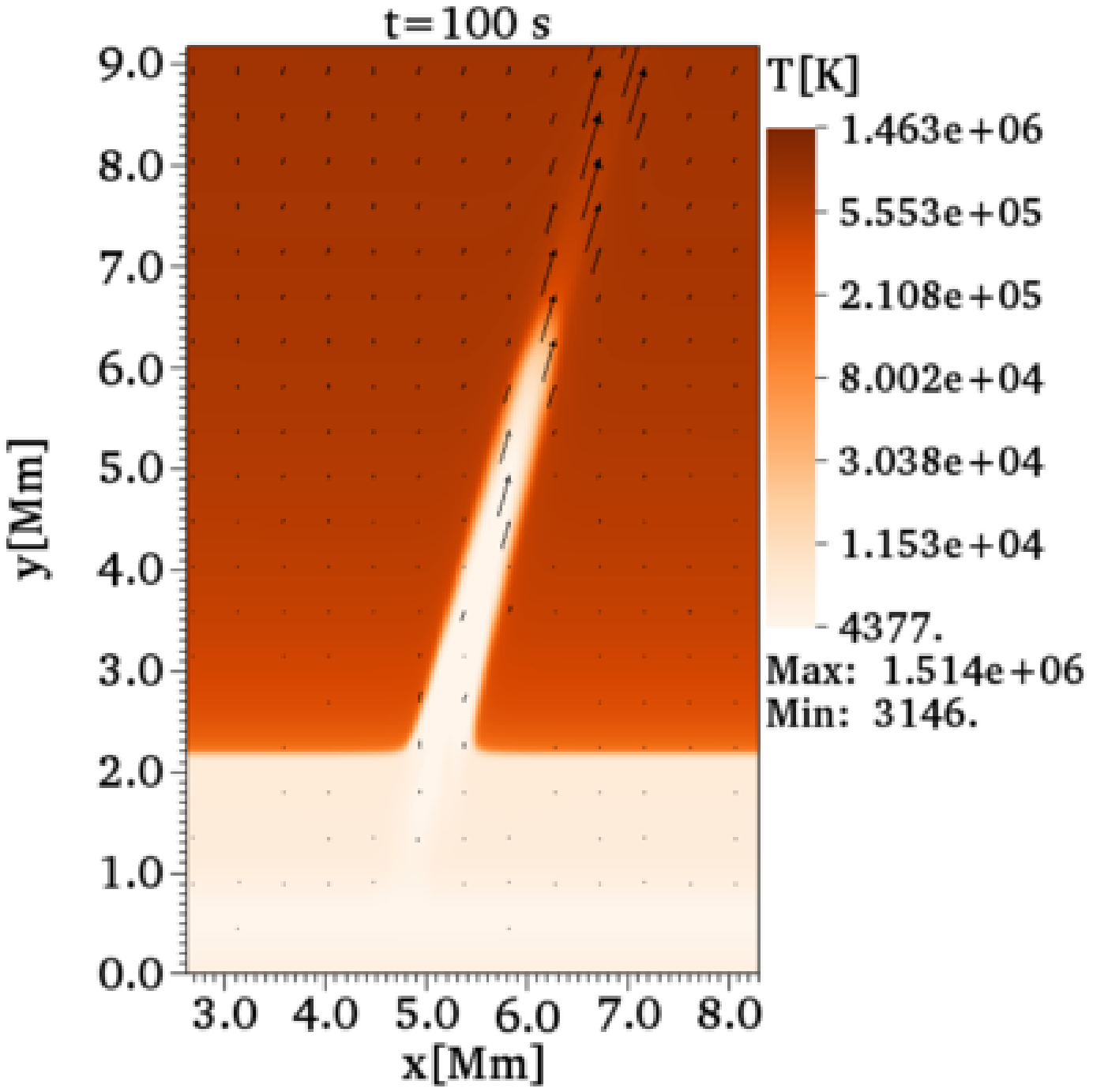}
\includegraphics[width=4.3cm,height=5.5cm]{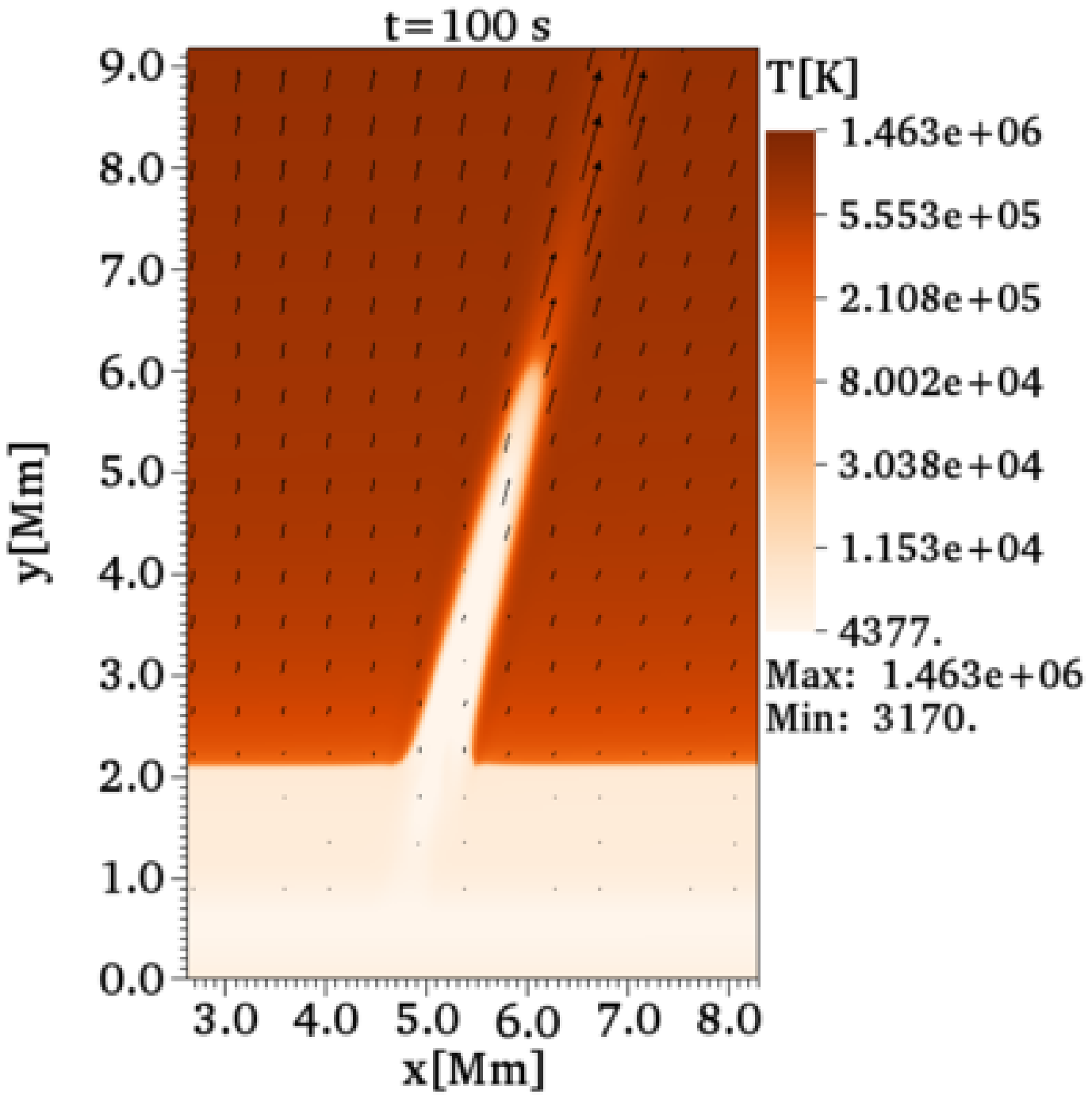}
\includegraphics[width=4.3cm,height=5.5cm]{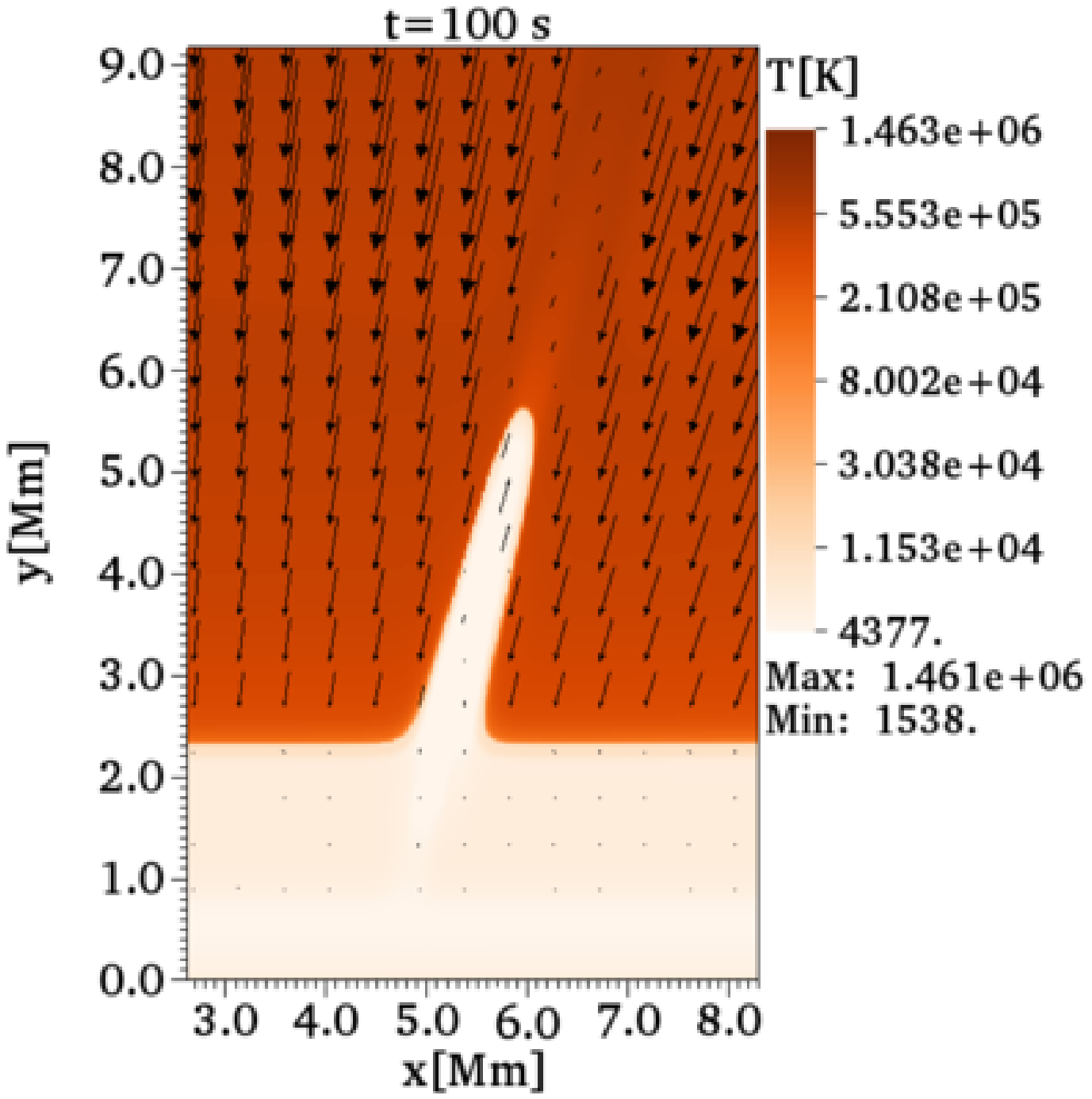}
\includegraphics[width=4.3cm,height=5.5cm]{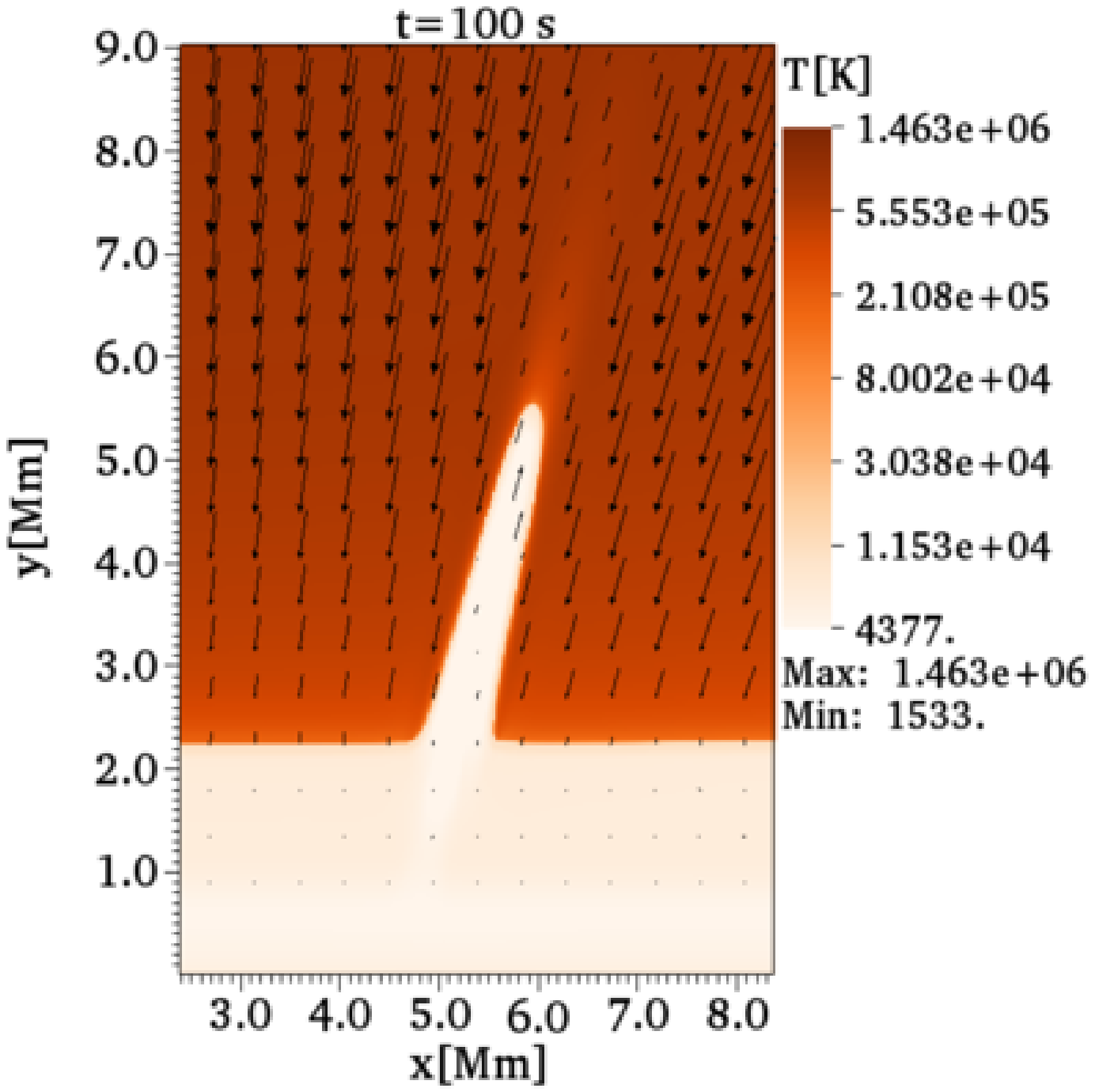}\\
\includegraphics[width=4.3cm,height=5.5cm]{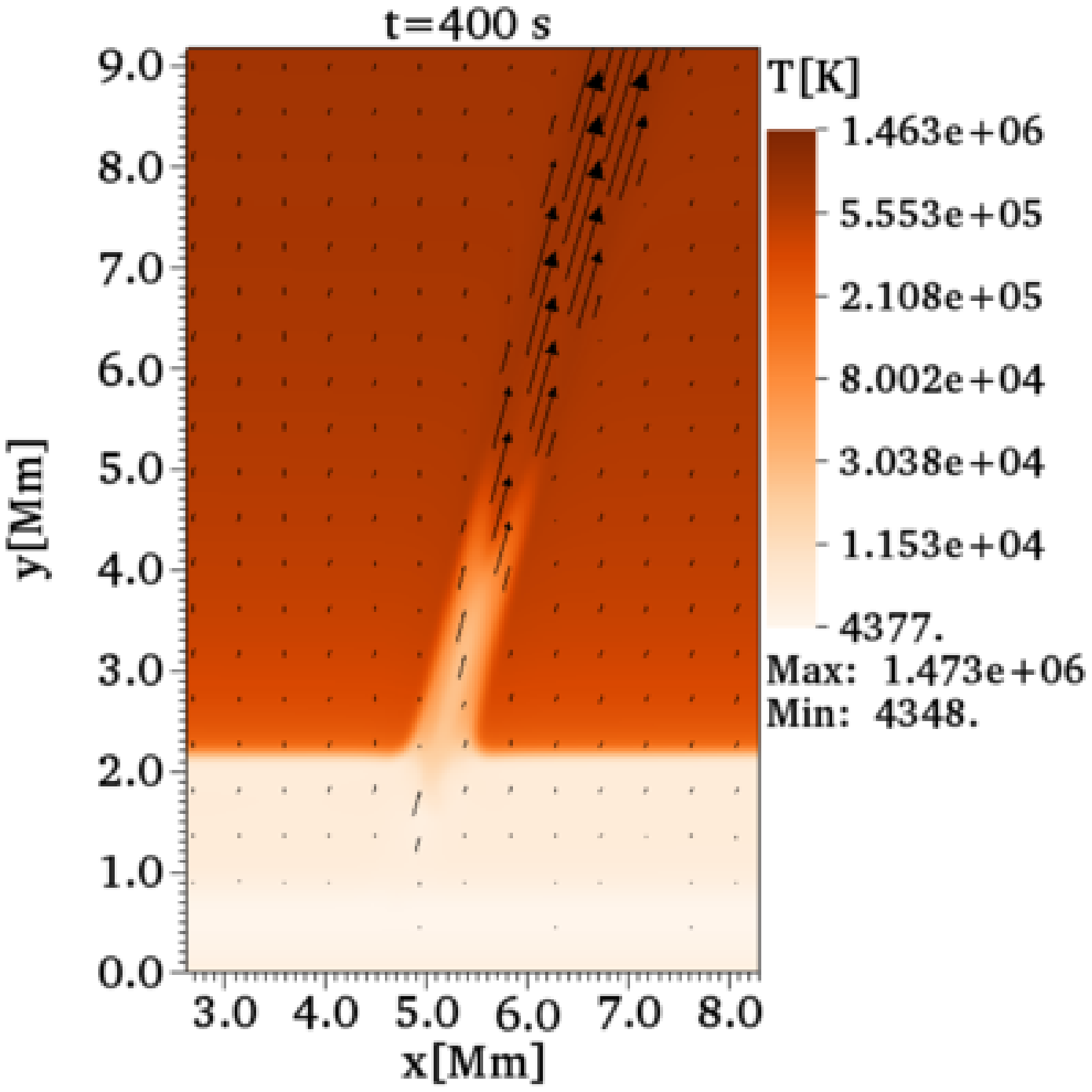}
\includegraphics[width=4.3cm,height=5.5cm]{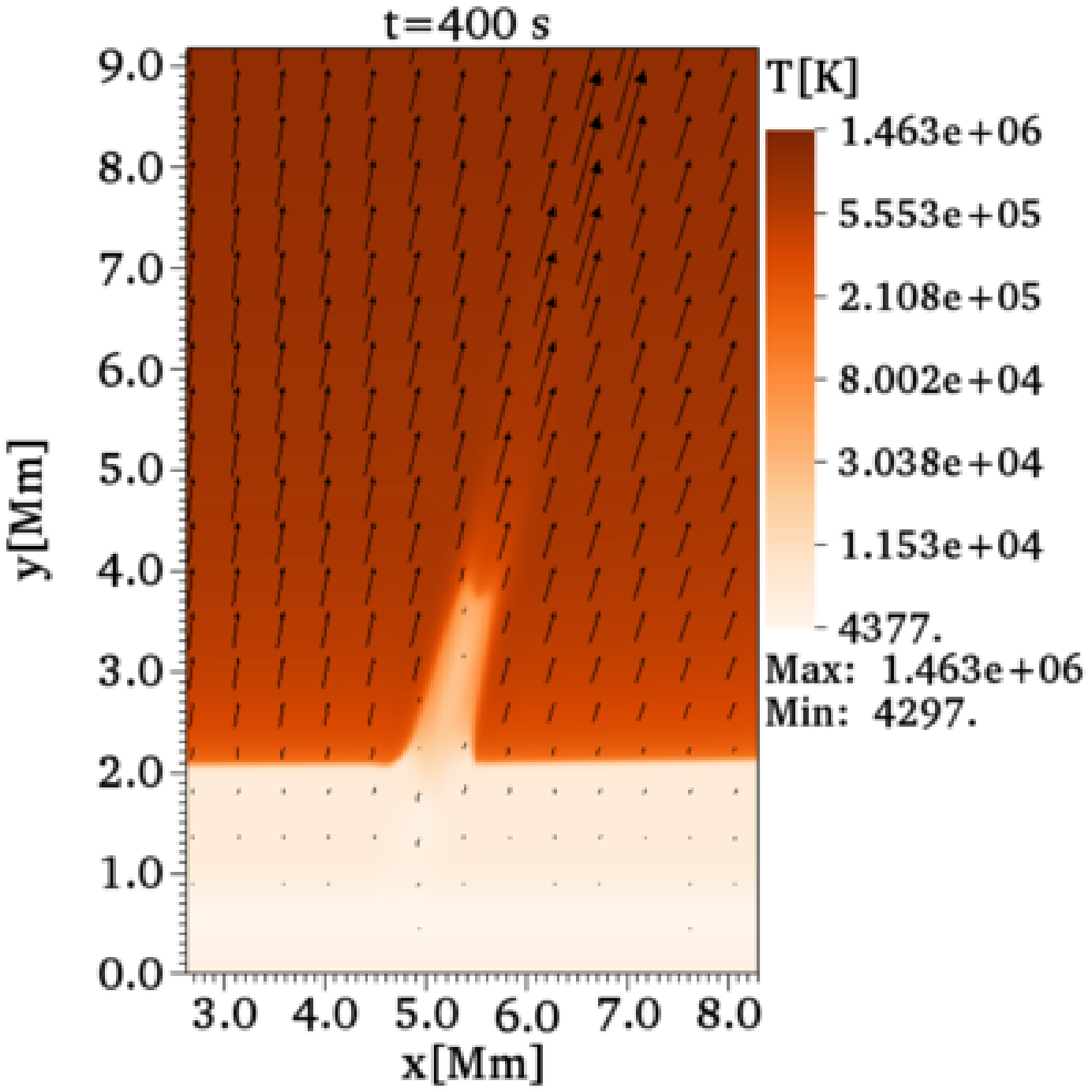}
\includegraphics[width=4.3cm,height=5.5cm]{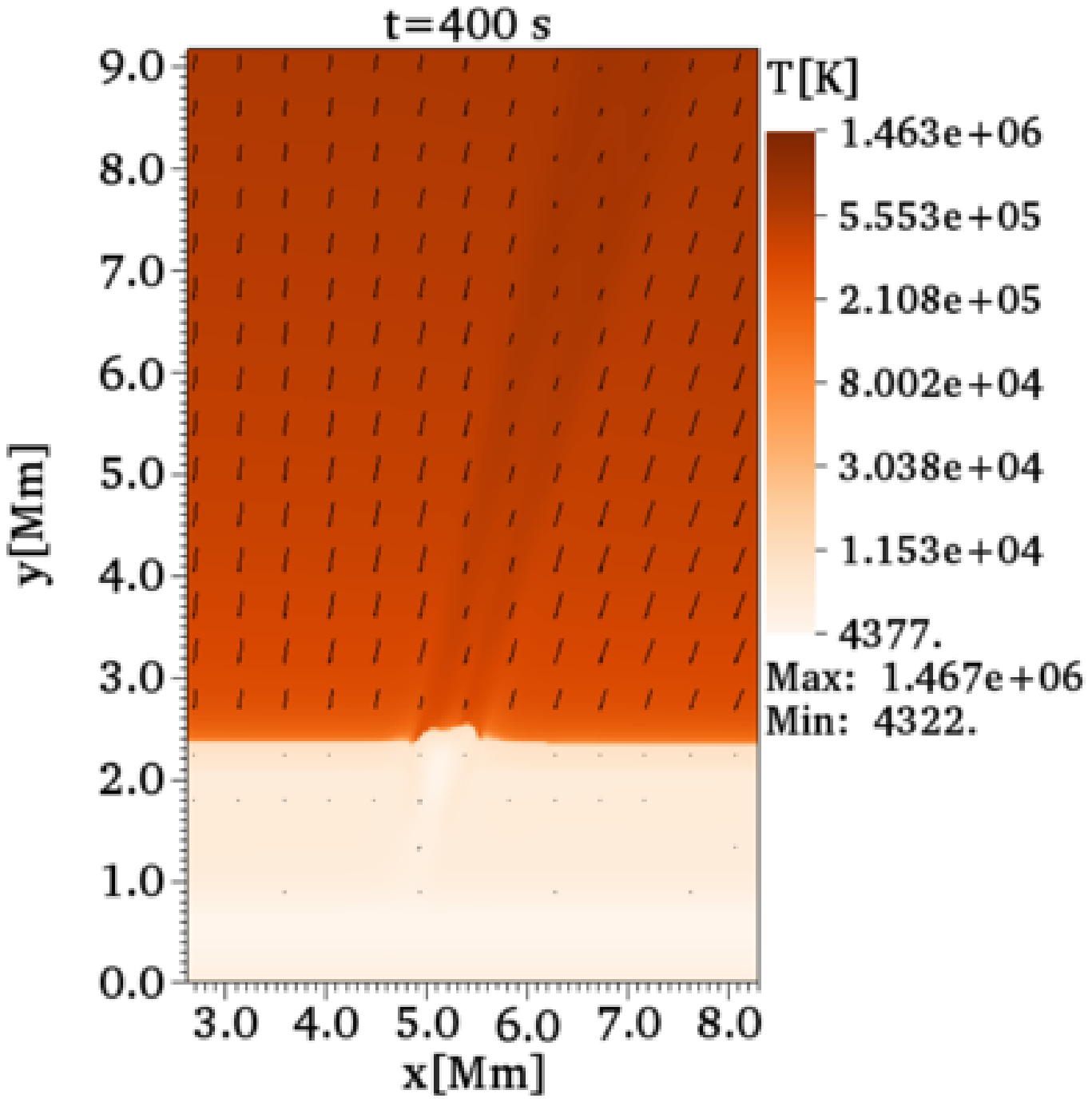}
\includegraphics[width=4.3cm,height=5.5cm]{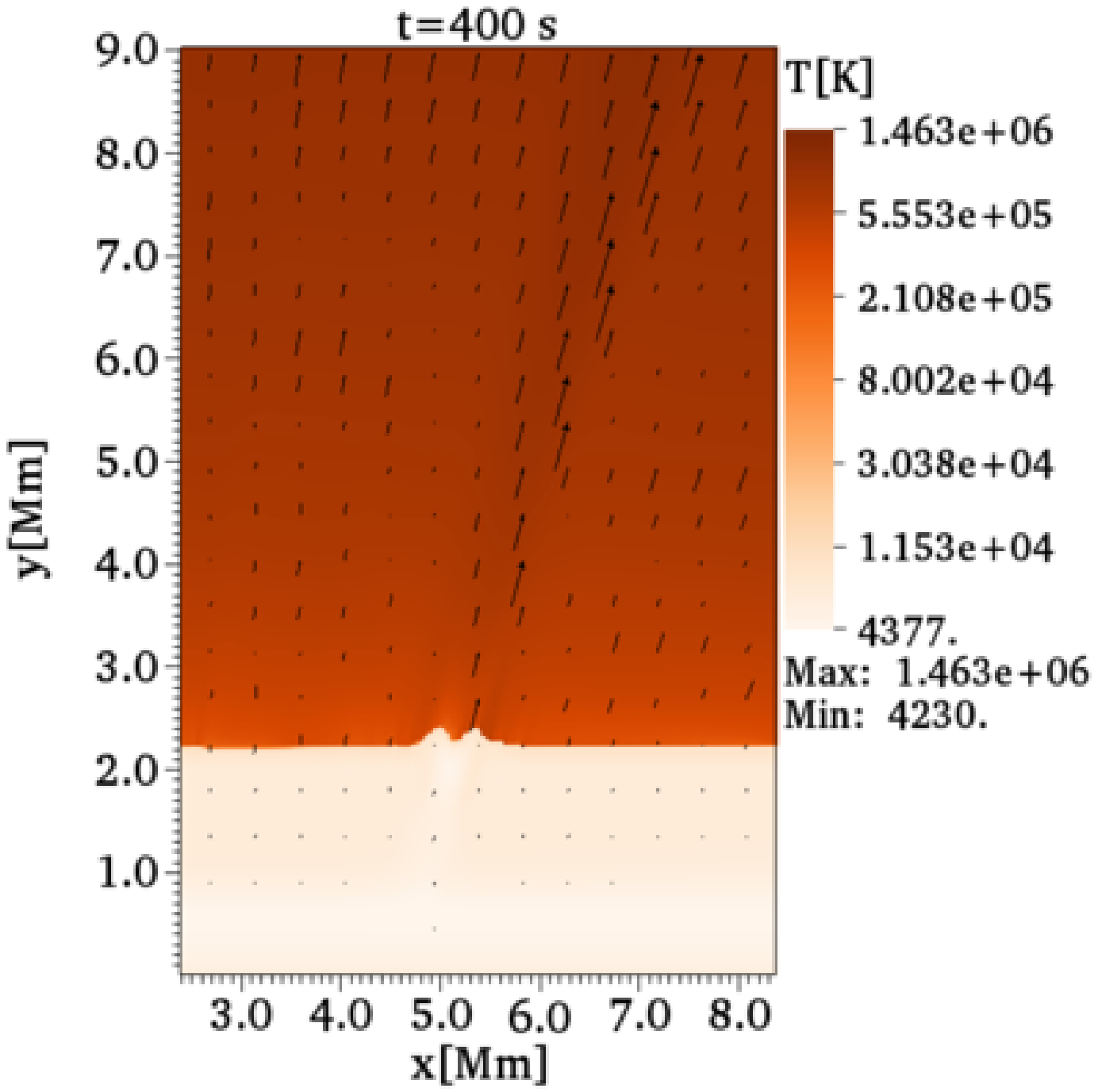}\\
\includegraphics[width=4.3cm,height=5.5cm]{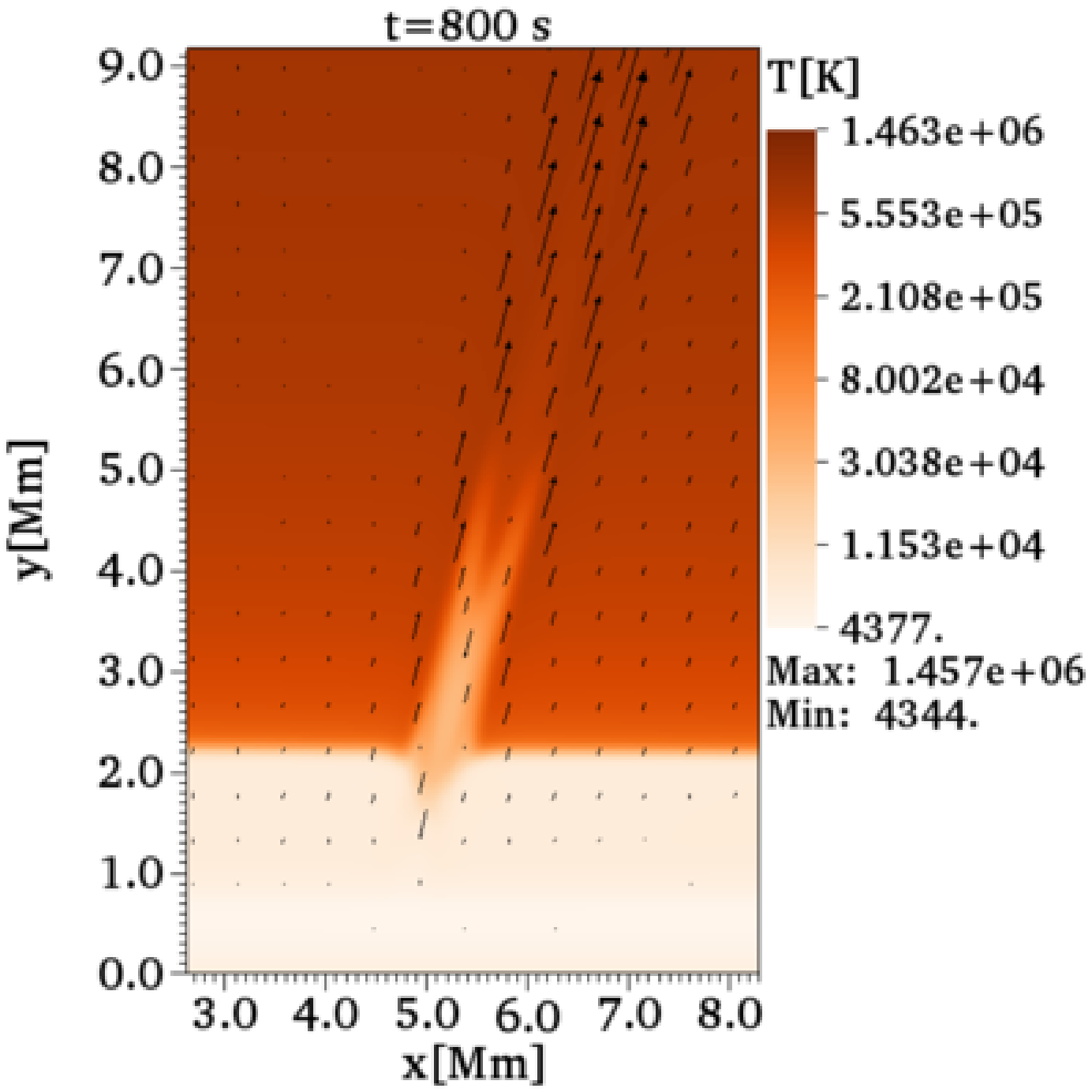}
\includegraphics[width=4.3cm,height=5.5cm]{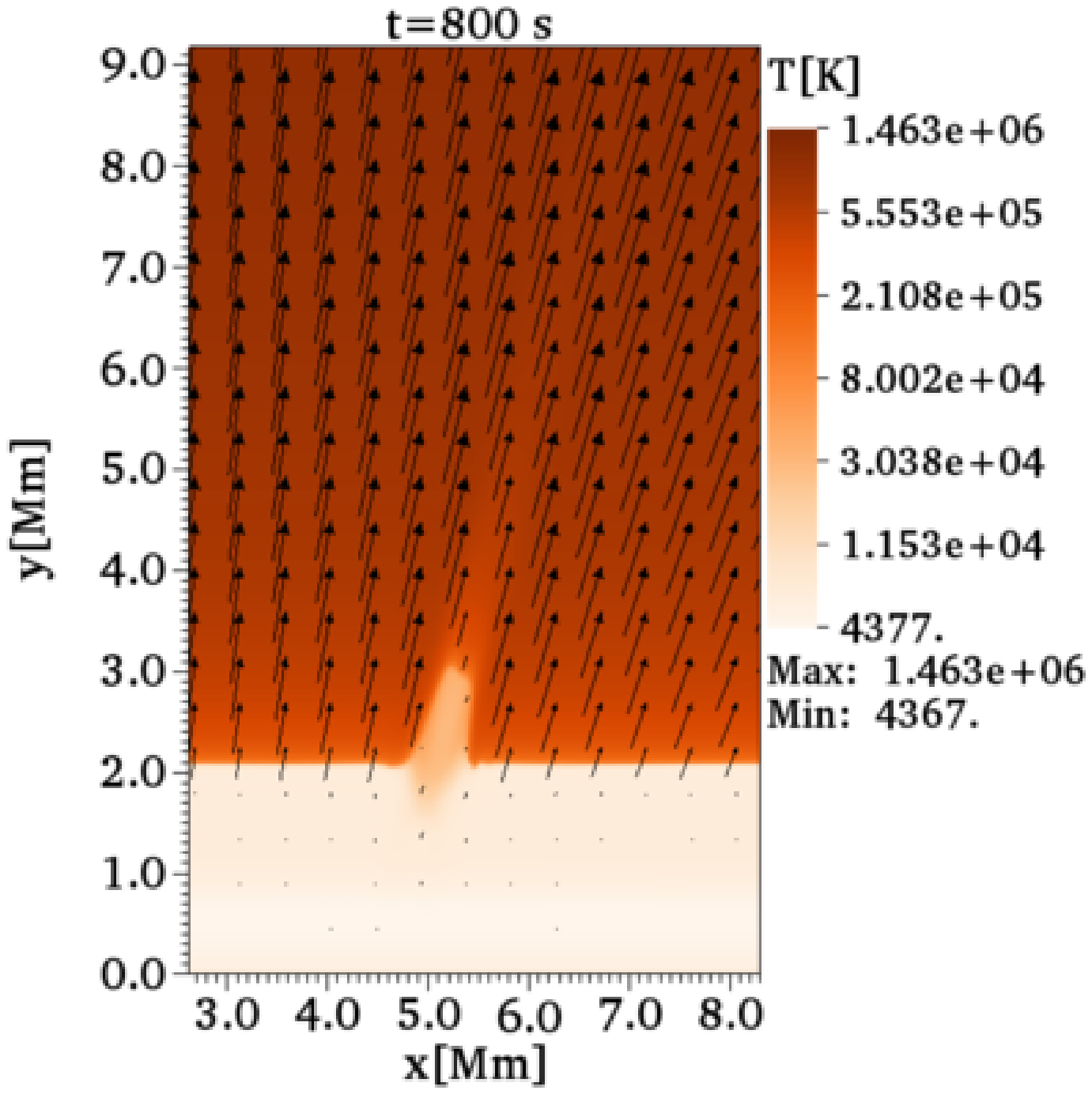}
\includegraphics[width=4.3cm,height=5.5cm]{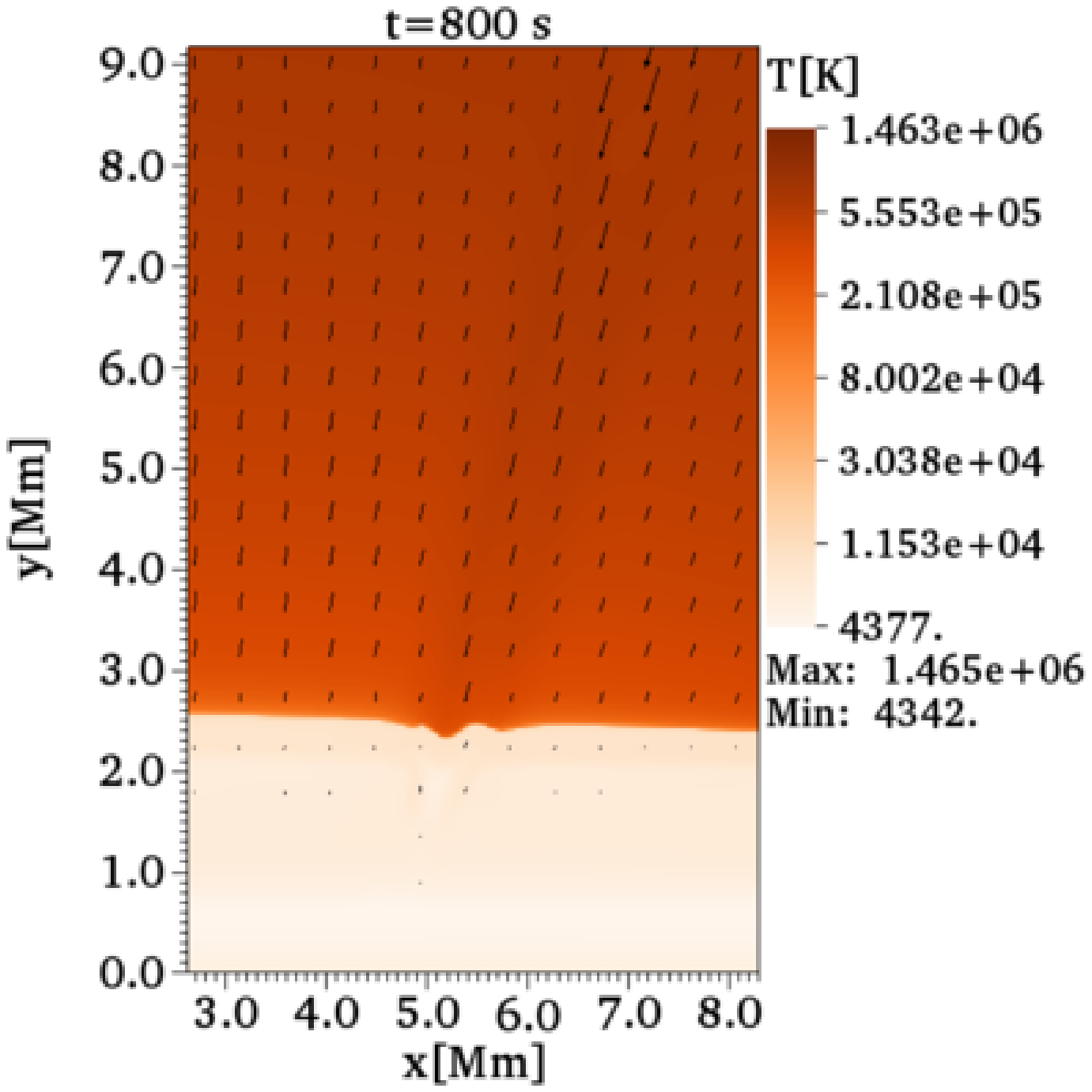}
\includegraphics[width=4.3cm,height=5.5cm]{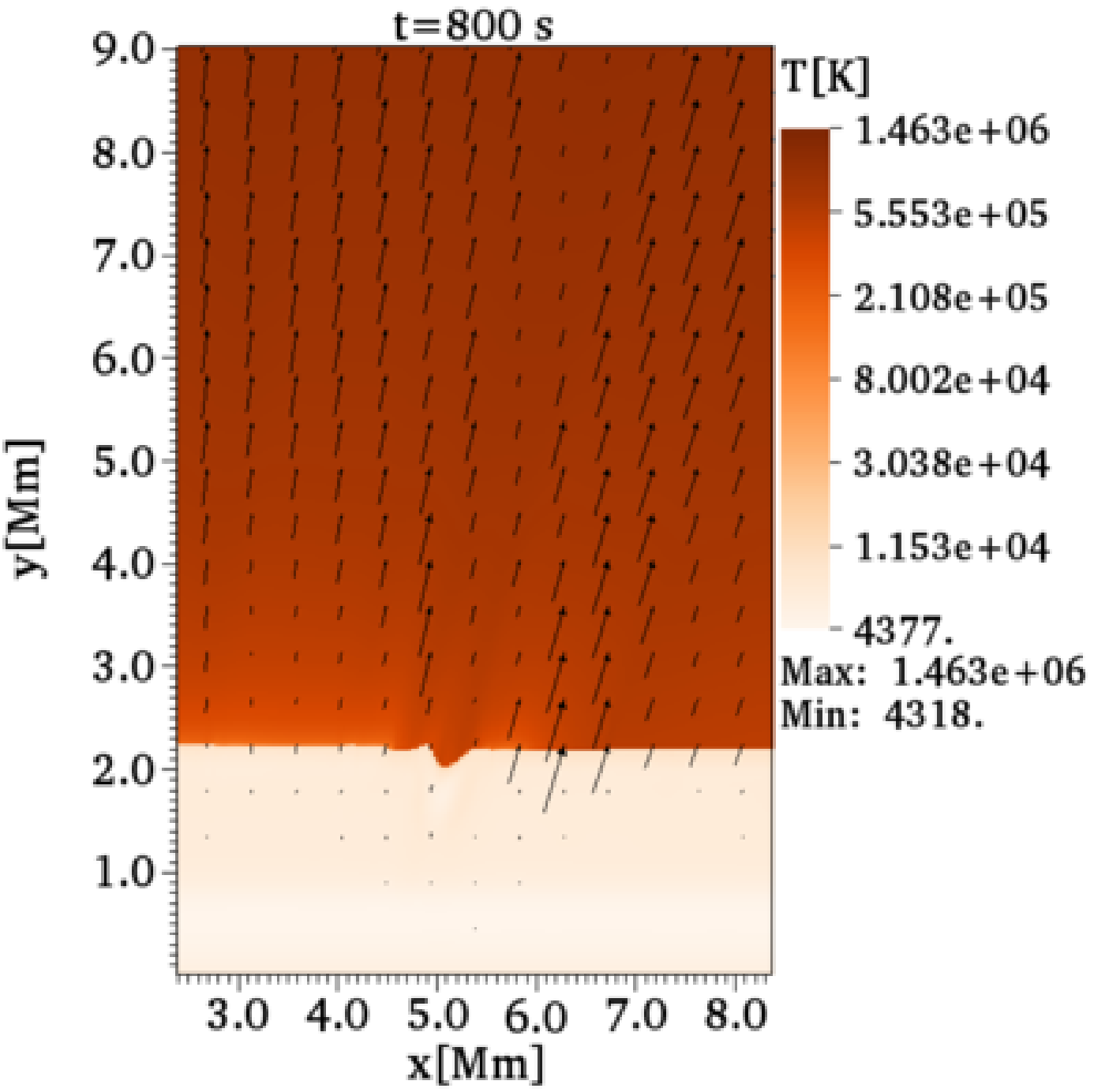}
\caption{Spatial profiles of temperature (in kelvin) and velocity vectors (in arbitrary units) at $t=100$ s, $t=400$ s and $t=800$ s for the adiabatic case (first column), thermal conduction (second column), radiative cooling (third column), and thermal conduction + radiative cooling (fourth column) cases.}
\label{fig:temp_evolution_comparison}
\end{figure*}

\begin{figure*}
\centering
\includegraphics[width=5.0cm,height=6.0cm]{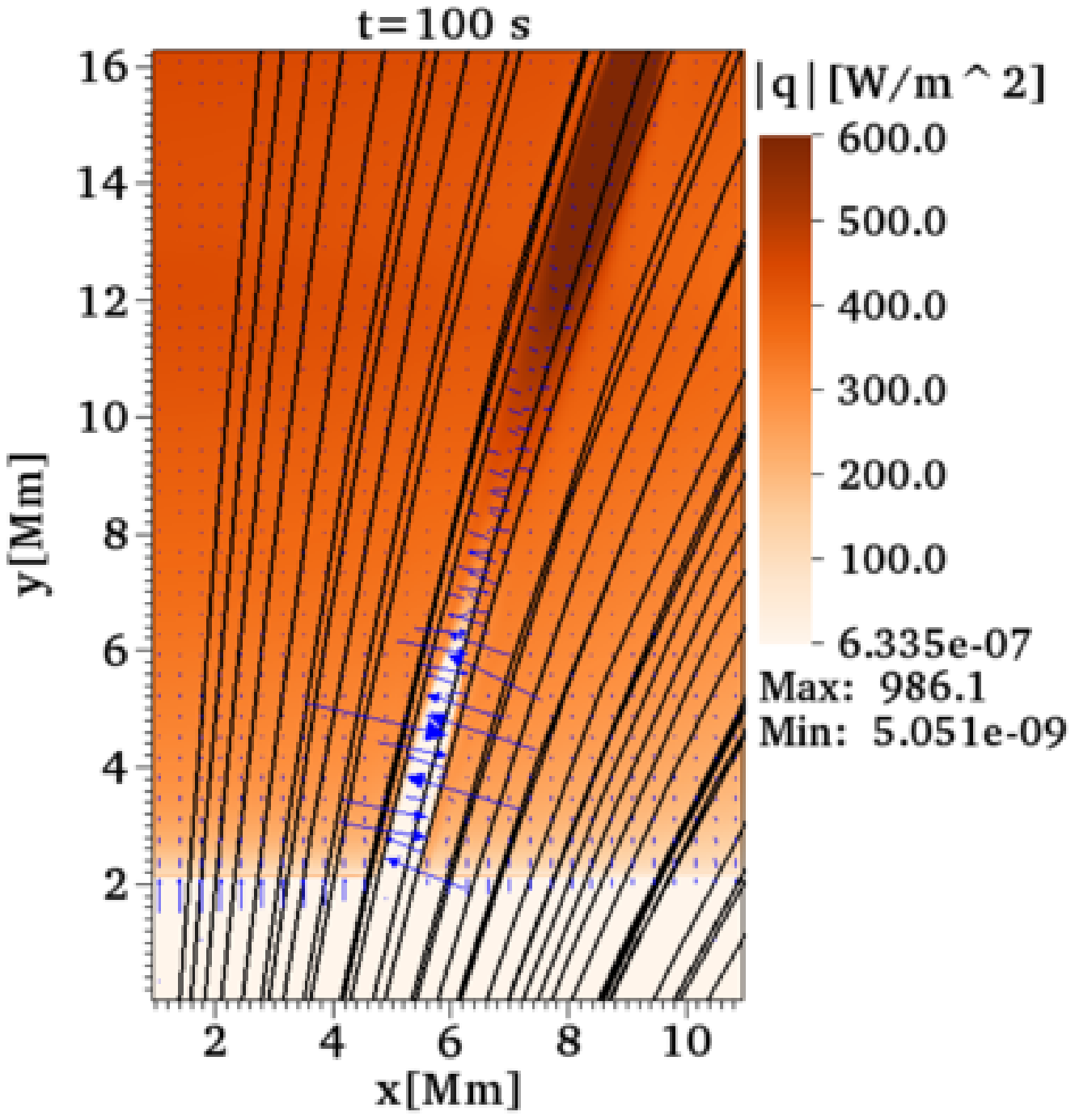}
\includegraphics[width=5.0cm,height=6.0cm]{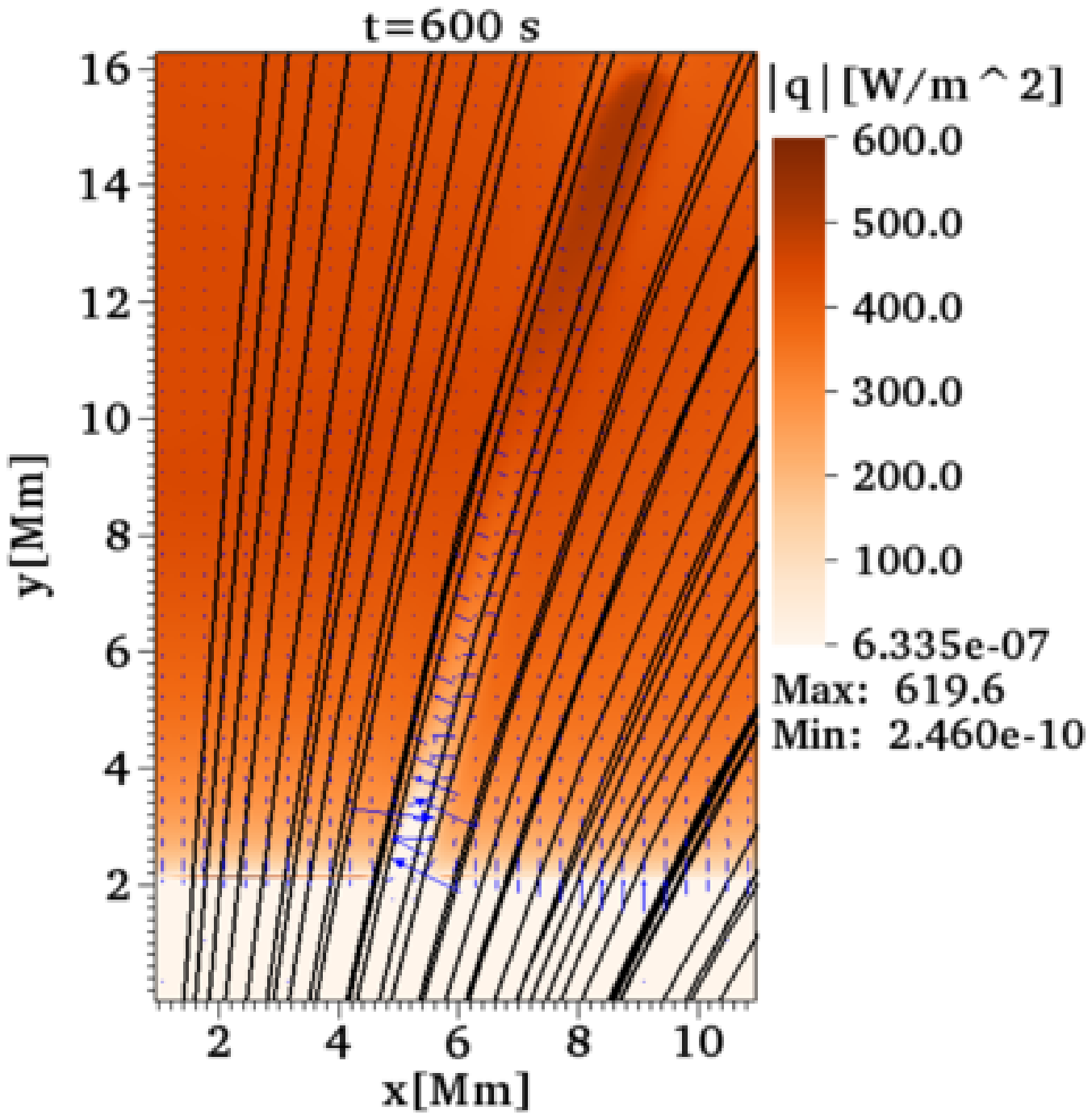}
\includegraphics[width=5.0cm,height=6.0cm]{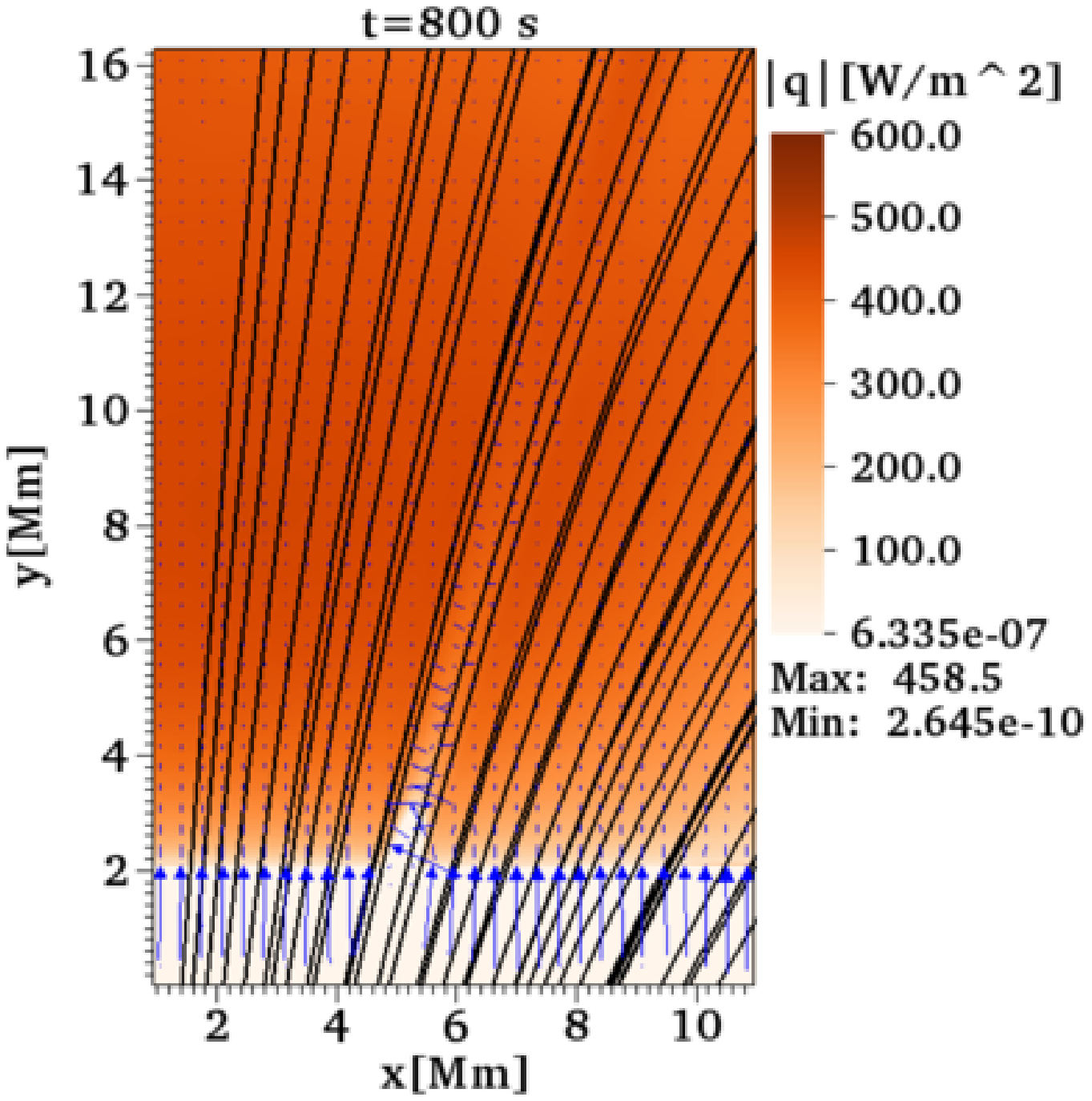}
\caption{{\bf Thermal conduction case}. Snapshots of the magnitude of heat flux $|{\bf q}|$, in W m$^{-2}$, with the vector field of $\nabla T$ represented by blue arrows and the magnetic field lines in black at $t=$ 100, 600, and 800 s.}
\label{fig:mag_q_thermal_conduction_case}
\end{figure*}

\begin{figure*}
\centering
\includegraphics[width=5.0cm,height=6.0cm]{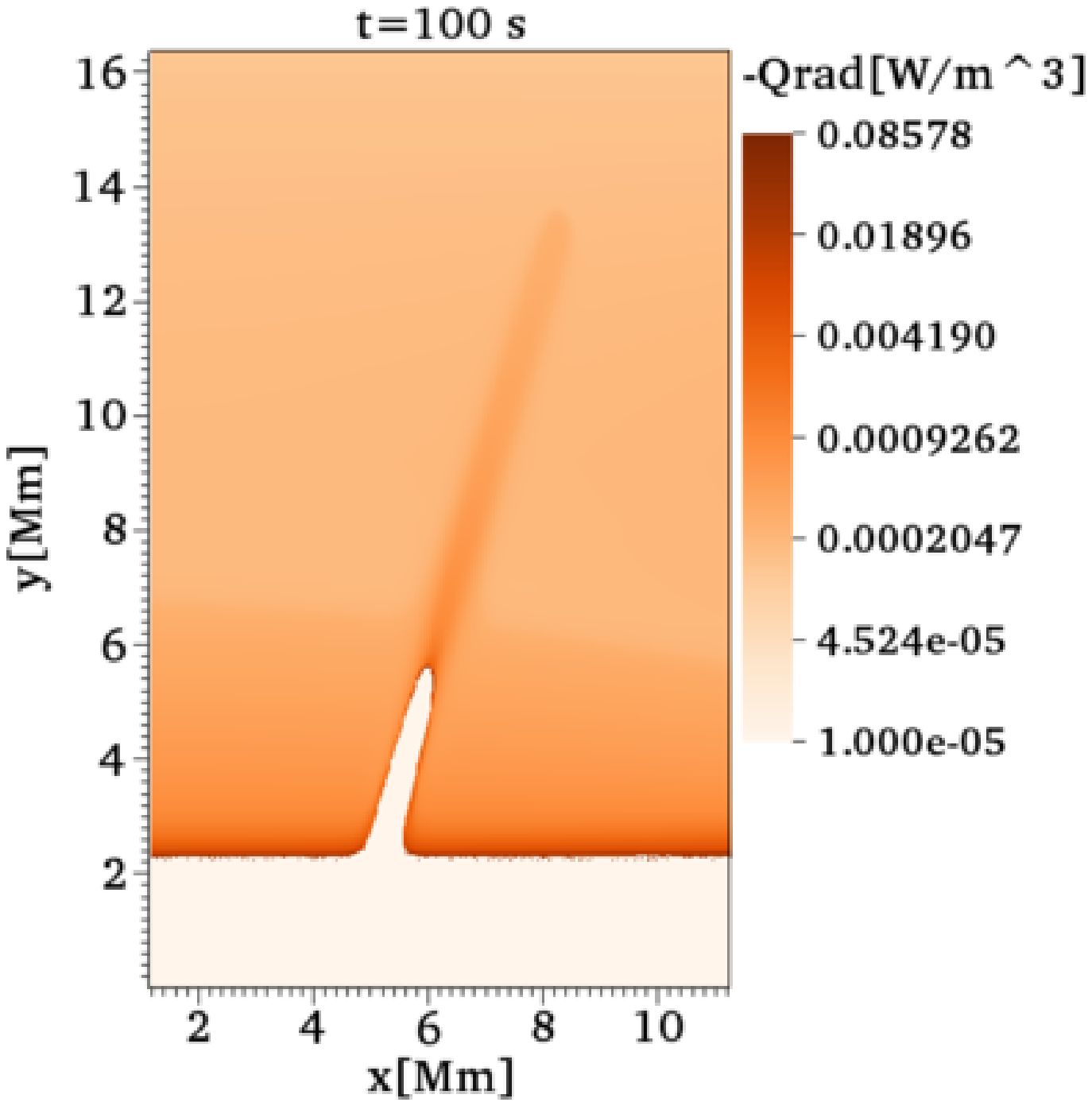}
\includegraphics[width=5.0cm,height=6.0cm]{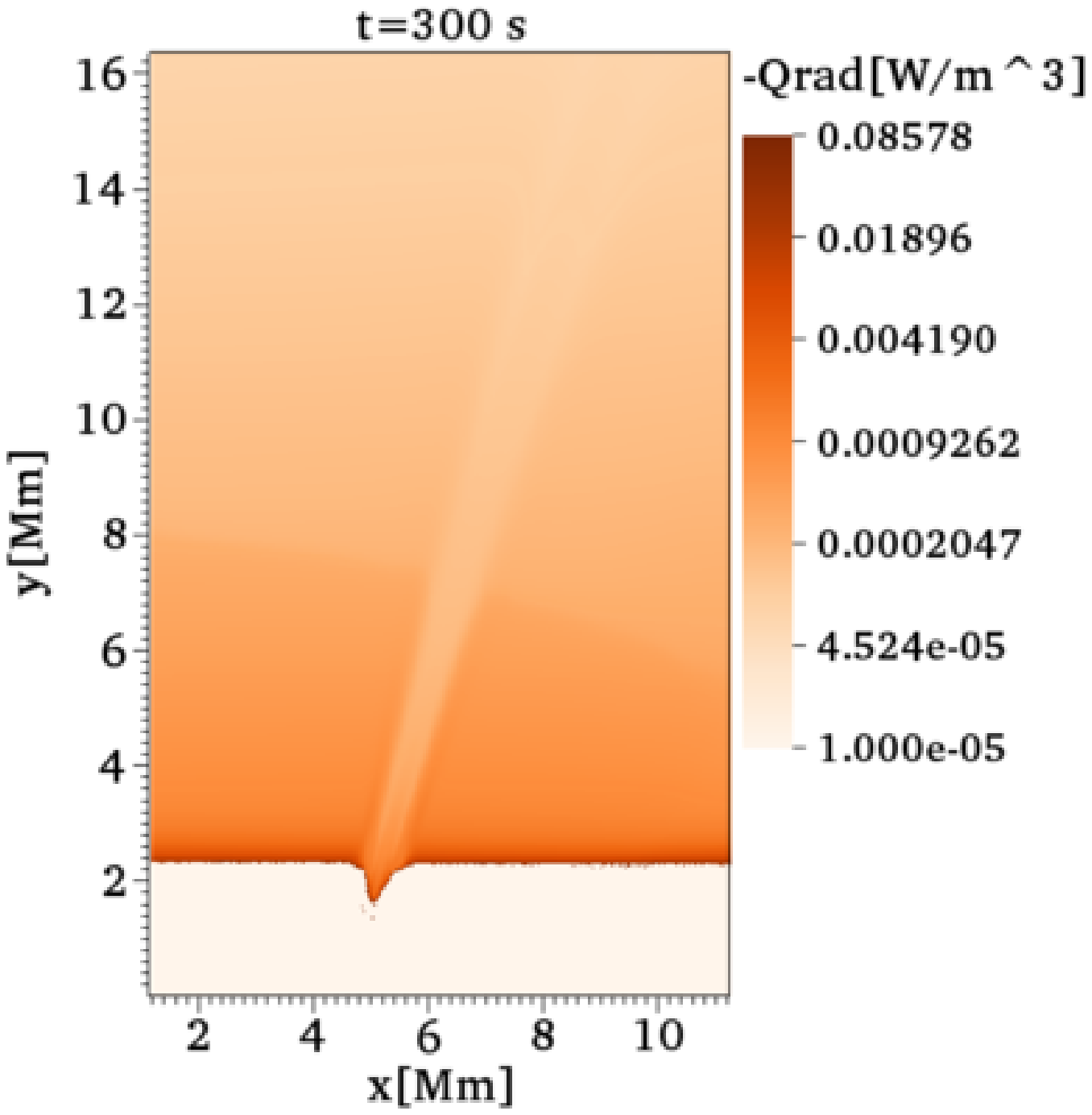}
\includegraphics[width=5.0cm,height=6.0cm]{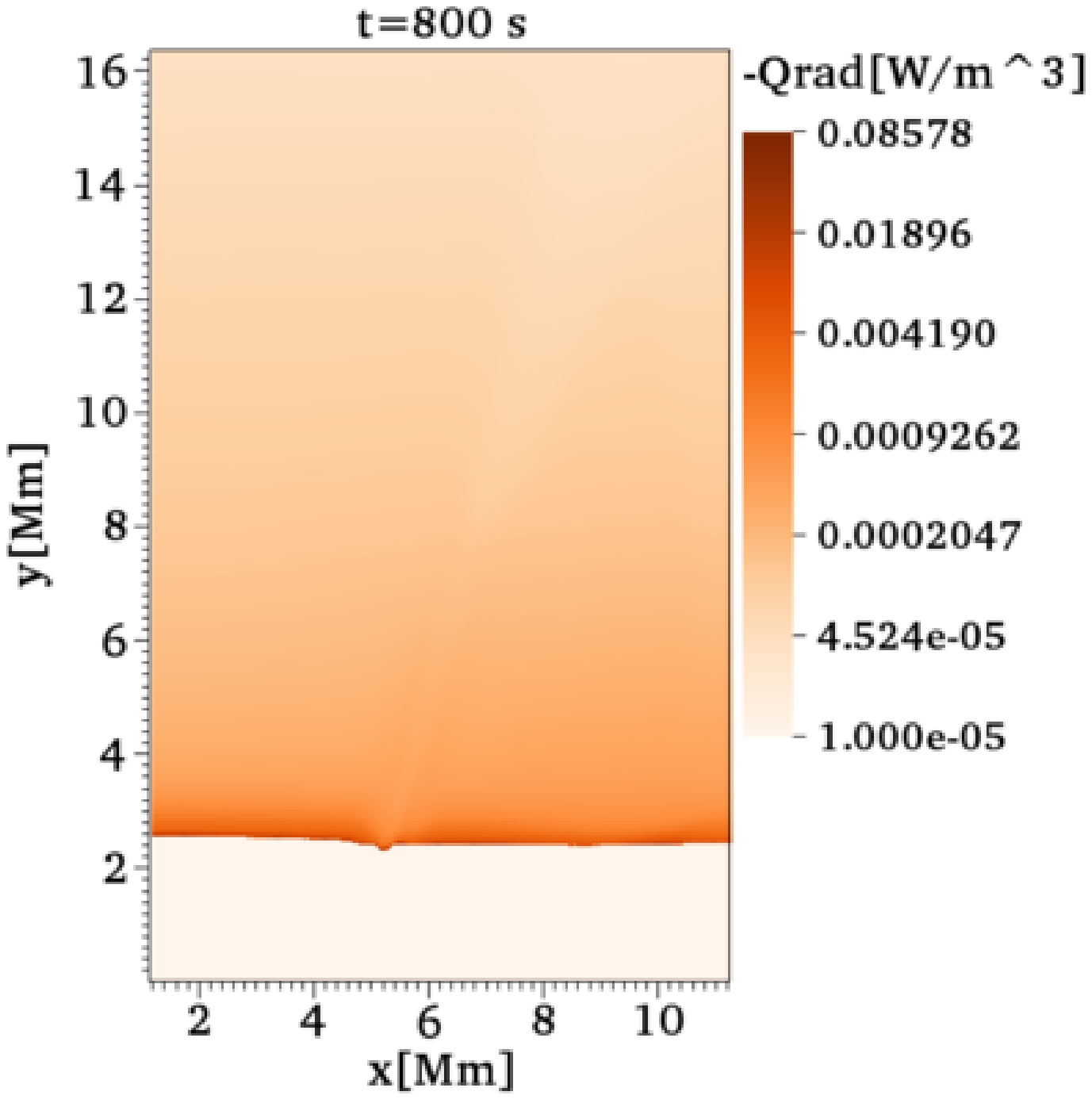}\\
\includegraphics[width=5.0cm,height=6.0cm]{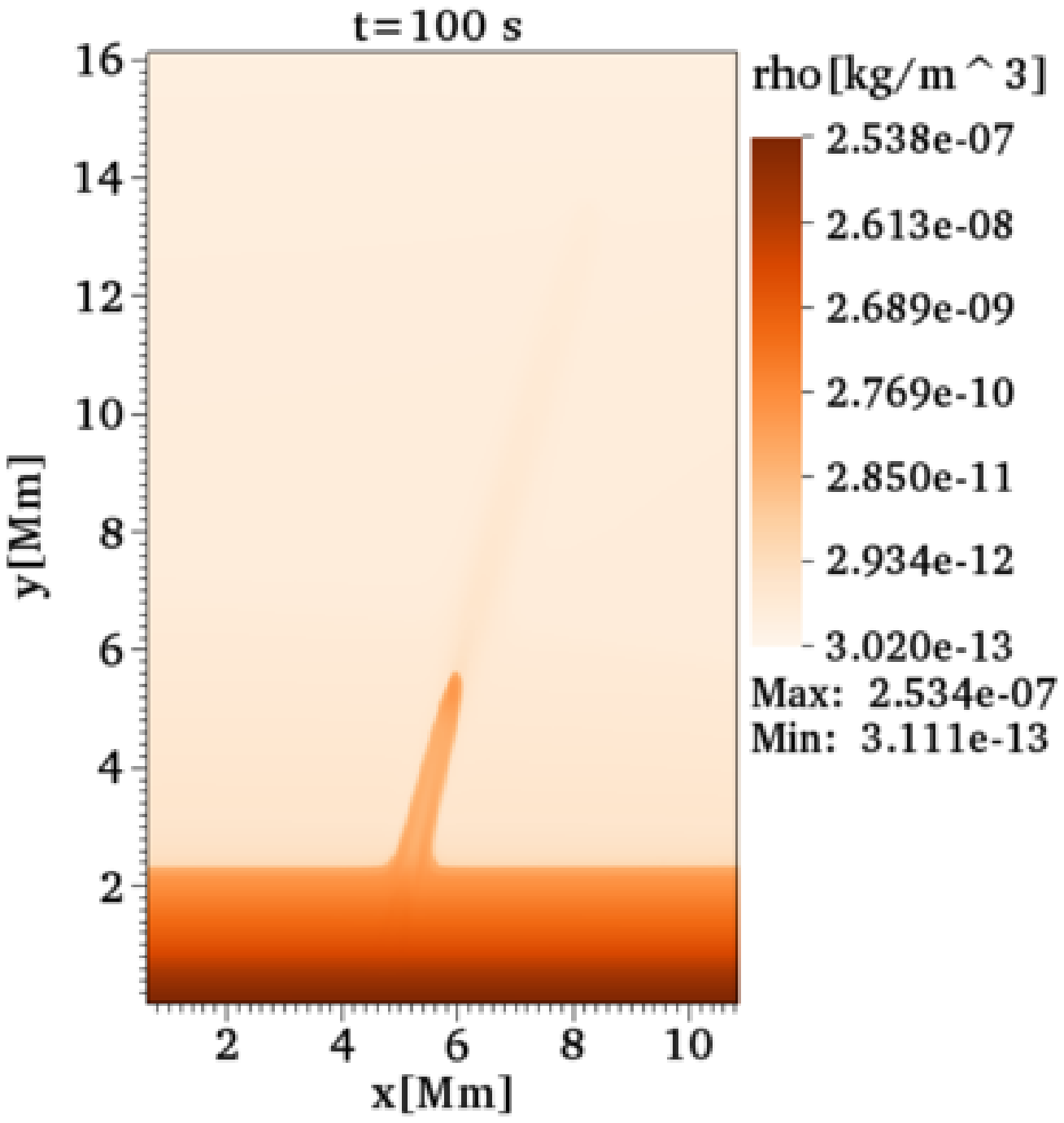}
\includegraphics[width=5.0cm,height=6.0cm]{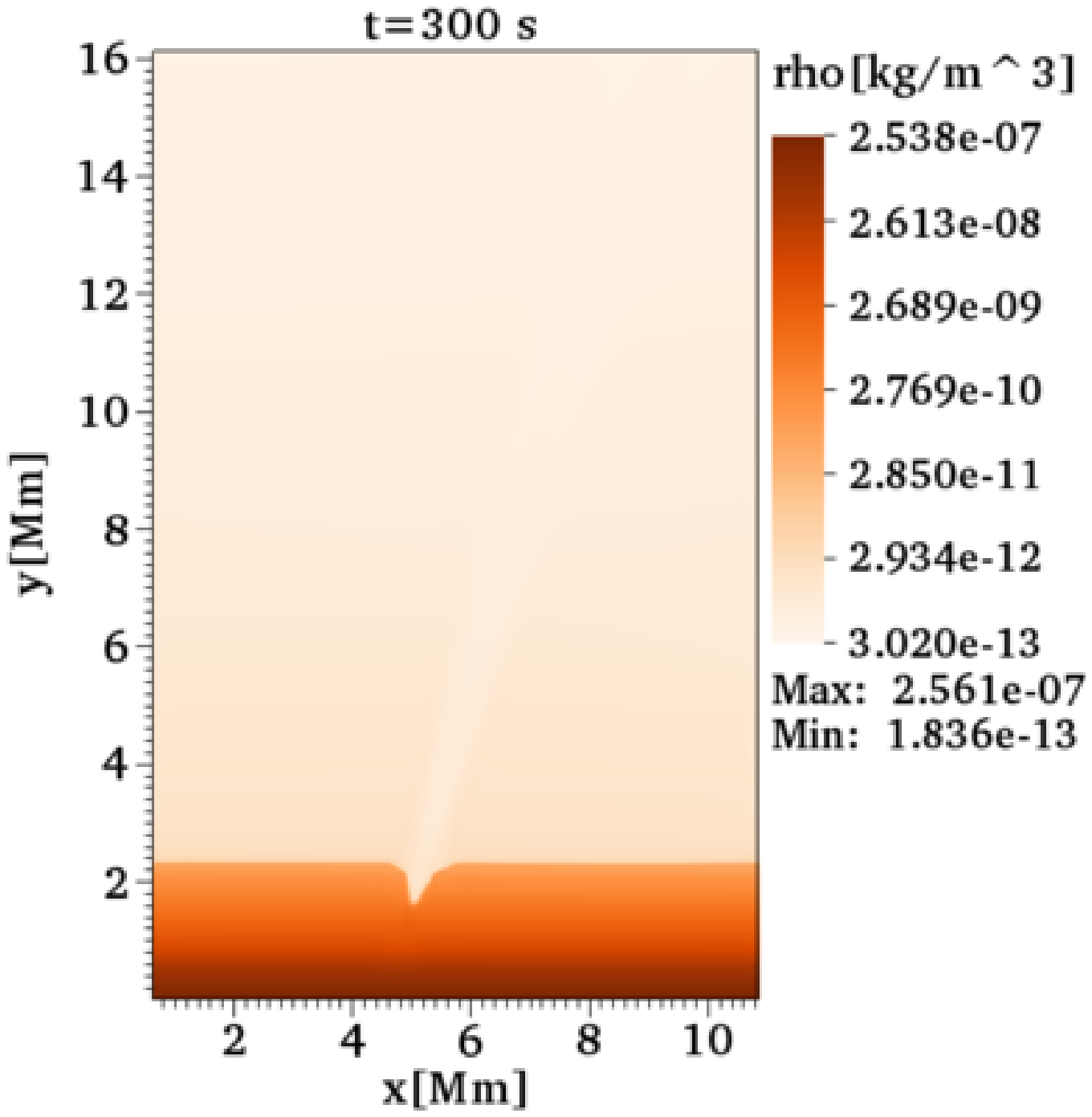}
\includegraphics[width=5.0cm,height=6.0cm]{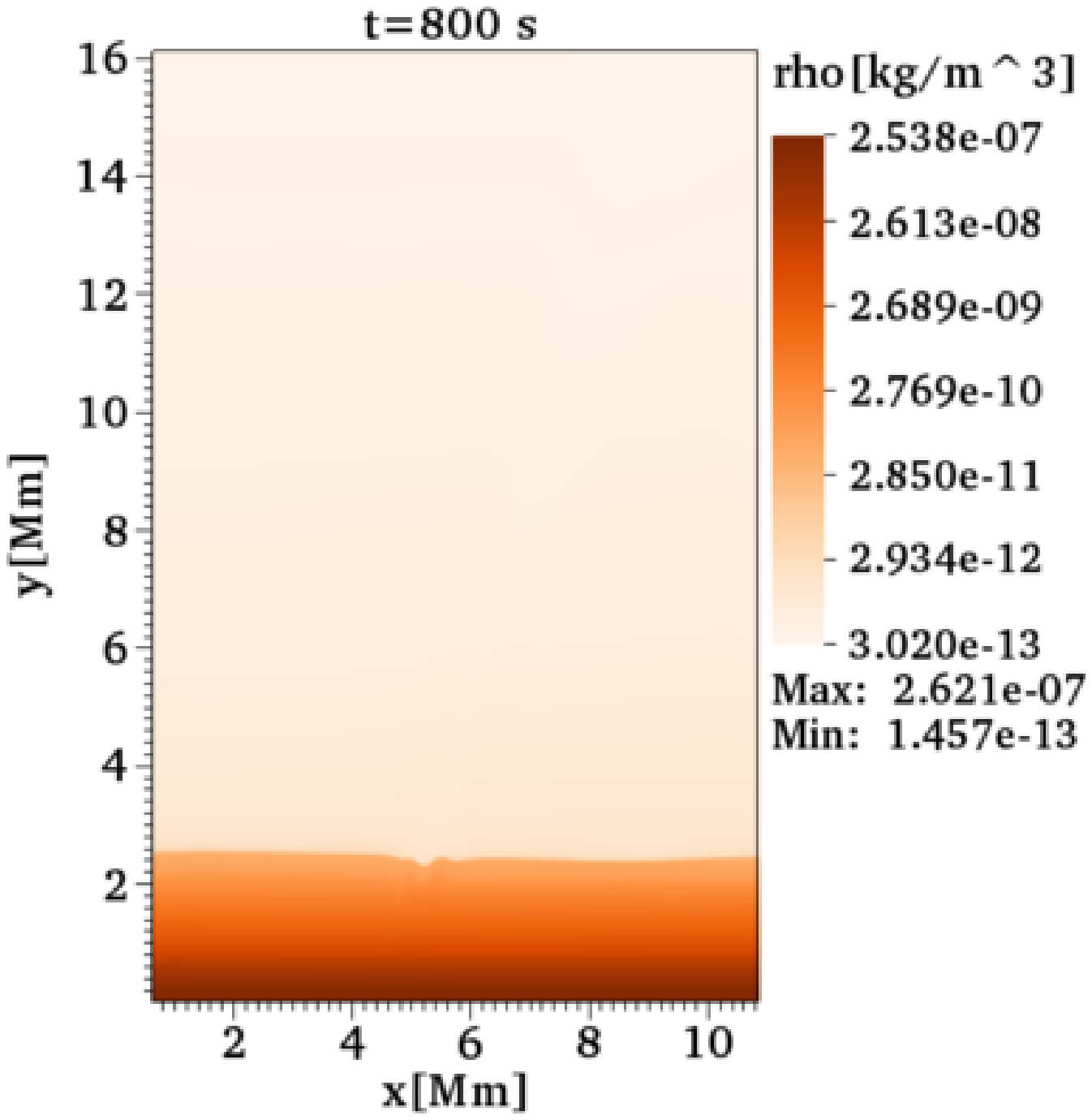}
\caption{{\bf Radiative cooling case}. Snapshots of $-Q_{rad}$ in W m$^{-3}$ (top), and of the mass density, $\varrho$, in kg m$^{-3}$ (bottom), at $t=$ 100, 300, and 800 s.}
\label{fig:Qrad_mass_density_rad_cooling_case}
\end{figure*}

In this test case, we include both terms thermal conduction and radiative cooling effects defined as described in previous subsections to test the most significant flow in the scenario of an energy imbalance background atmosphere. In the fourth column (d) of Figure \ref{fig:temp_evolution_comparison}, we show snapshots of the temperature with the velocity vectors at the same three moments as in previous cases. For example, at $t=100$ s, we note that the jet has practically the same morphology as observed in the radiative cooling case. At $t=400$ s, we see a small jet with a visible double structure that remains just above the transition region. At $t=800$ s, the jet dissipates at the transition region, which is similar to the snapshot for the radiative cooling case at the same instance of time ($t=800$ s). Still, in this case, the jet reaches a slightly lower temperature ($\approx 4318$ K) than in the radiative cooling case. According to the morphology observed in these snapshots, radiative cooling dominates over thermal conduction since the observed jets' characteristics are similar to those simulated in the radiative cooling case. 

% -------------------------------------------
% ----->   SUB-SECTION     <-----
% -------------------------------------------
\subsection{Discussion}
\label{subsec:analysis_and_discussion}

\begin{figure*}
\centering
\includegraphics[width=4.0cm,height=5.0cm]{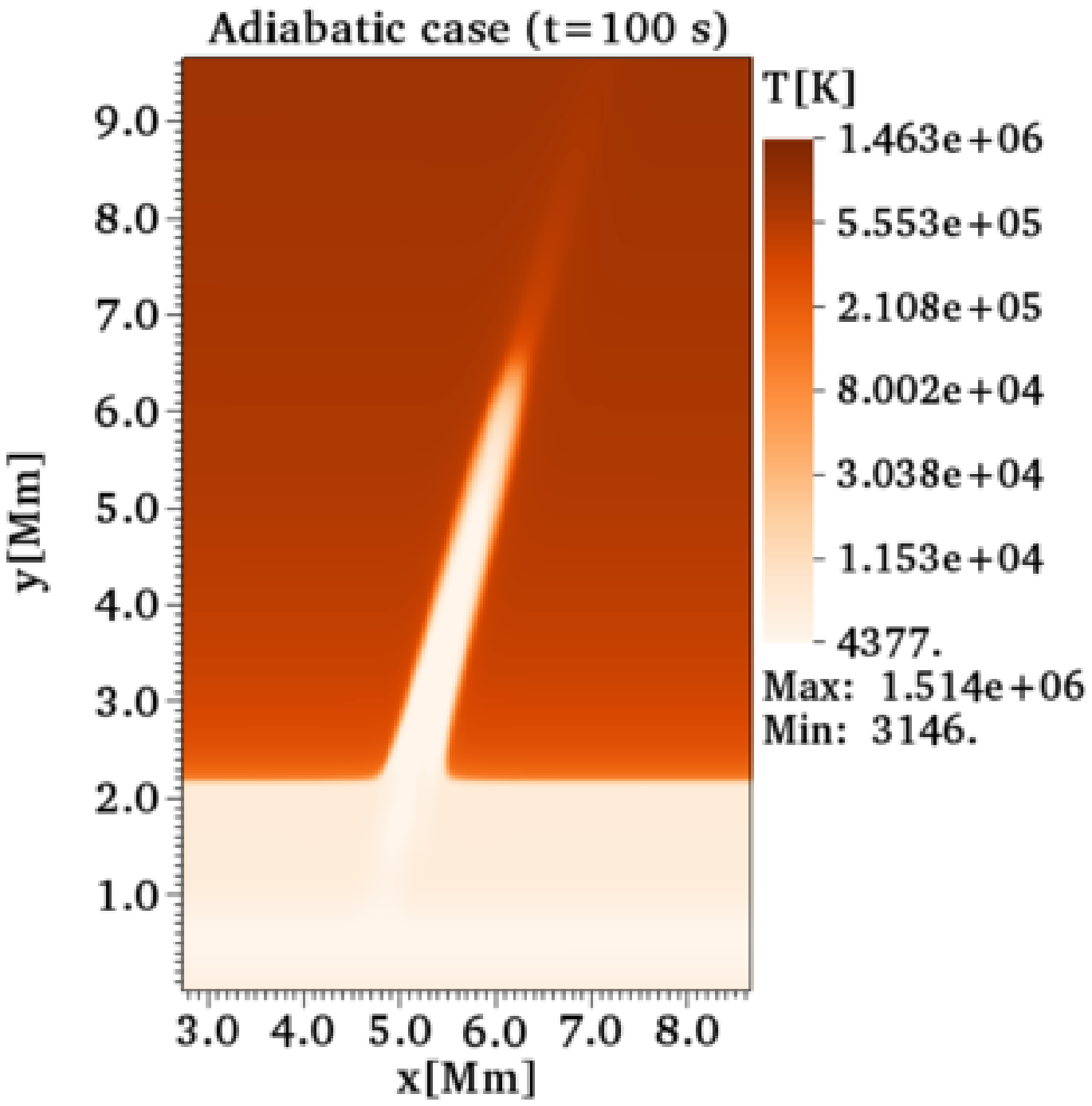}
\includegraphics[width=4.0cm,height=5.0cm]{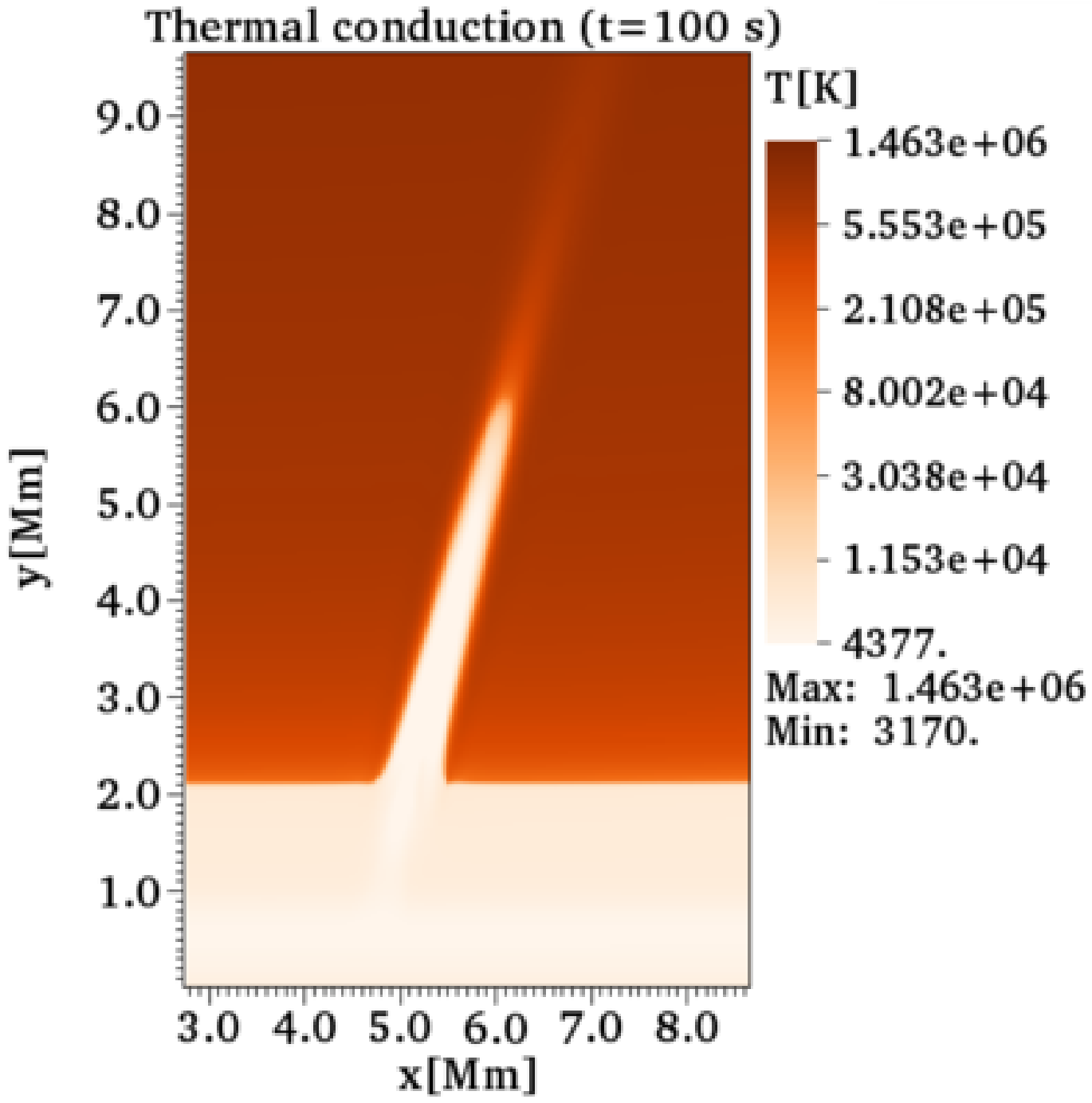}
\includegraphics[width=4.0cm,height=5.0cm]{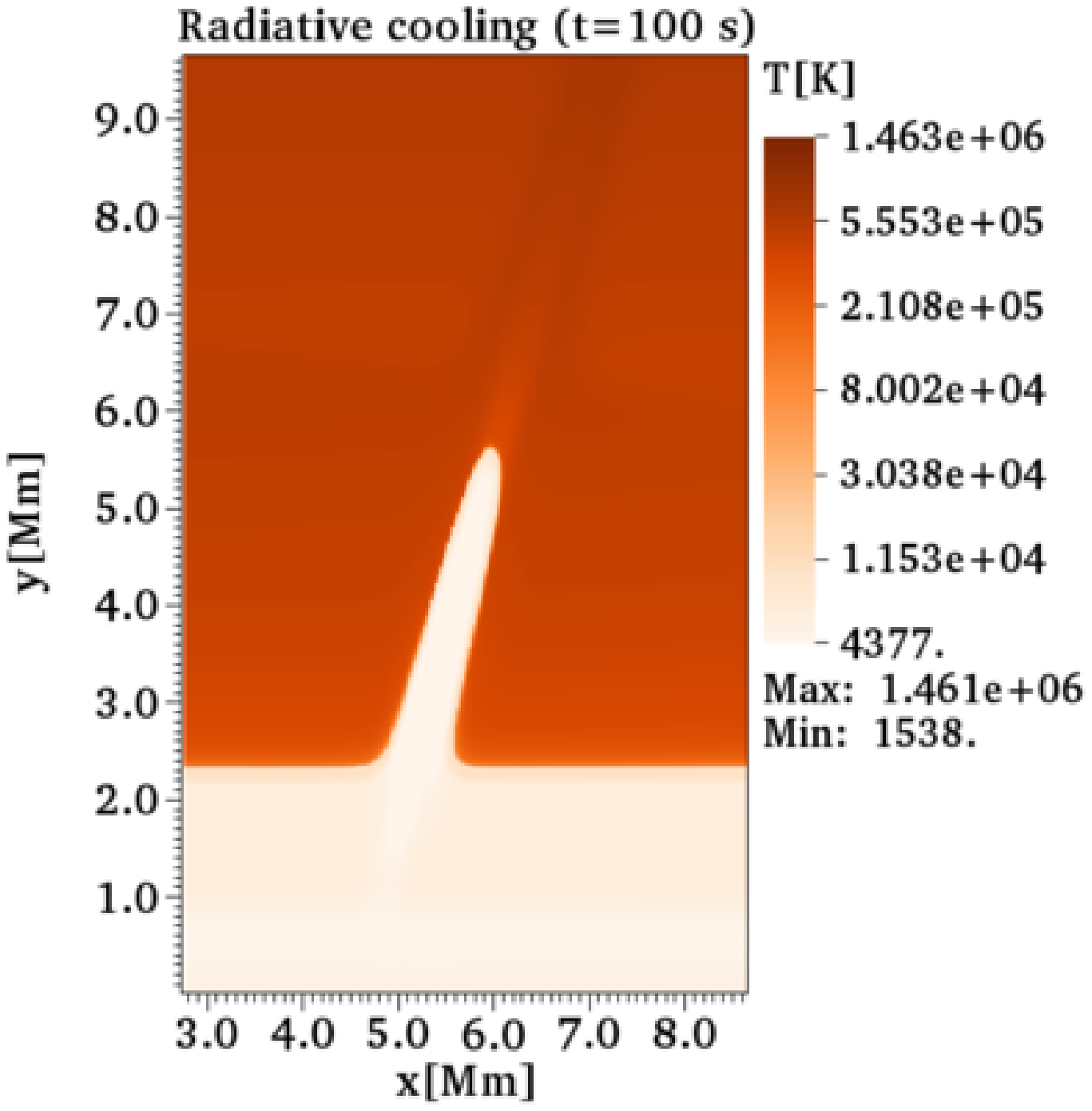}
\includegraphics[width=4.0cm,height=5.0cm]{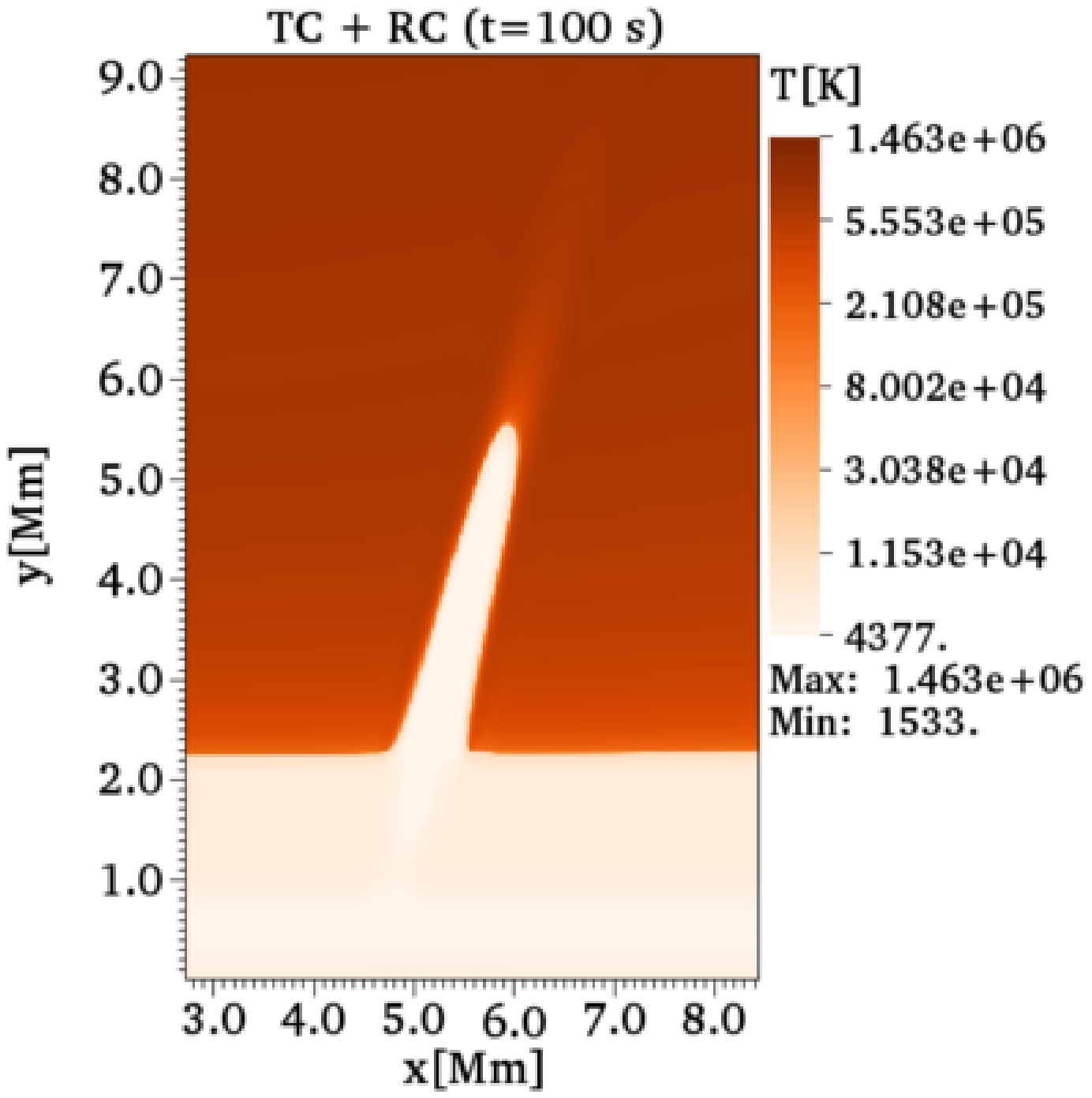}\\
\includegraphics[width=4.0cm,height=5.0cm]{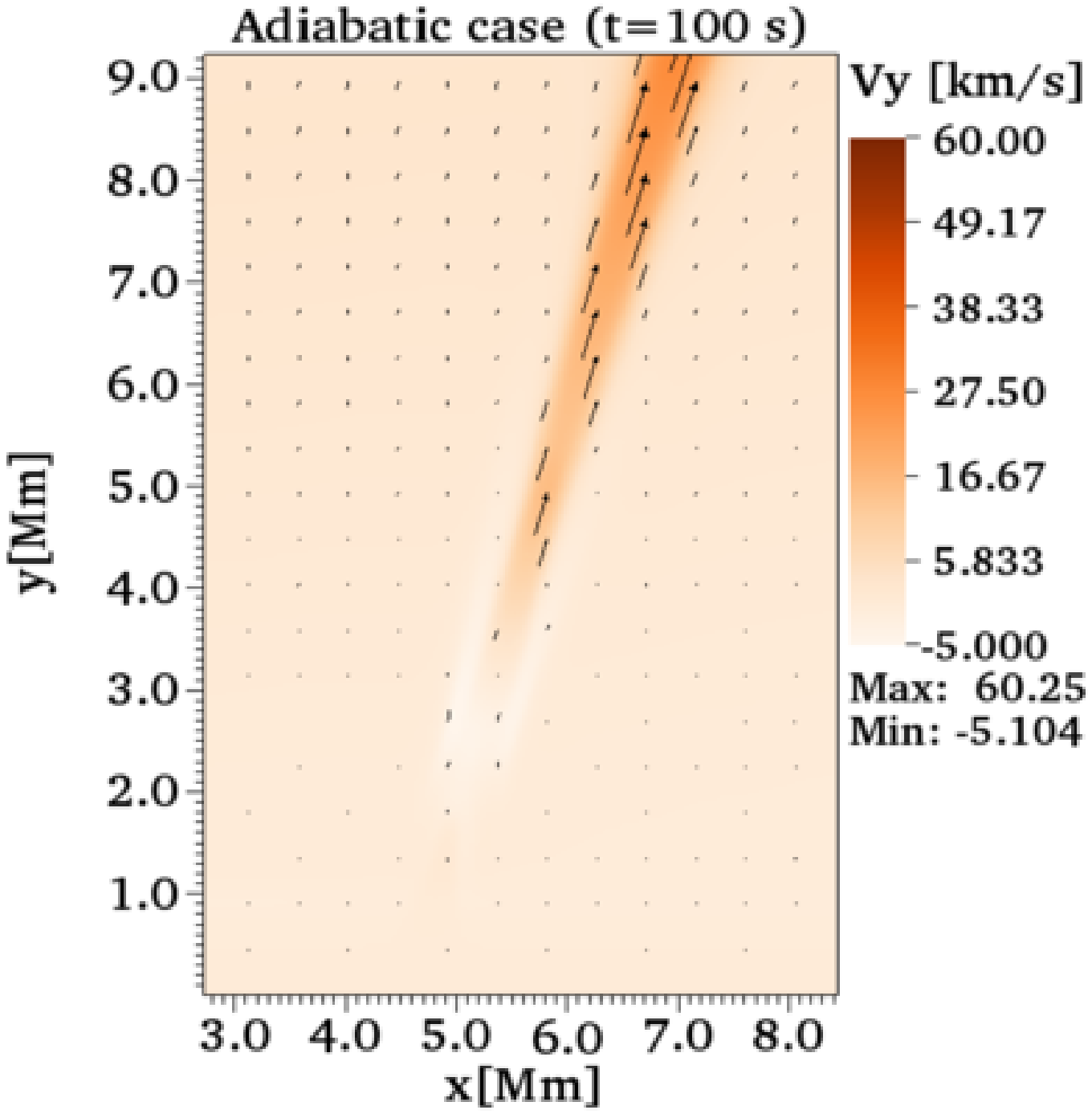}
\includegraphics[width=4.0cm,height=5.0cm]{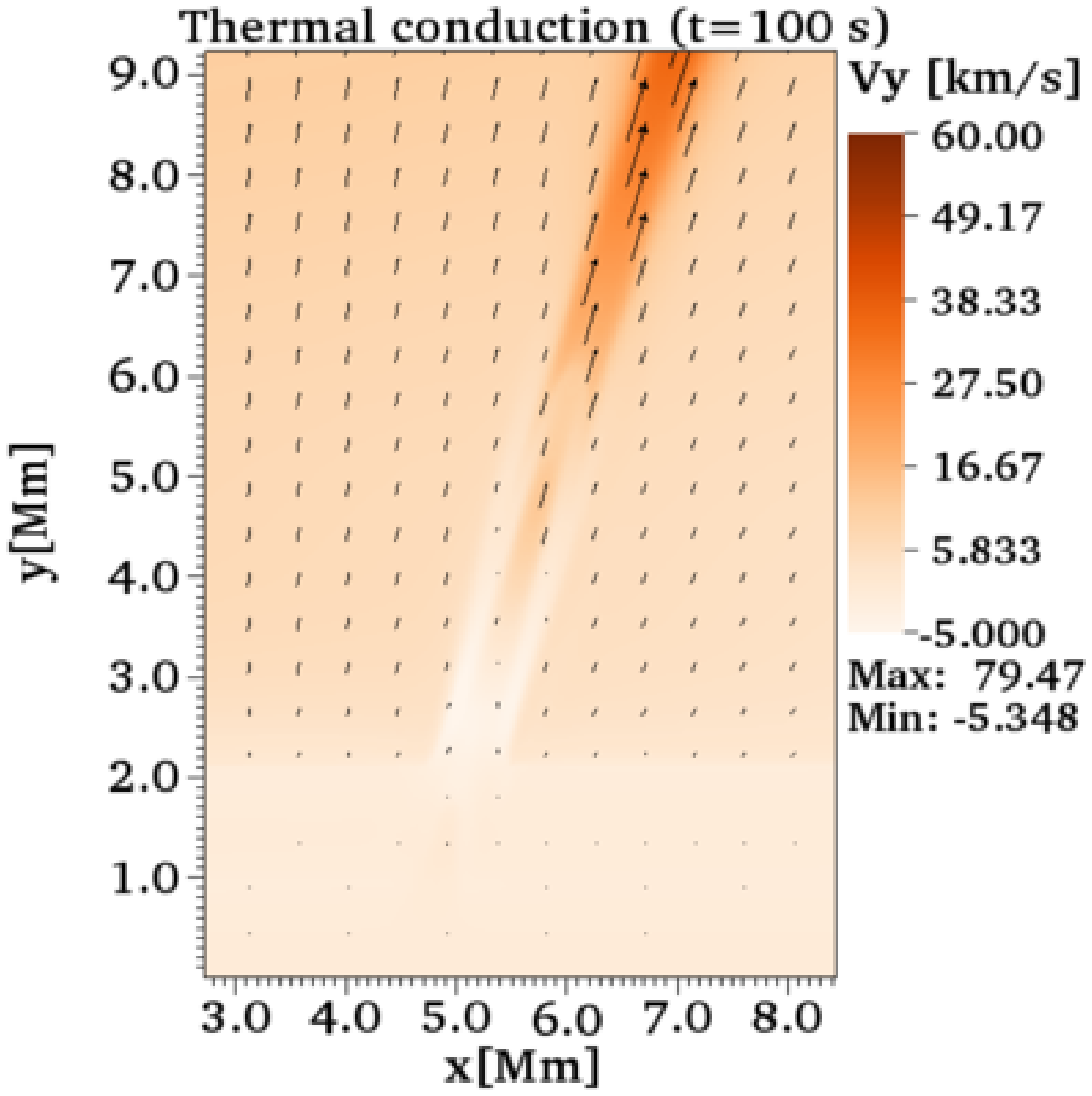}
\includegraphics[width=4.0cm,height=5.0cm]{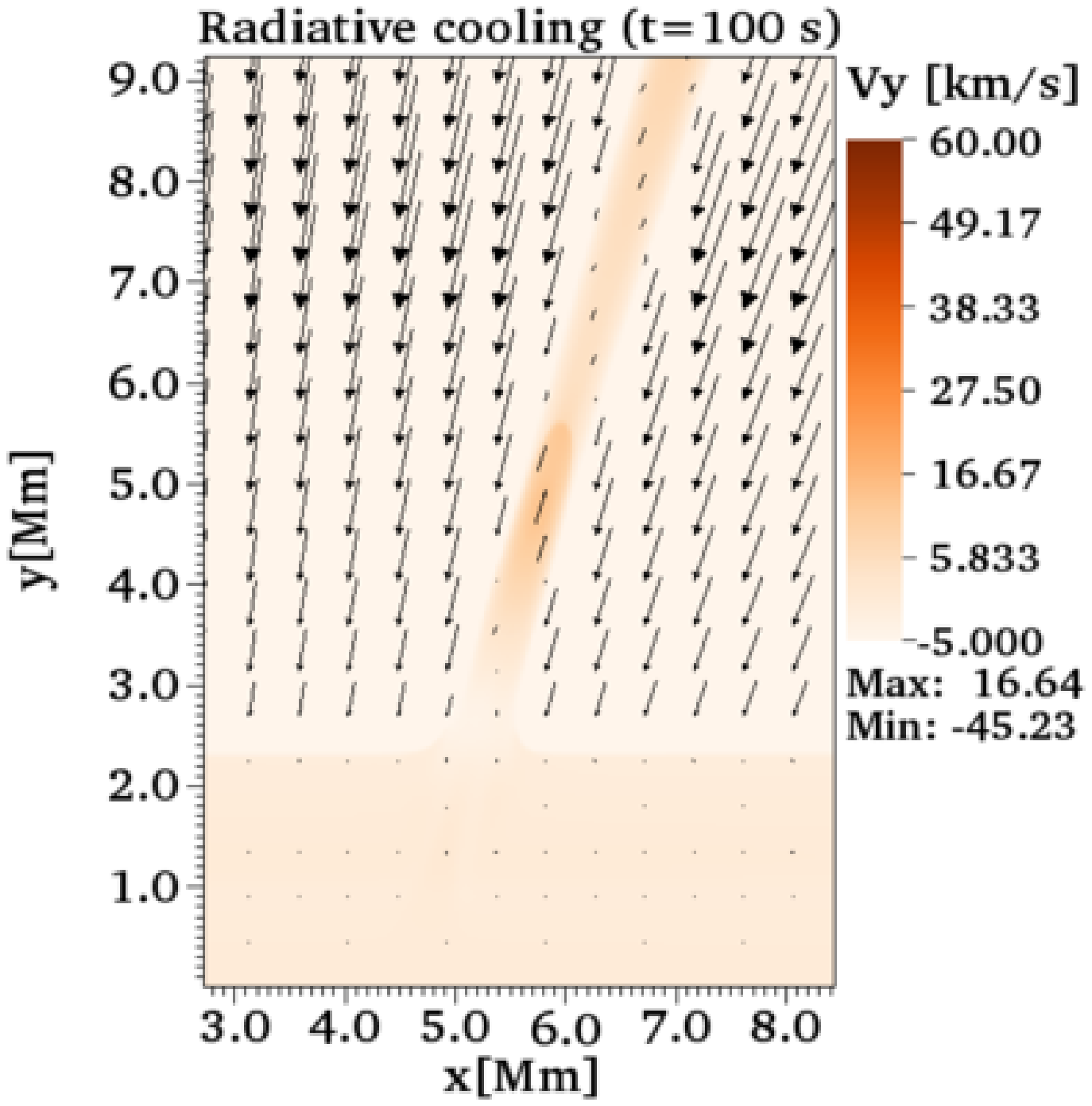}
\includegraphics[width=4.0cm,height=5.0cm]{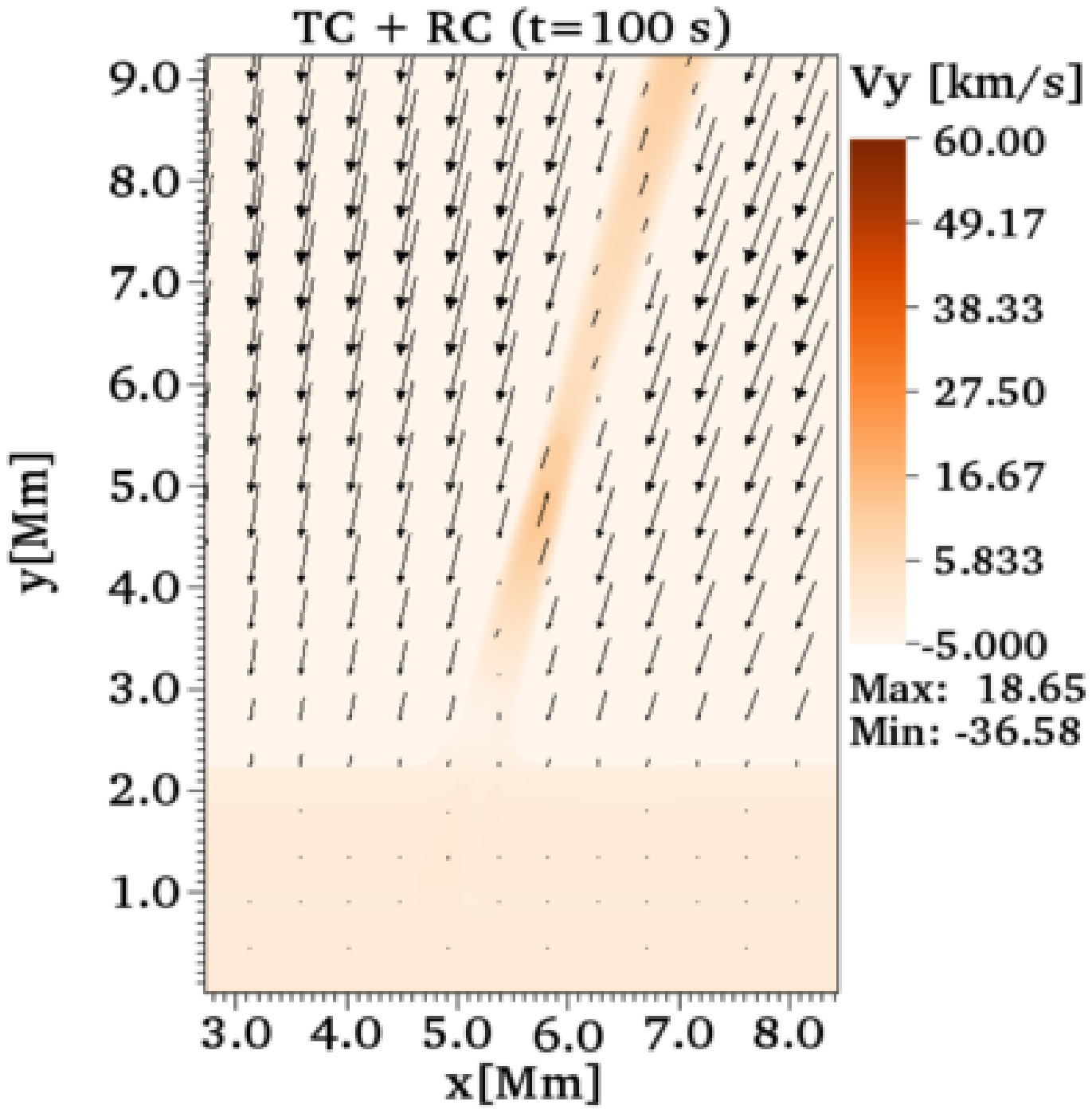}\\
\includegraphics[width=4.0cm,height=5.0cm]{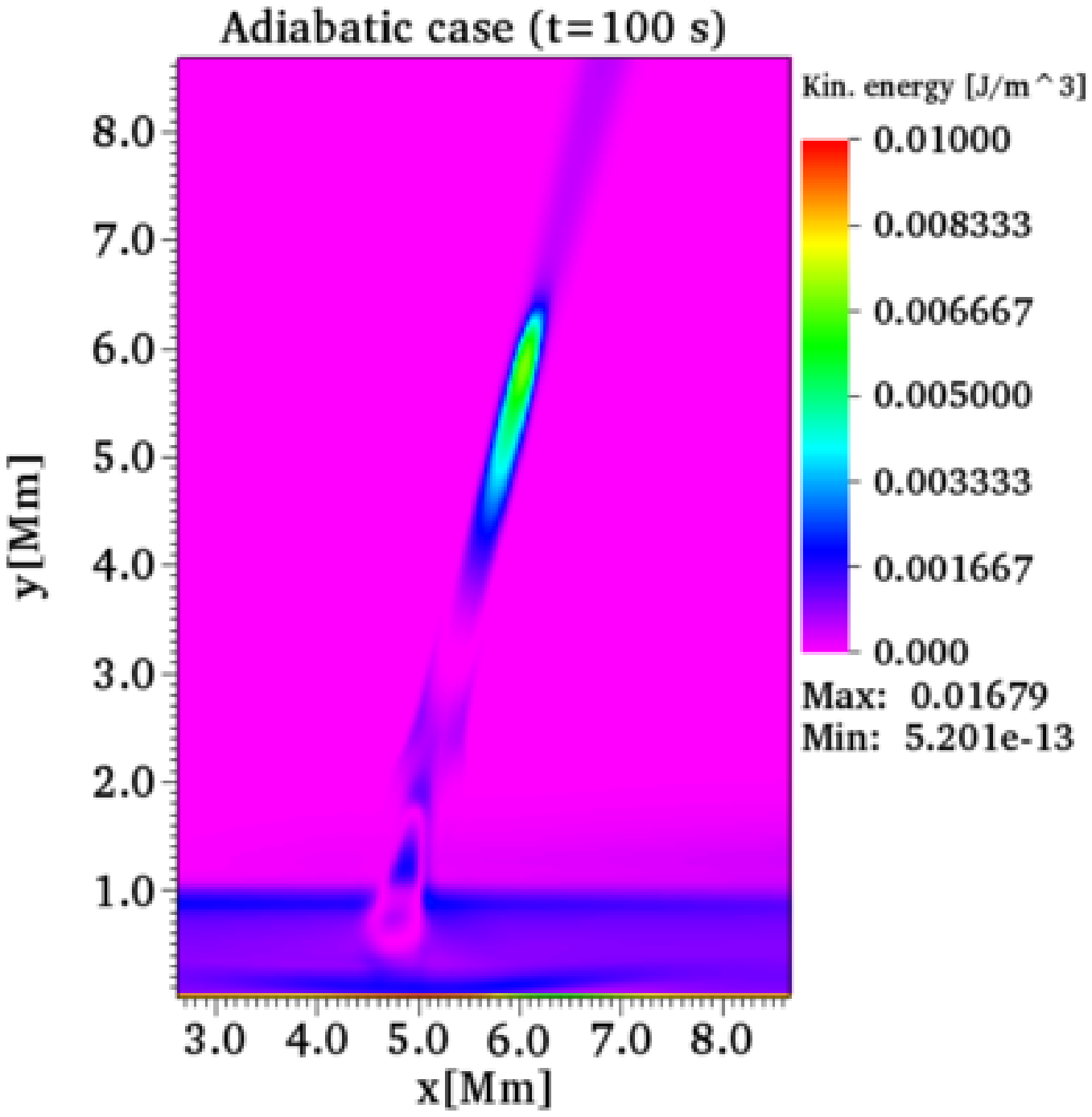}
\includegraphics[width=4.0cm,height=5.0cm]{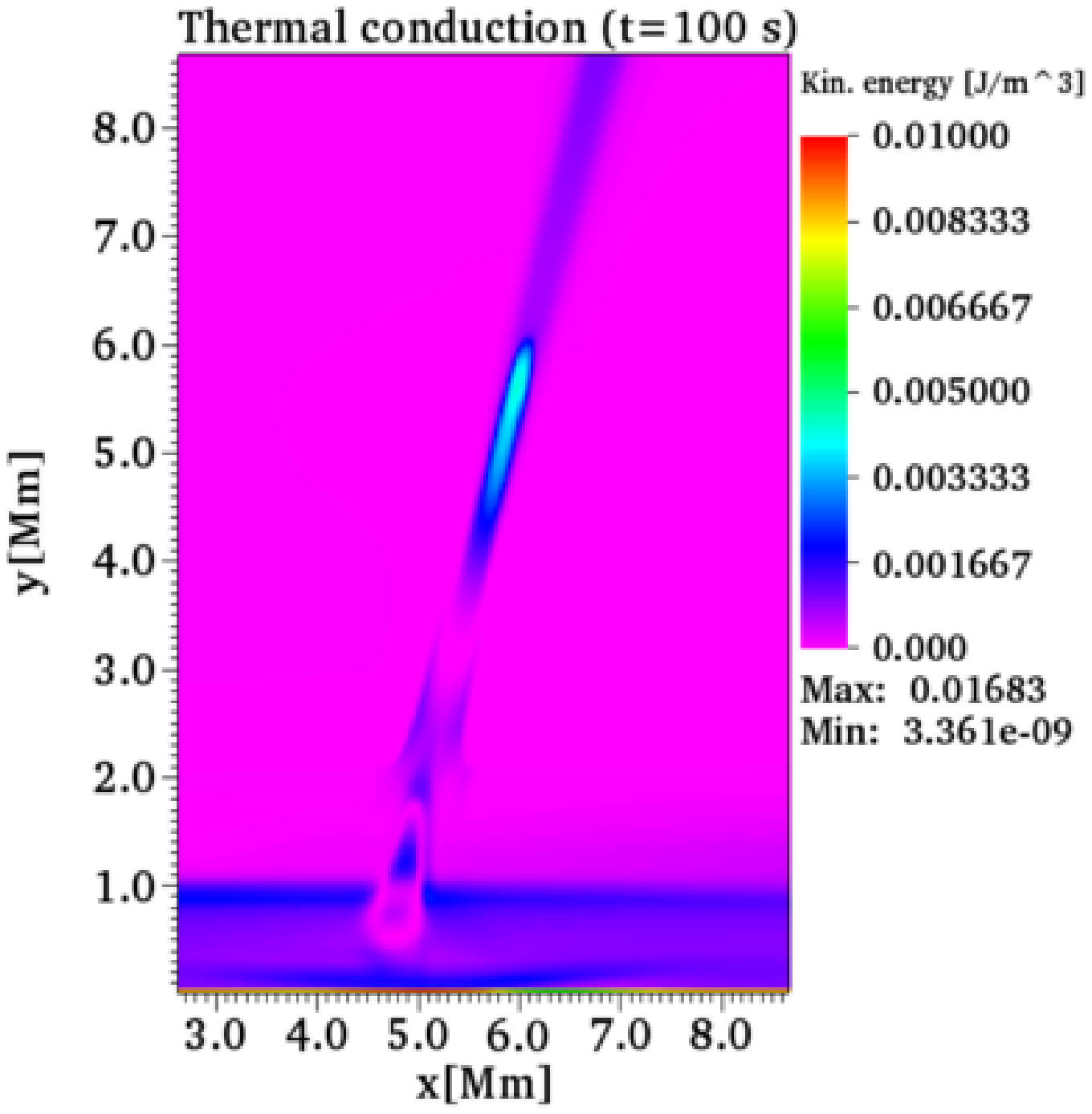}
\includegraphics[width=4.0cm,height=5.0cm]{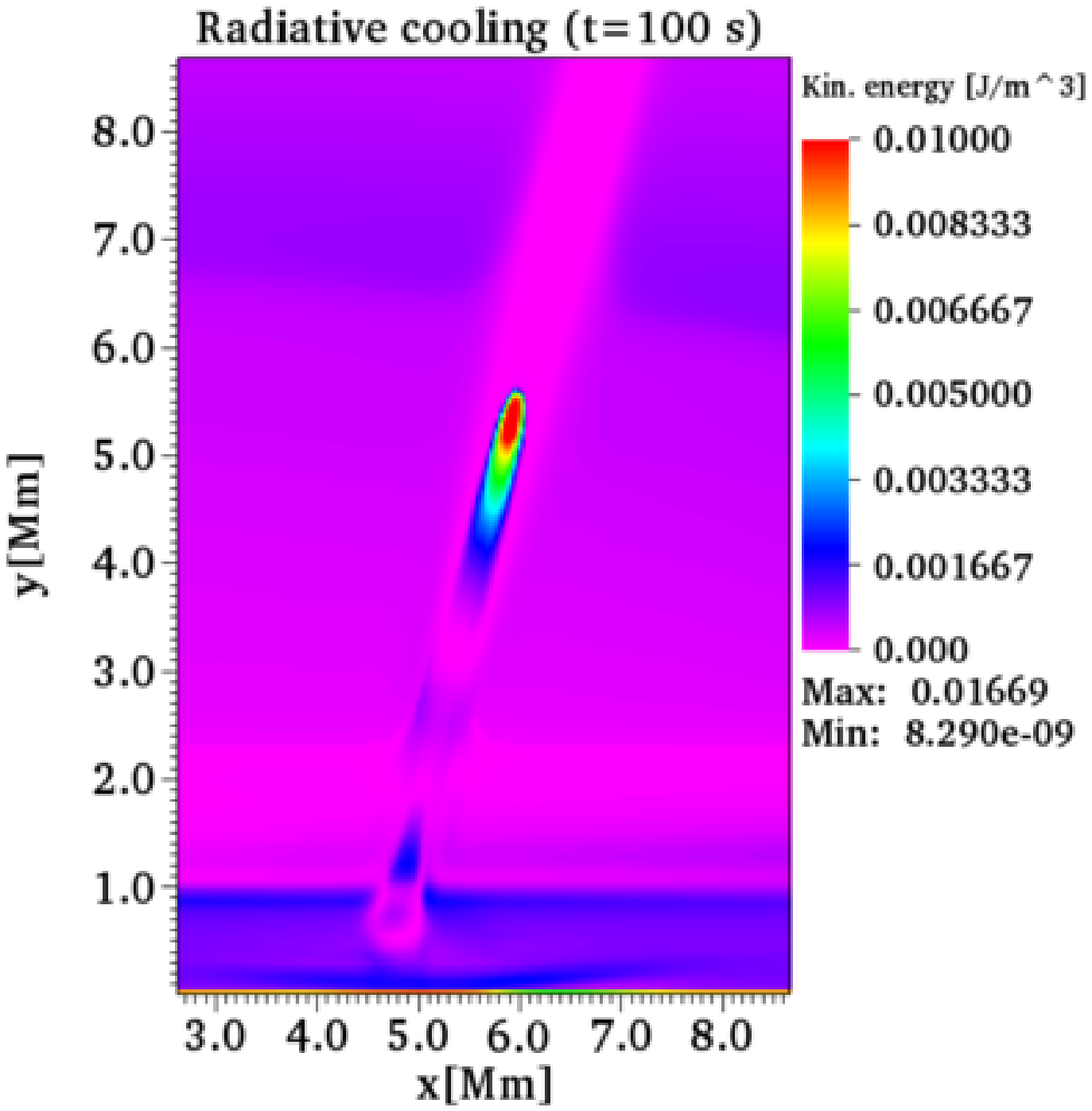}
\includegraphics[width=4.0cm,height=5.0cm]{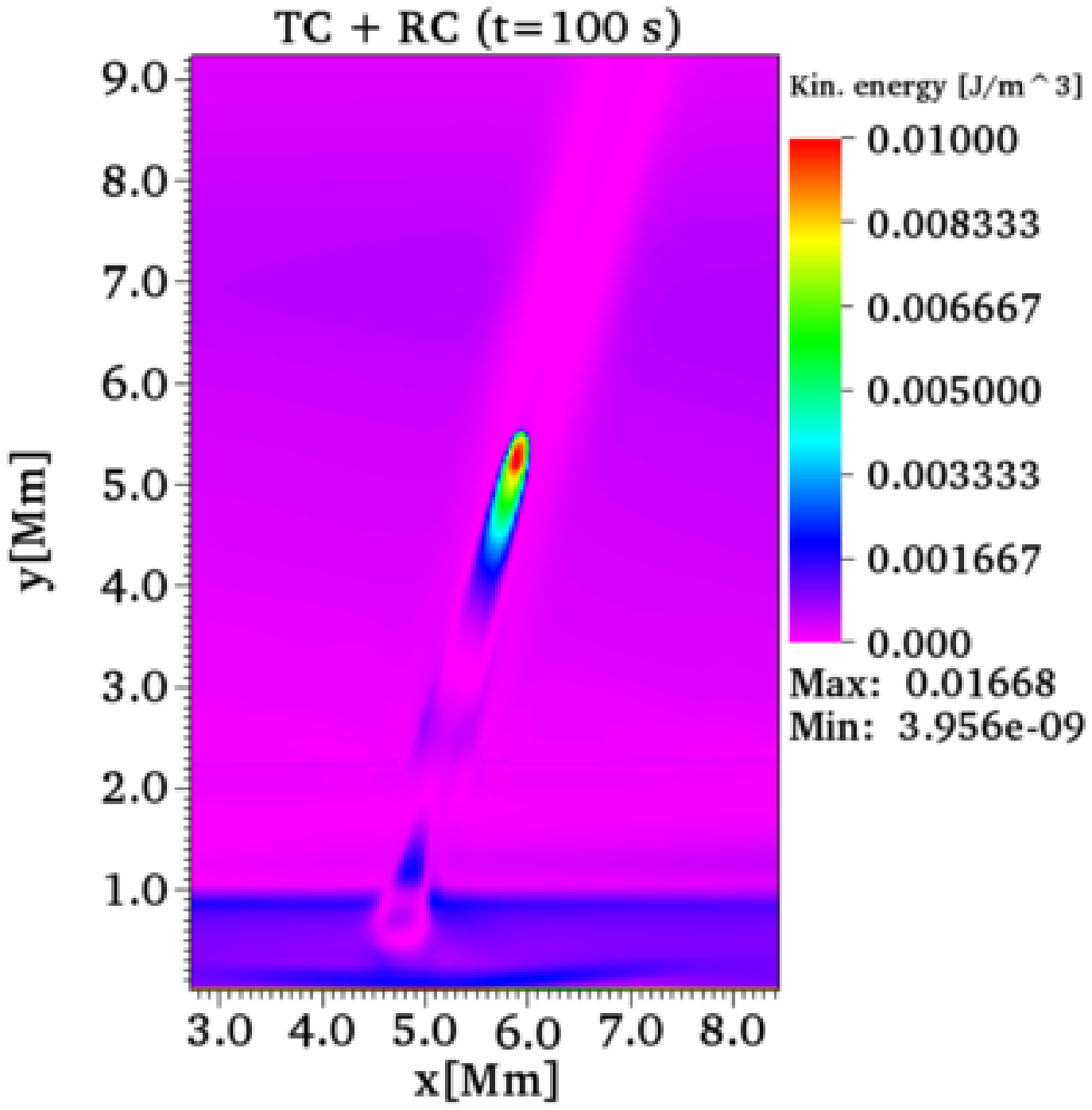}
\caption{(Top row) temperature in Kelvin for the cases: adiabatic (first column), thermal conduction (second column), radiative cooling (third column), and thermal conduction + radiative cooling (fourth column) at $t=100$ s. (Middle row) vertical component of velocity $v_{y}$, in km s$^{-1}$ with velocity field vector at $t=100$ s for the same four cases above. (Bottom row) kinetic energy density, in J m$^{-3}$, for the same cases above at $t=100$ s.}
\label{fig:temp_k_energy_and_vy_zoom_comparisson_t=100s}
\end{figure*}

This subsection presents a more detailed analysis and discussion of the results obtained in the four test cases described in the previous subsections. Specifically, we compare the temperature, vertical velocity, and kinetic energy density of the jets. For instance, at the top row of Figure \ref{fig:temp_k_energy_and_vy_zoom_comparisson_t=100s}, we show the temperature of the jets at $t=100$ s, which is the time when the jets approximately reach the maximum height and the effect of radiative cooling does not strong act on the background atmosphere. Specifically, we see the collimated jet going a height of about 7 Mm in the adiabatic case. This jet has a width of about 0.5 Mm. In the thermal conduction case, we observe a smaller jet that reaches a height of about 6 Mm, but its width remains similar as in the adiabatic case. On the other hand, the jet formed in the radiative cooling case is smaller (5.5 Mm), colder and broader than in the two previous cases. In the simulation that includes thermal conduction and radiative cooling, we obtain a jet with practically the same characteristics as the radiative cooling case. 

In the middle of Figure \ref{fig:temp_k_energy_and_vy_zoom_comparisson_t=100s}, we display the vertical velocity $v_{y}$ together with the velocity vector field at $t=100$ s. In the adiabatic case, the jet reaches a maximum velocity of about 50 km s$^{-1}$; note that flow represented by the velocity vector is dominant along with it. The vertical velocity and vector field's behavior in the thermal conduction is similar to the adiabatic case. Nevertheless, we can observe a higher flow in jet surroundings, and the shock produced by the jet reaches a higher speed ($\approx$ 80 km s$^{-1}$). Unlike in the two previous cases, the jet's speed behaves very differently in the radiative cooling and thermal conduction with radiative cooling cases. Since we can see that the jet reaches a maximum vertical velocity of about 10 km s$^{-1}$, following by a negative flow, as shown in panels (c) and (d) of Figure \ref{fig:temp_evolution_comparison}. In the bottom panels of Figure \ref{fig:temp_k_energy_and_vy_zoom_comparisson_t=100s}, we illustrate the kinetic energy density generated by the propagation of jets at $t=100$ s for the different test cases. In the adiabatic case, we can distinguish that kinetic energy is higher at the shock front, while along with the jet, it is of the order of $10^{-3}$ W m$^{-3}$. In the thermal conduction case, we observe a similar distribution of the kinetic energy density, both in the shock and throughout the jet. It is also important to mention that the kinetic energy density in the jets' surroundings, particularly in the corona, is minimal. The kinetic energy density in the radiative cooling and thermal conduction with radiative cooling cases is higher at the shock front, but it is minimal along with the jet. In these cases, the kinetic energy dissipates in the solar corona remains significant due to the high flow shown in the velocity vector field.     

Another way to diagnostic the differences between the four test cases is to measure scalars as a function of time. In particular, we estimate the vertical velocity $v_{y}$, the temperature, and the kinetic energy density at two different detection points $(x=5.6, y=4)$ Mm and $(x=8.5, y=14)$ Mm. These points are in two regions of interest; for example, the point $(x=5.6, y=4)$ Mm is localized in a region that covers inside the jet, while the point  $(x=8.5, y=14)$ Mm is in a region where the shock front locates. In the top-left panel of Figure \ref{fig:vy_kin_energy_temp_vs_time_comparisson}, we display the vertical velocity as a function of time at point $(x=5.6, y=4)$ Mm; here, we see that jet reaches a maximum velocity of about 90 km s$^{-1}$ for the adiabatic and thermal conduction cases just after the beginning of the simulations. The jet reaches a maximum velocity near 45 km s$^{-1}$ in the radiative cooling and thermal conduction with radiative cooling cases. After this impulsive stage, the velocity decreases in the four cases at around 300 s. It is clear that in the radiative cooling case, the velocity is predominantly negative, reaching a value of about -60 km s$^{-1}$. After $t=300$ s, the jet starts to rise again, getting a velocity of the order of 30 km s$^{-1}$ in the adiabatic and thermal conduction cases. In contrast, the radiative cooling case's velocity remains negative, and in the thermal conduction with radiative cooling, the velocity reaches a positive value of about 20 km s$^{-1}$. From $t=350$ s, the jet propagates with a velocity of about 10 km s$^{-1}$ for the adiabatic, thermal conduction, and thermal conduction with radiative cooling cases. In the radiative cooling case, the velocity is about -30 km s$^{-1}$. In the top-right panel of Figure \ref{fig:vy_kin_energy_temp_vs_time_comparisson}, we show the time history of vertical at point $(x=8.5, y=14)$ Mm, which is the same point where \cite{Murawski_et_al_2011} collected the velocity to see the behavior of the shocks front. At this point, we can see that the shock front shows an oscillatory behavior in the four cases. The shock front arrives at the detection point at about $t=150$ s.
In contrast, the second shock arrives at about $t=450$ s, which seems to have a period of $\approx$ 300 s. The shocks fronts oscillate with a more significant positive amplitude in the adiabatic, thermal conduction, and thermal conduction with radiative cooling cases. In the radiative cooling case, the oscillations have a predominantly negative amplitude. The adiabatic results, thermal conduction, and thermal conduction with radiative cooling cases are similar to those shown in Figure 6 of \cite{Murawski_et_al_2011}. However, the oscillations seem to have a dominant period of about 400 s in their analysis. In the middle-left panel of Figure \ref{fig:vy_kin_energy_temp_vs_time_comparisson}, we display the kinetic energy density produced by the jets measured at the point $(x=5.6, y=4)$ Mm; in this case, we see that the energy is maximum at $t\approx50$ s, that coincides with the beginning impulsive stage of the jets. After that time, the energy decreases until $t\approx250$ s, and then it growing again. From $t=300$ s and following times, the energy is practically negligible in all four cases. In particular, in the thermal conduction and thermal conduction with radiative cooling cases, the energy reaches higher values at the initial impulsive time. At the following stage, the energy growth is more remarkable in the radiative cooling and thermal conduction with radiative cooling cases; however, the negative flow produces that energy. In the middle-right panel of Figure \ref{fig:vy_kin_energy_temp_vs_time_comparisson}, we show the kinetic energy density collected at the point $(x=8.5, y=14)$ Mm, where the shock front produced in the thermal conduction case results in a higher dissipation of energy at $t\approx150$ s. After this moment, the energy oscillates in the four cases, consistently with the vertical velocity behavior. 

To schematize how the temperature behaves in the jet and shocks front; at the bottom-left of Figure \ref{fig:vy_kin_energy_temp_vs_time_comparisson}, we show the temperature measured at the point $(x=5.6, y=4)$ Mm. At this location, we can observe that the jets' temperature is hotter at the initial impulsive stage. At the same time, they are colder ($\approx 5\times10^{3}$ K) at $t=150$ s, which is the time when approximately the jets reach the maximum height, as shown in Figure \ref{fig:temp_evolution_comparison}. At about $t=300$ s, the temperature increases by getting a value of around $7\times10^{5}$ K, which maintains the radiative cooling, thermal conduction, and thermal conduction with radiative cooling cases for latter times. In contrast, in the adiabatic case, the temperature decreases to approximately $10^{5}$ K. On the other hand, the temperature observed in the radiative cooling and the thermal conduction with radiative cooling cases after 300 s is the coronal temperature not directly associated with the jet structure since the jet already dissipated shown in Figure \ref{fig:temp_evolution_comparison}. In the bottom-right panel of Figure \ref{fig:vy_kin_energy_temp_vs_time_comparisson}, the temperature collected at the point $(x=8.5, y=14)$ Mm, shows that the shock's front produced in the thermal conduction and thermal conduction with radiative cooling cases reaches higher temperatures ($\approx 1.3\times10^{6}$ K). On the other hand, the shocks have cooler temperatures in the adiabatic case and even colder in the radiative cooling case. In all three cases, the temperature oscillates, consistent with the shocks' behavior at this location shown in the vertical velocity time series. We can relate the shocks shown in the previous time-series to magnetoacoustic shocks, which can drive the non-purely ballistic motion observed in macrospicules \citep{Loboda&Bogachev_2019}.      
To have a more general idea of how the non-adiabatic terms represented by thermal conduction and radiative cooling can affect the jets' dynamics in the scenario of an energy imbalance background. We calculate the distance-time diagrams of the logarithm of mass density $\varrho$ in kg m$^{-3}$ and the vertical velocity $v_{y}$ in km s$^{-1}$ in a line along with the jet for the four study cases, as shown in Figure \ref{fig:mass_density_velocity_along_line}. For example, in the top-left of Figure \ref{fig:Distance-time_diagrams_density}, we see that the jet reaches a maximum height $\approx$ 7 Mm at $t\approx160$ s in the adiabatic case. It is also evident in the mass density diagram shows that the jet follows a parabolic path that covers about 300 s, which is consistent with the observed behavior of chromospheric jet structures such as type I spicules, dynamics fibrils, and mottles \citep{Kuzma_et_al_2017a, Hansteen_et_al_2006}. After $t=300$ s, the jet becomes smaller, reaching a height of about 4 Mm and not following a parabolic behavior. In the top-left of Figure \ref{fig:Distance-time_diagrams_velocity}, the distance-time diagram of the vertical velocity for the adiabatic case shows an impulsive stage at the initial time where the velocity is high. The temporal evolution of the velocity also indicates that the jet presents an oscillation behavior until $t=300$ s; however, the oscillations tend to have small amplitudes later.

In the top-right panel of Figure \ref{fig:Distance-time_diagrams_density}, the diagram of density for the thermal conduction case shows that the jet reaches a maximum height $\approx$ 6.2 Mm at $t\approx150$ s, which is faster but smaller than in the adiabatic case. It is also clear that the jet follows a parabolic path, which is evident at the first 300 s of the jet's evolution. In this case, we note more clearly that the jet becomes smaller following a less marked parabolic path. In the top-right of Figure \ref{fig:Distance-time_diagrams_velocity}, the temporal evolution of the vertical velocity shows consistency with the behavior of mass density for the thermal conduction. In particular, we can see the generation of a shock wave following by a contact wave at initial times that delineates the jet's dynamics behavior later. There is also an evident difference compared to the adiabatic case since the velocity shows higher positive values during evolution. In the bottom-left of Figure \ref{fig:Distance-time_diagrams_density}, we see that temporal evolution of mass density in the radiative cooling case shows that jet reaches a maximum height $\approx$ 6 Mm, which is smaller than in the two previous cases. Besides, the jet clearly shows a parabolic path that covers exactly until $t=300$ s; after this time, the jet becomes smaller $\approx$ 2.5 Mm, which indicates the jet's dissipation. In the bottom-left of Figure \ref{fig:Distance-time_diagrams_velocity}, the distance-time diagram of velocity shows an agreement with the density behavior; in particular, it is clear how the velocity becomes negative after 300 s, which in turns dissipates the jet. We see the shock appearance, the contact discontinuity, and the rarefaction wave in the three cases. Finally, in the bottom-right of Figure \ref{fig:Distance-time_diagrams_density} we show the temporal evolution of mass density in the thermal conduction with radiative cooling case.
Here, we observe that behavior is similar to the radiative cooling case; but, the jet is smaller from $t=300$ s. In the bottom-right of Figure \ref{fig:Distance-time_diagrams_velocity}, we see that temporal evolution of the vertical velocity in the thermal conduction with radiative cooing case is similar to the radiative cooling one until $t\approx300$ s. Still, after that time, we see regions with flows that have positive amplitude, which is related to thermal conduction. The distance-time diagrams shown above are comparable with those shown in Figure 5 of \cite{Kuzma_et_al_2017a}; however, in this paper, we focus on macrospicules, and we display a longer evolution time of the jets. We can see how the shocks' behavior is in all the distance-time diagrams of the mass density. Expressly, we can state that when the shock penetrates the corona, the chromospheric material follows the siphon-like mechanism (i.e., the pressure behind the pulse is smaller than in front). We can observe these features in Figure  \ref{fig:Distance-time_diagrams_density}, where the first line of density difference is due to this shock. It may be worth saying that a rarefaction wave follows a shock, and a jet by itself corresponds to a thermal mode as, along with it, only mass density experiences a jump.  

% -------------------------------------------
% ----->   SUB-SUBSECTION     <-----
% -------------------------------------------
\subsubsection{{\bf Effects of the energy imbalance on the background atmosphere model}}
\label{sub-subsec:effects_energy_conditions}

To show how the thermal conduction and radiative cooling can affect the solar background atmosphere in a scenario of energy imbalance, we decided to run the system without any explicit perturbation in velocity or any other variable but considering the magnetic field configuration and gravity. This experiment can give us a more general idea of how the energy fluxes affect the system even without jet excitation. For instance, in Figure \ref{fig:temp_background_test}, we show temporal snapshots of the plasma temperature at times $t=100$ s and $t=800$ s corresponding to the four test cases analyzed in this paper. In particular, at $t=100$ s, we can see a stable solar background atmosphere for the adiabatic and thermal conduction cases; in these two cases, the coronal temperatures remain close to the initial maximum temperature ($1.463\times10^{6}$ K). At $t=100$ s, we can see a slight variation of temperature near the transition region and the far-right domain for the radiative cooling and thermal conduction with radiative cooling cases. We can relate the variation to the radiative cooling since it depends on mass density, which is high at the upper chromosphere and transition region. Despite the variations, the corona's temperature maintains its values close to the initial maximum value ($1.463\times10^{6}$ K), which indicates that radiative cooling is not cooling down the corona. At $t=800$ s, the background temperature for the adiabatic and thermal conduction cases is stable and only shows minor variations respecting the initial state. However, for the radiative cooling case at $t=800$ s, we can see that temperature decreases over the transition region and part of the low corona ($\approx 5$) Mm. Also, the background is not as stable as in the adiabatic and thermal conduction cases. Even the radiative cooling effect is evident; the temperature remains between $4352-1.465\times10^{6}$ K, which implies that cooling does not decrease temperature significantly and dissipates the background. Finally, the plasma temperature for the thermal conduction with the radiative cooling case at $t=800$ s shows some similarities to the radiative cooling case itself, however in this case; it seems that thermal conduction moderates the strong effect of the radiative cooling, and the temperature remains in a range of $4318-1.463\times10^{6}$ K.

To have a better perspective of how the background atmosphere's temperature varies, in Figure \ref{fig:temp_vs_height_background_test}, we show 1D cuts as a function of the vertical coordinate $y$ to the four test cases at three different times. We calculate the 1D profiles in a vertical line traced from $y=0$ to $y=30$ Mm localized at the middle of the domain ($x=15$ Mm). At $t=0$ s, we show that all temperature profiles are the same. At $t=100$ s, the background model's plasma temperature remains close in the adiabatic, TC, and T+RC cases; but in the RC case, the temperature decreases. At $t=400$ s, the tendency of temperature decreasing is persistent in the radiative cooling case.
In contrast, in the thermal conduction and thermal conduction with radiative cooling cases, the temperature slightly increases, respecting the adiabatic case. The snapshots at $t=800$ s describe the background temperature's final state, particularly the radiative cooling; in fact, it cools down the plasma, and the thermal conduction slightly warms it up. Even though the background's plasma temperature was affected, it did not suffer substantial variations that make it collapse under its weight.

In conclusion, from this experiment, we can state that the background solar atmosphere model is affected by thermal conduction and radiative cooling when there is not an energy balance. Therefore, if we perturb the background with a velocity perturbation, we expect that the energy flows generated affect the jet's behavior. We verify that flows are mainly present with this simple experiment because the solar background atmosphere is not in energy balance. However, these flows do not dominate over the whole system, especially at the solar corona; therefore, we can test whether the background flows resulting from an imbalance between thermal conduction and radiative cooling influence the jets' behavior.

% -------------------------------------------
% ----->   SUB-SECTION     <-----
% -------------------------------------------
\subsection{Comparison with the observations of macrospicules}
\label{subsec:observation_comparison}

Despite the energy imbalance scenario of the solar background model, the previous subsections show the jets' formation that exhibits macrospicules' features. To prove the consistency of our simulations, in this subsection, we present some observational evidence of macrospicules, which are helpful to compare and contrast with our simulated jets for the various test cases of this paper. 

\begin{figure*}
\centering
\includegraphics[width=8.0cm,height=6.0cm]{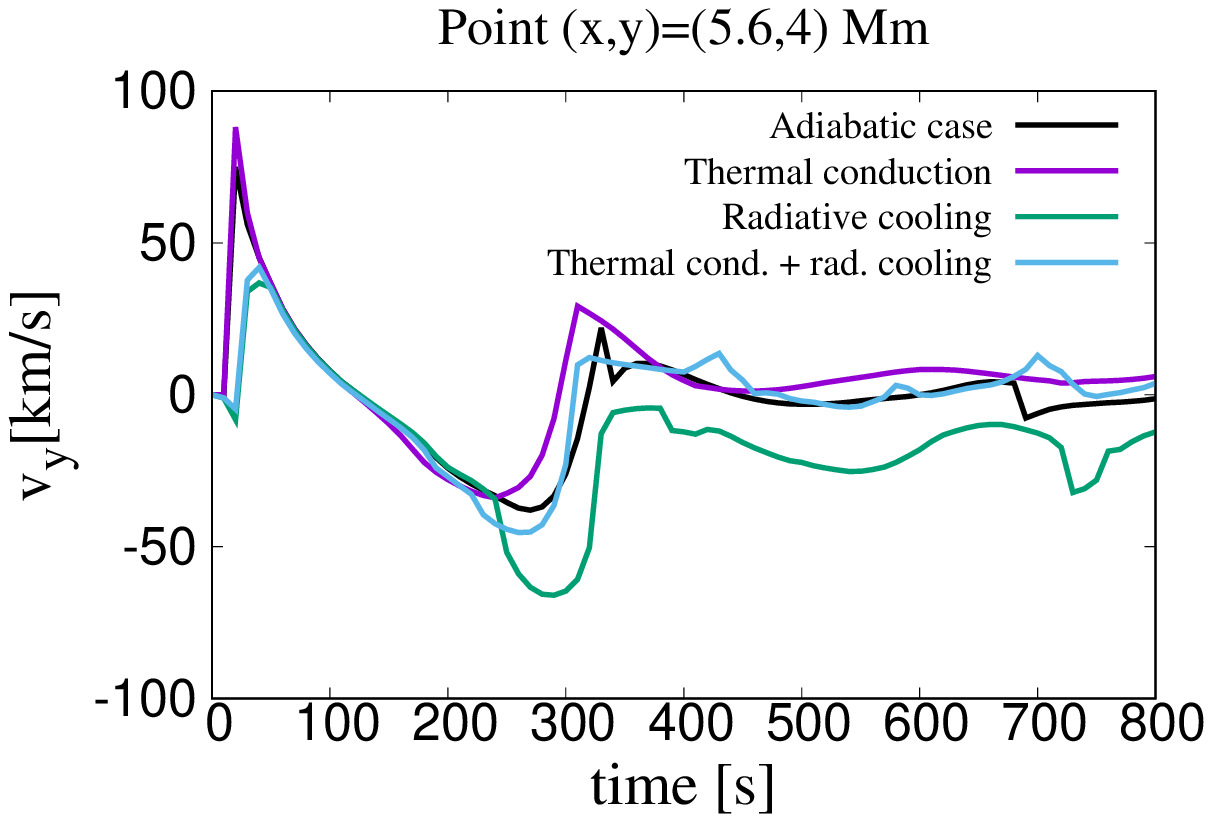}
\includegraphics[width=8.0cm,height=6.0cm]{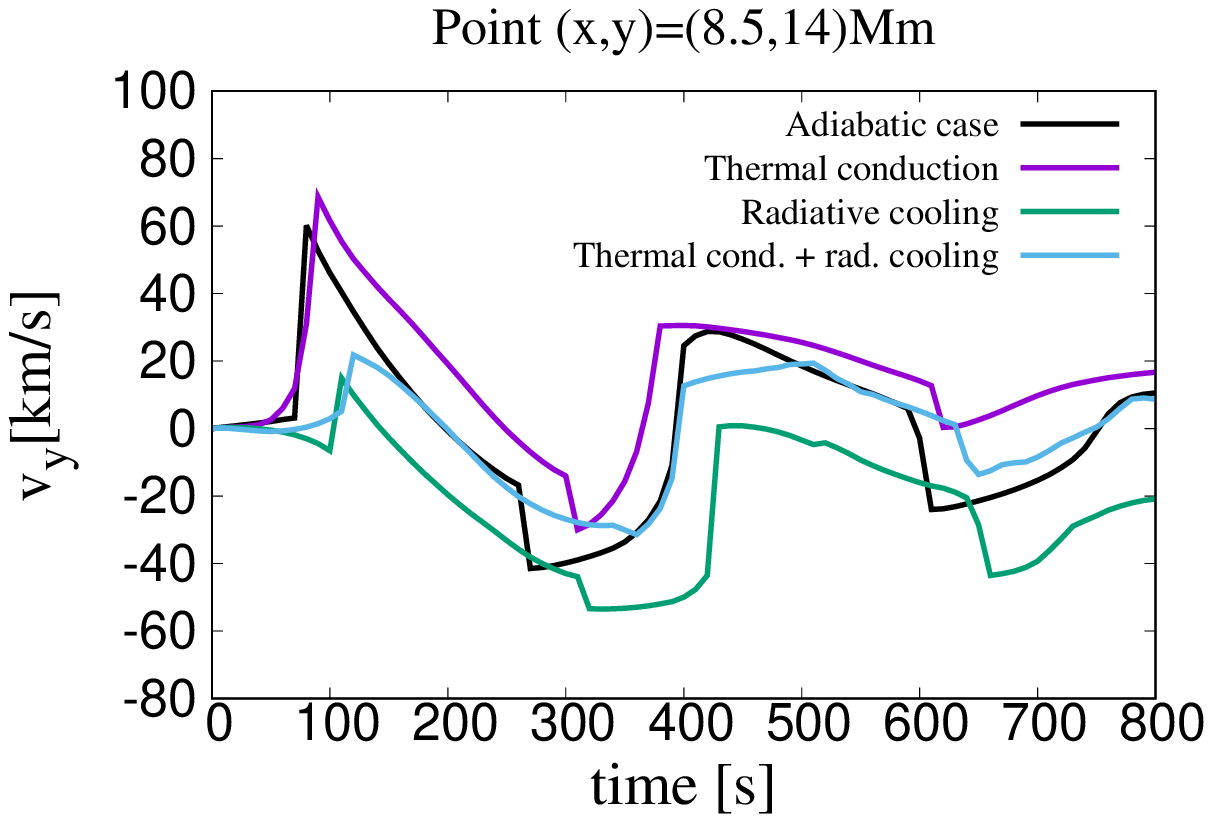}\\
\includegraphics[width=8.0cm,height=6.0cm]{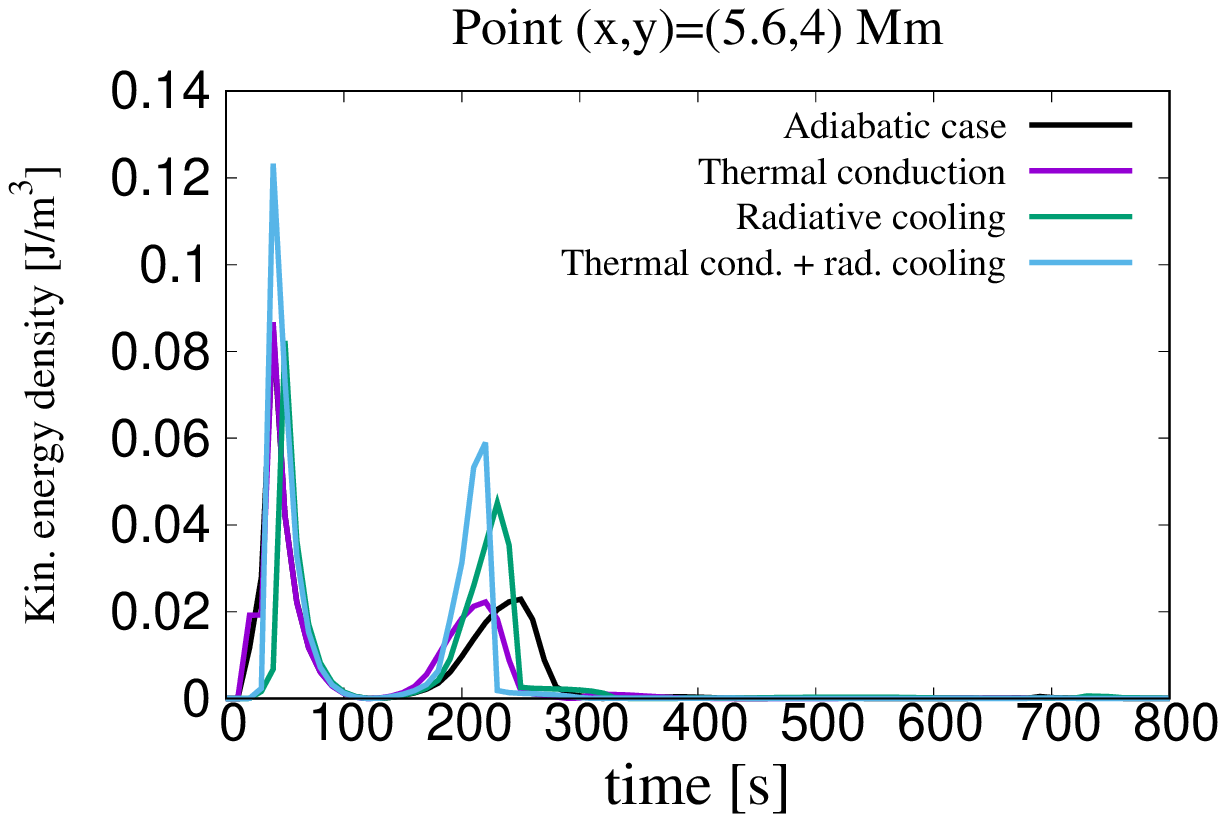}
\includegraphics[width=8.0cm,height=6.0cm]{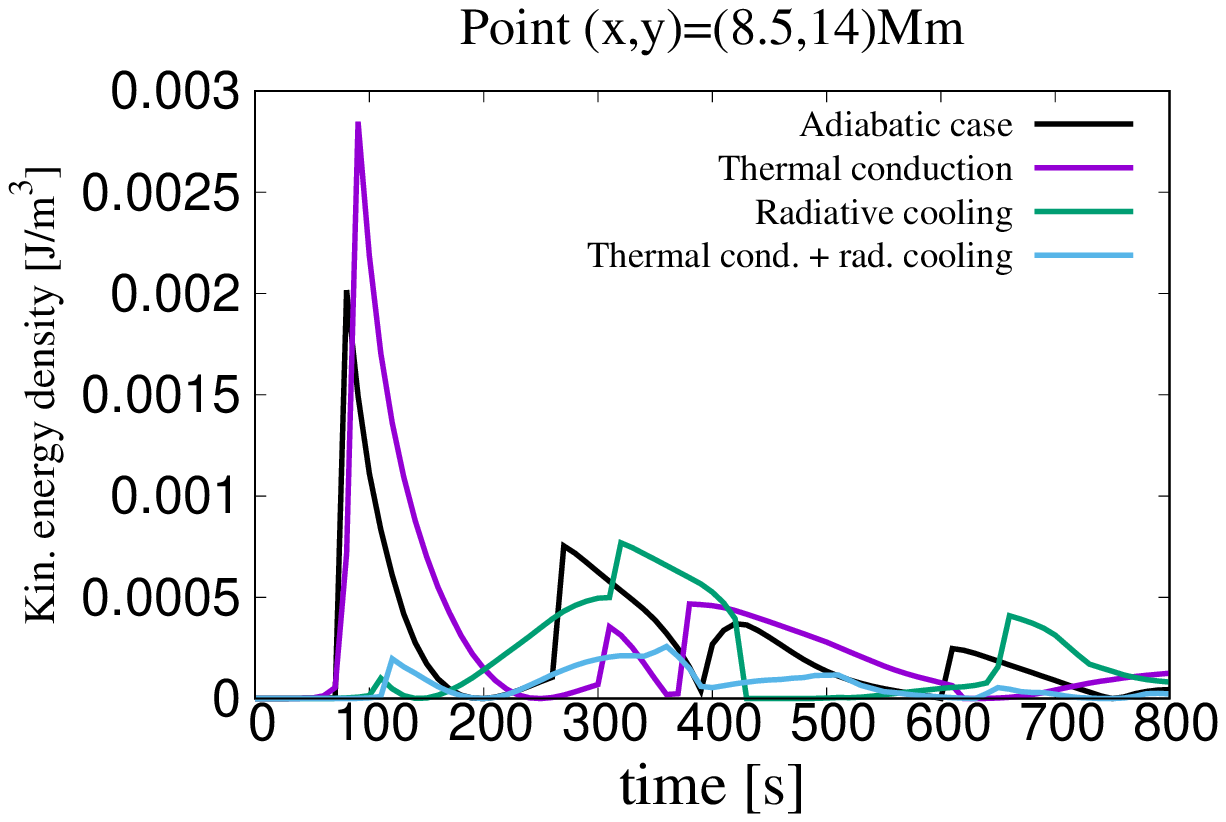}\\
\includegraphics[width=8.0cm,height=6.0cm]{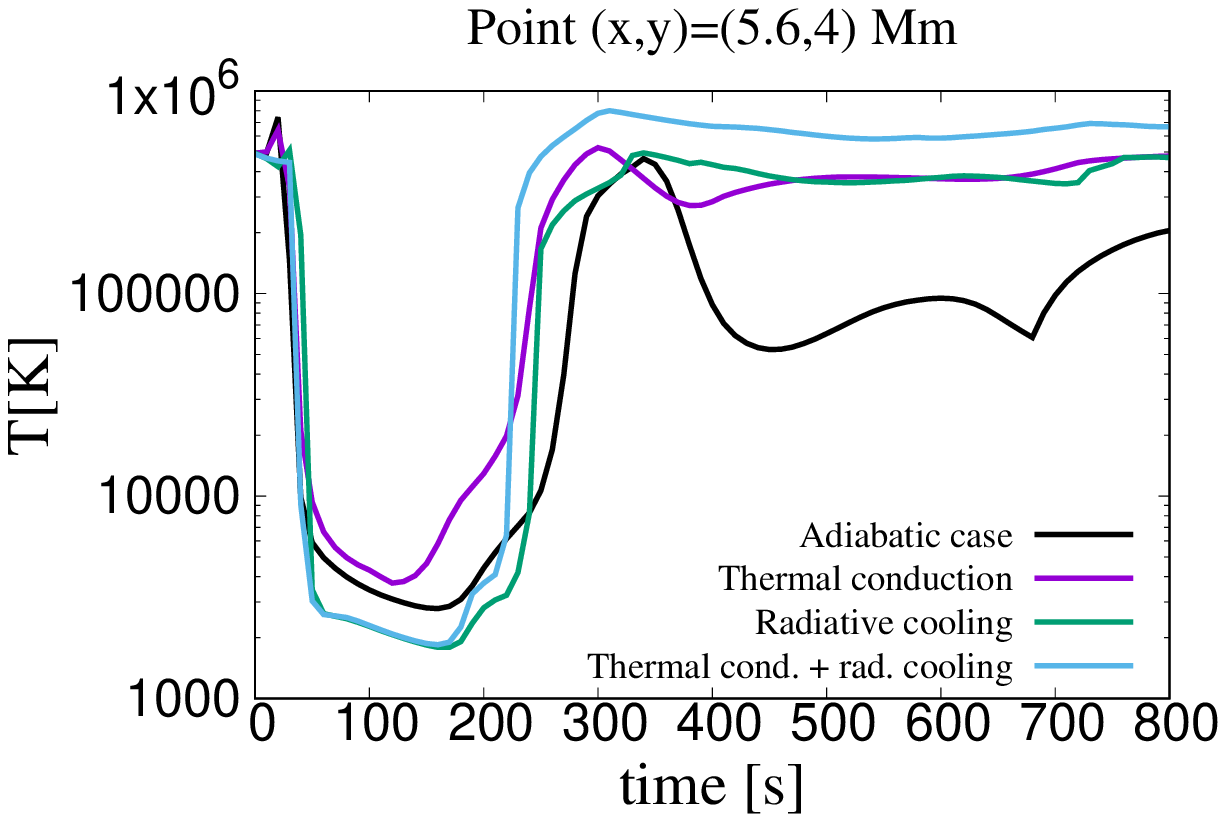}
\includegraphics[width=8.0cm,height=6.0cm]{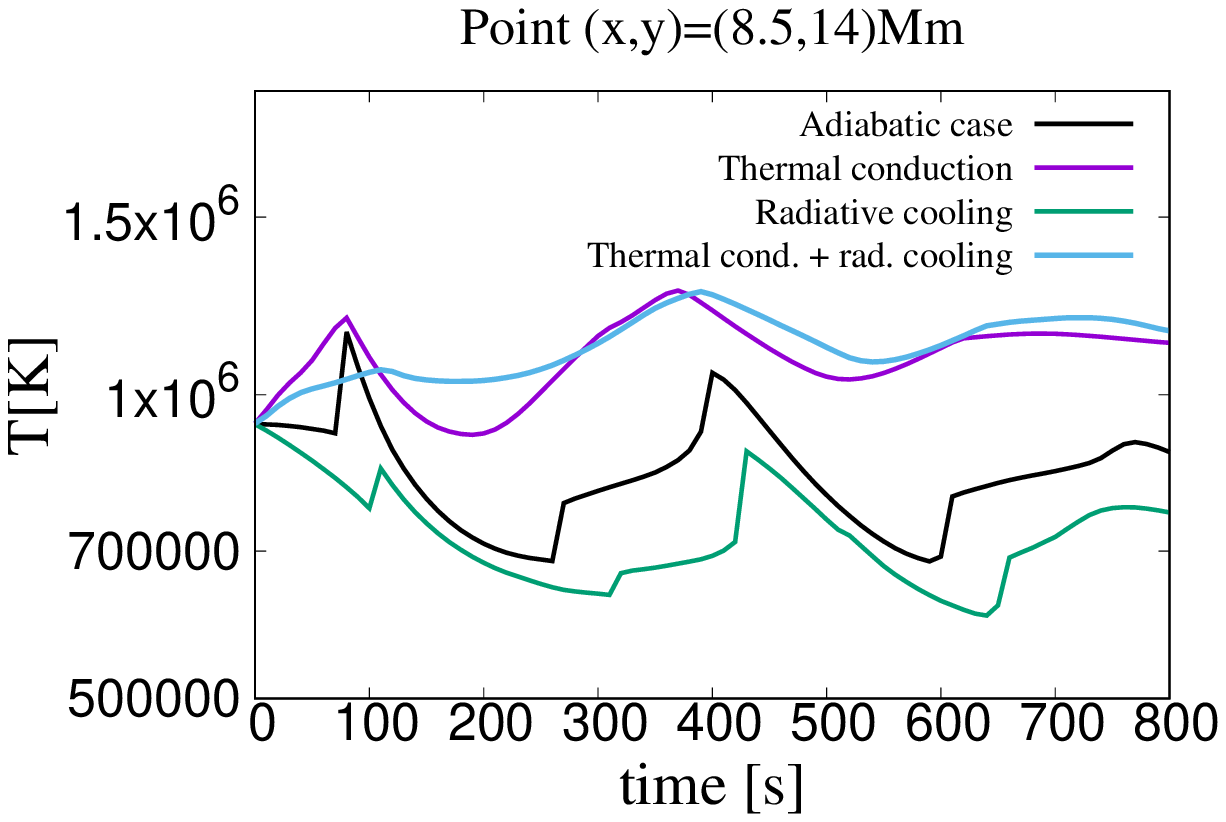}
\caption{Time signatures of the vertical velocity, $v_{y}$, in km s$^{-1}$ (top), kinetic energy density, in J m$^{-3}$ (middle), and logarithm of temperature, in Kelvin (bottom), collected at the points ($x=5.6$, $y=4$) Mm (left) and ($x=8.5$, $y=14$) Mm (right).}
\label{fig:vy_kin_energy_temp_vs_time_comparisson}
\end{figure*}

\begin{figure*}
\centering
\includegraphics[width=4.5cm,height=5.5cm]{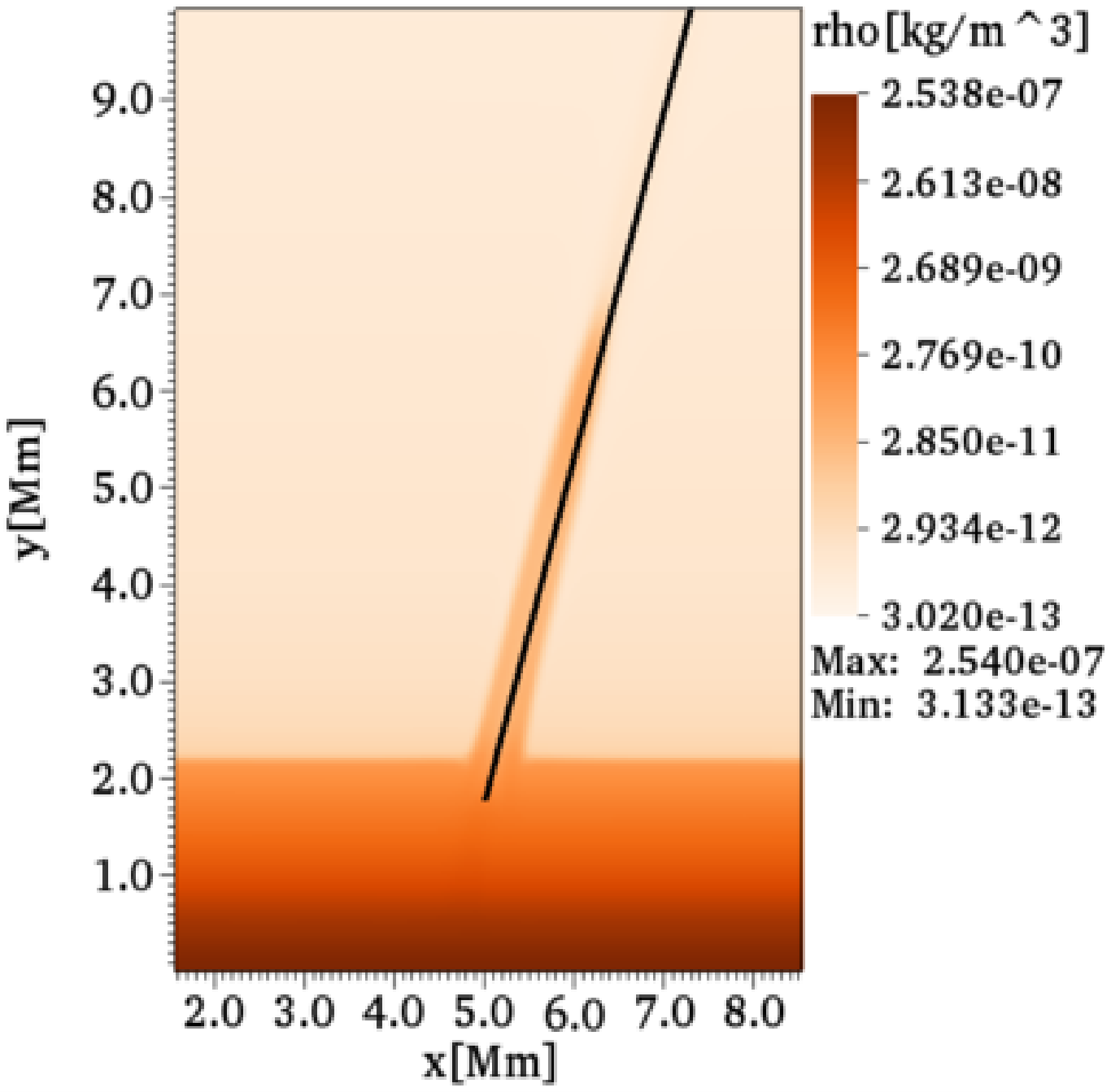}
\includegraphics[width=4.5cm,height=5.5cm]{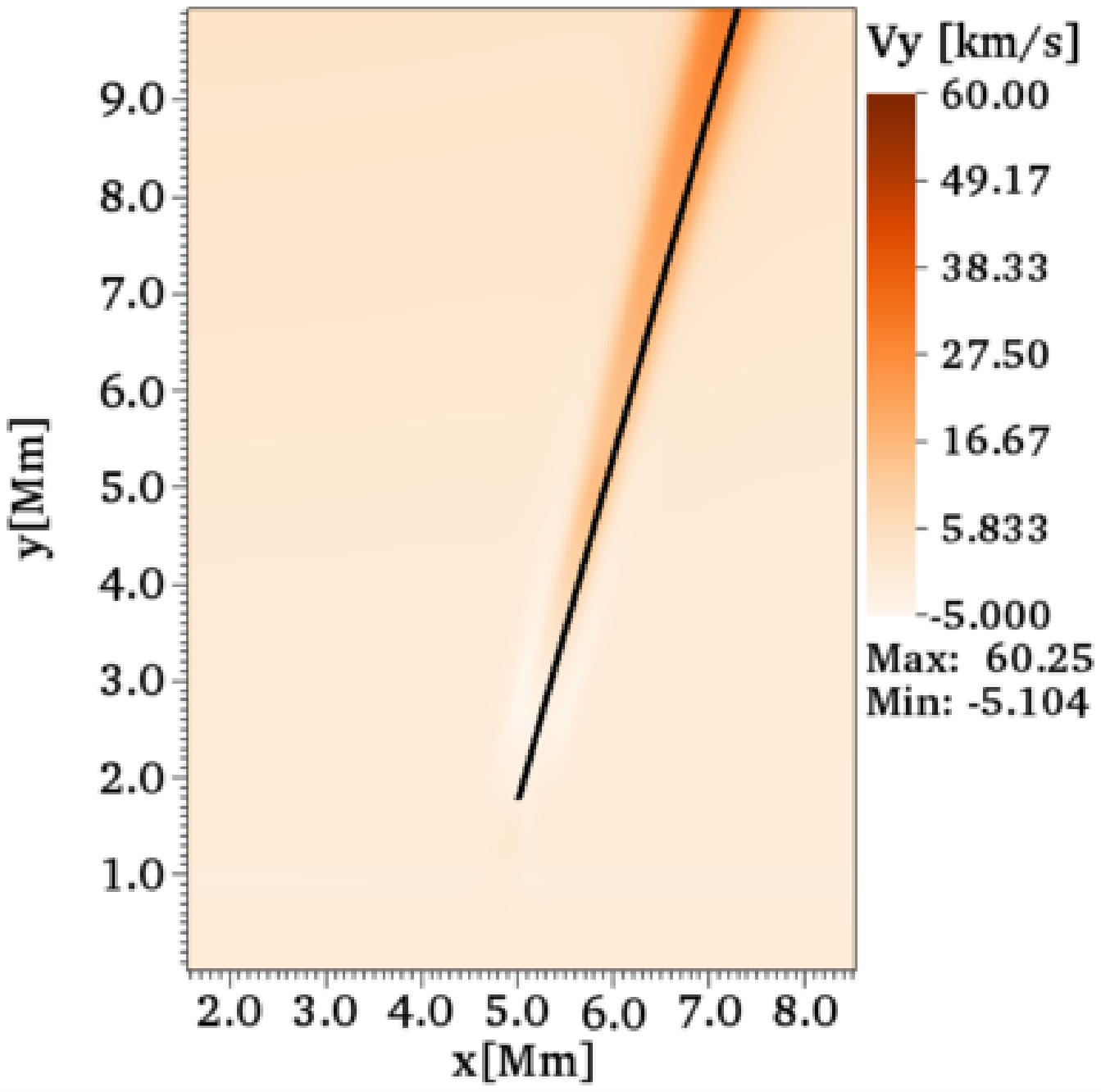}
\caption{(Left) The logarithm of mass density, $\varrho$, in kg m$^{-3}$, and vertical velocity, $v_y$, in km s$^{-1}$ (right) with the drawing of the line in black, where we calculate the distance-time diagrams.}
\label{fig:mass_density_velocity_along_line}
\end{figure*}

\begin{figure*}
\centering
\includegraphics[width=7.0cm,height=4.5cm]{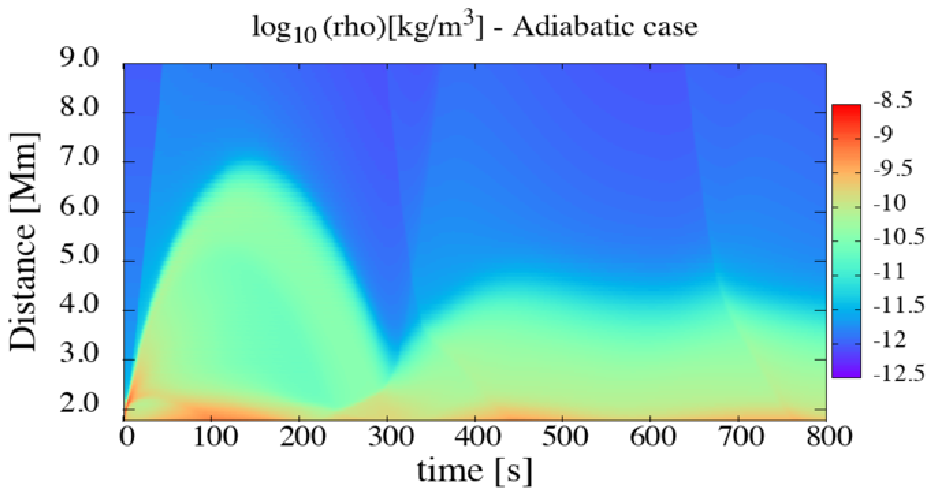}
\includegraphics[width=7.0cm,height=4.5cm]{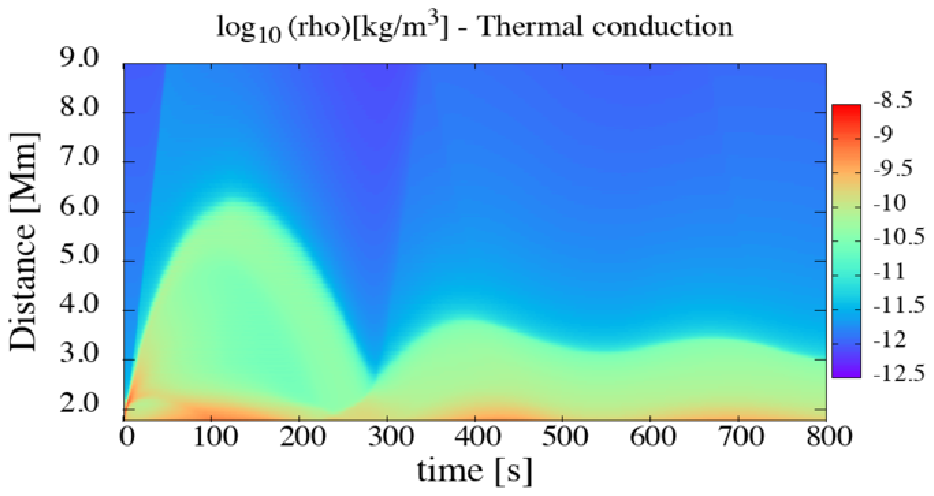}\\
\includegraphics[width=7.0cm,height=4.5cm]{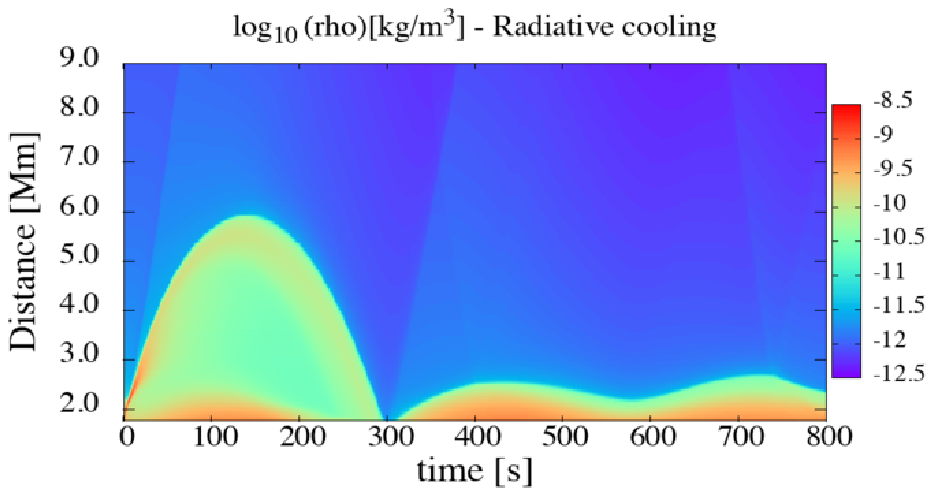}
\includegraphics[width=7.0cm,height=4.5cm]{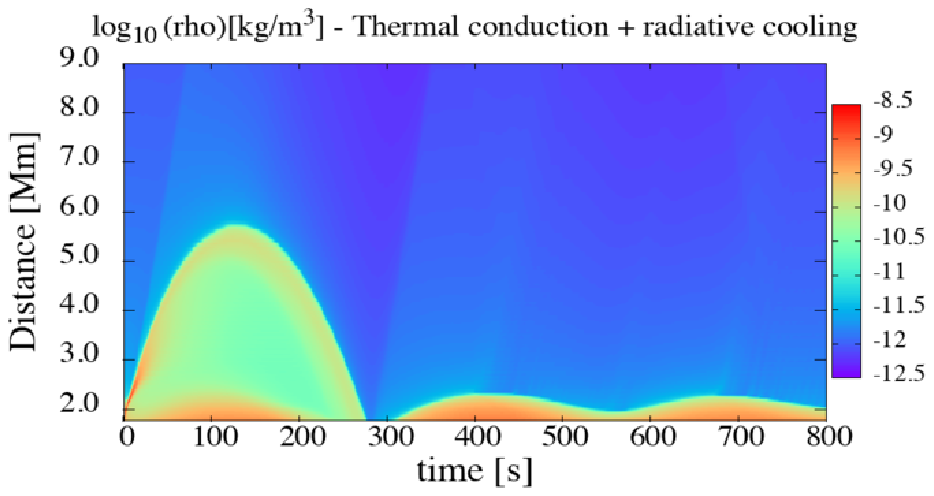}
\caption{Distance-time diagrams of the logarithm of mass density, $\varrho$, in kg m$^{-3}$, corresponding to the adiabatic case (top-left), thermal conduction (top-right), radiative cooling (bottom-left), and thermal conduction + radiative cooling (bottom-right) cases.}
\label{fig:Distance-time_diagrams_density}
\end{figure*}

For instance, the spatial profiles of temperature display in the panels (a) of Figure \ref{fig:temp_evolution_comparison} for the adiabatic case show a jet that reaches a maximum height of about 7 Mm. Instead, some of the observed macrospicules at the north of a polar region using the 304 {\AA} filter of the Atmospheric Imaging Assembly (AIA) onboard the Solar Dynamics Observatory (SDO) shows maximum heights of $\approx 12$ Mm \citep{Murawski_et_al_2011}. Therefore, we can note that our simulated jet is smaller than the observed one but is at the lower range limit; since the macrospicules might reach 7-40 Mm above the limb \citep{Sterling_2000}. Regarding the lifetime, in the adiabatic case, the simulated jet is stable until the total simulation time ($t=800$ s), indicating that the jet can last $\approx 13$ min. This lifetime is consistent with the statistical estimation of the lifetime ($\approx 15$ min) of the observed macrospicules in the He II 304 {\AA} line using the AIA \citep{Loboda&Bogachev_2019}. Besides, the width of our simulated jet is $\sim 1$ Mm, which is approximately three times thinner than the observed width of macrospicules $\sim 3$ Mm \citep{Bohlin_et_al_1975,Dere_et_al_1989,Loboda&Bogachev_2019}. The maximum velocity $\sim 90$ km s$^{-1}$ of the simulated jet as shown in the time series of the vertical velocity at the point (5.6,4) Mm display in Figure \ref{fig:vy_kin_energy_temp_vs_time_comparisson} is consistent with the rise velocities (70-140 km) in the macrospicules observed in the He II 304 {\AA} line \citep{Loboda&Bogachev_2019}. 

\begin{figure*}
\centering
\includegraphics[width=7.0cm,height=4.5cm]{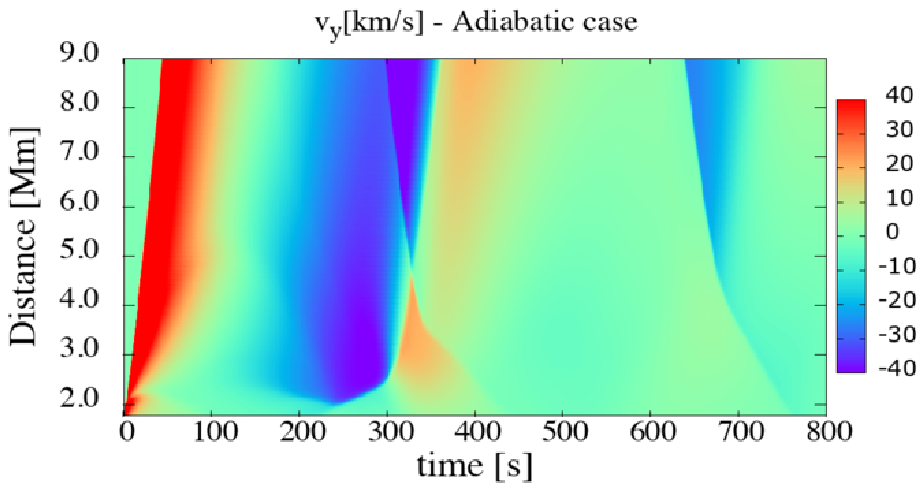}
\includegraphics[width=7.0cm,height=4.5cm]{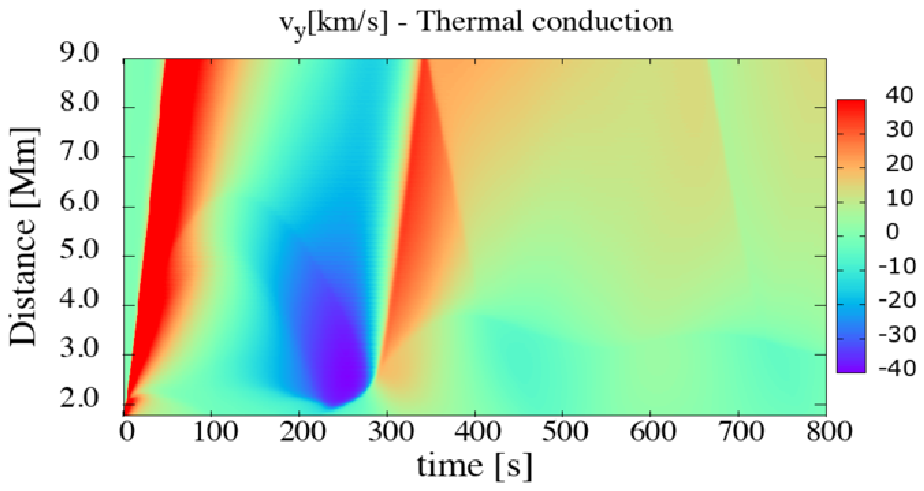}\\
\includegraphics[width=7.0cm,height=4.5cm]{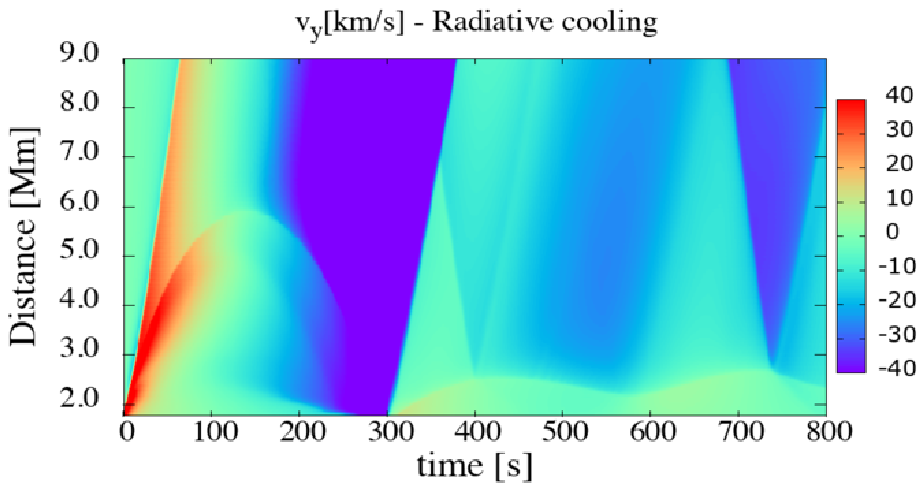}
\includegraphics[width=7.0cm,height=4.5cm]{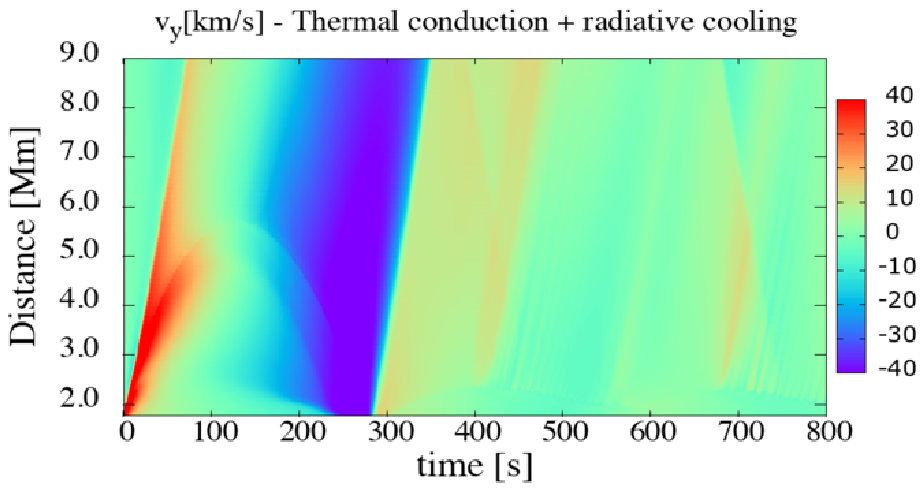}
\caption{Distance-time diagrams of vertical velocity $v_{y}$, in km s$^{-1}$, corresponding to the adiabatic case (top-left), thermal conduction (top-right), radiative cooling (bottom-left), and thermal conduction + radiative cooling (bottom-right) cases.}
\label{fig:Distance-time_diagrams_velocity}
\end{figure*}

\begin{figure*}
\centering
\centerline{\Large \bf   
      \hspace{0.055\textwidth}  \color{black}{\normalsize{(a) Adiabatic}}
      \hspace{0.1\textwidth}  \color{black}{\normalsize{(b) Thermal conduction}}
      \hspace{0.06\textwidth}  \color{black}{\normalsize{(c) Radiative cooling}}
      \hspace{0.1\textwidth}  \color{black}{\normalsize{(d) TC + RC}}
         \hfill}
\includegraphics[width=4.3cm,height=5.5cm]{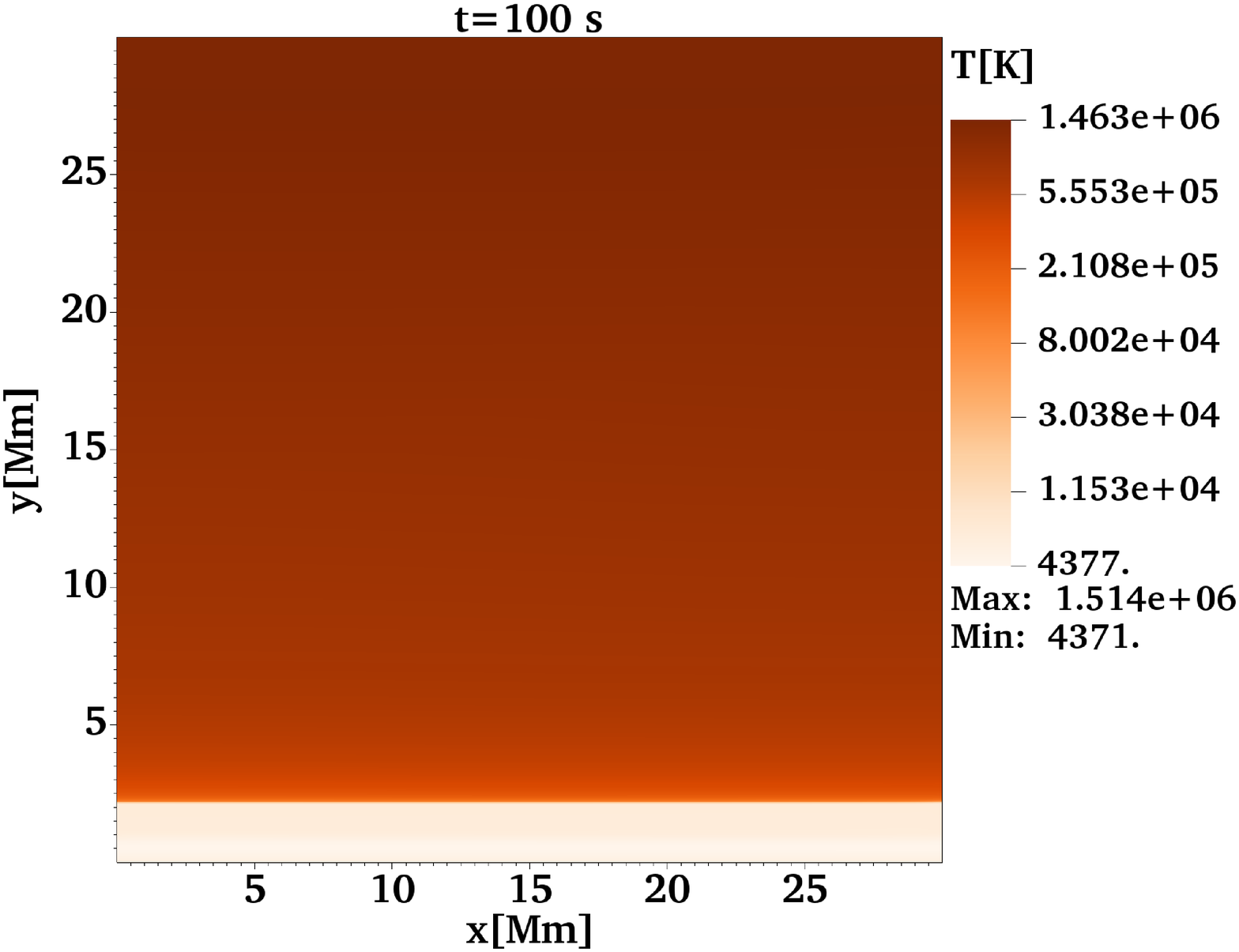}
\includegraphics[width=4.3cm,height=5.5cm]{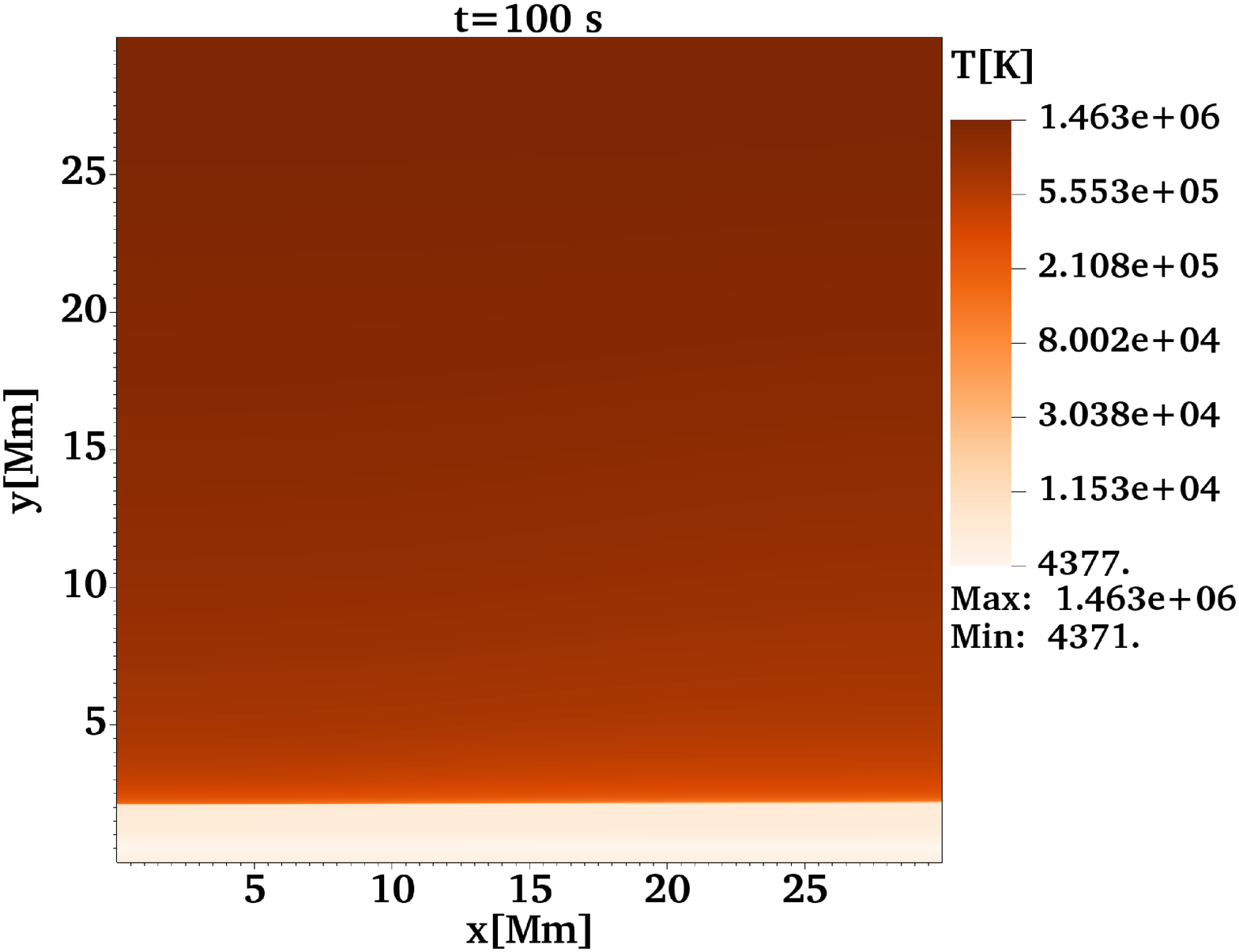}
\includegraphics[width=4.3cm,height=5.5cm]{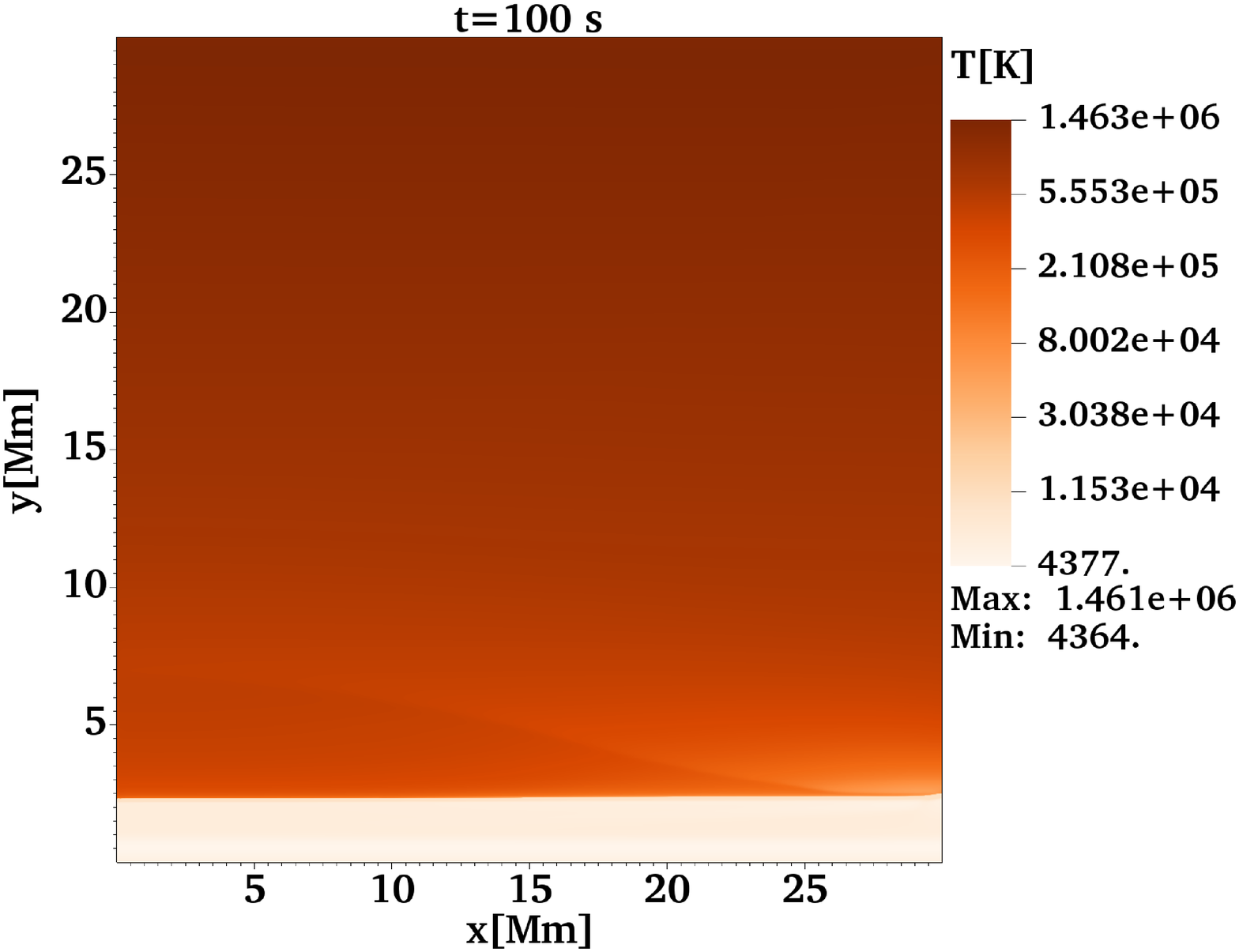}
\includegraphics[width=4.3cm,height=5.5cm]{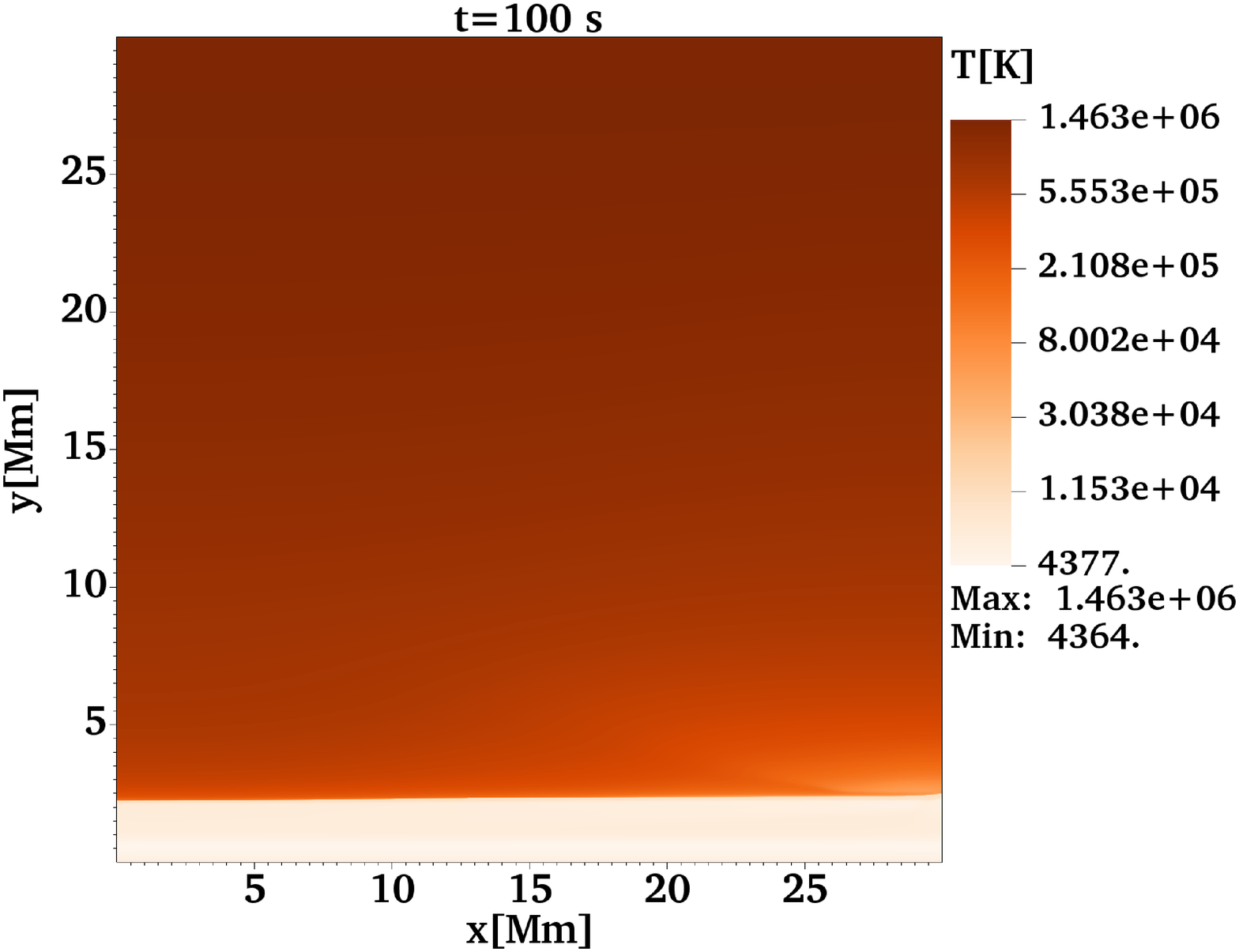}\\
\includegraphics[width=4.3cm,height=5.5cm]{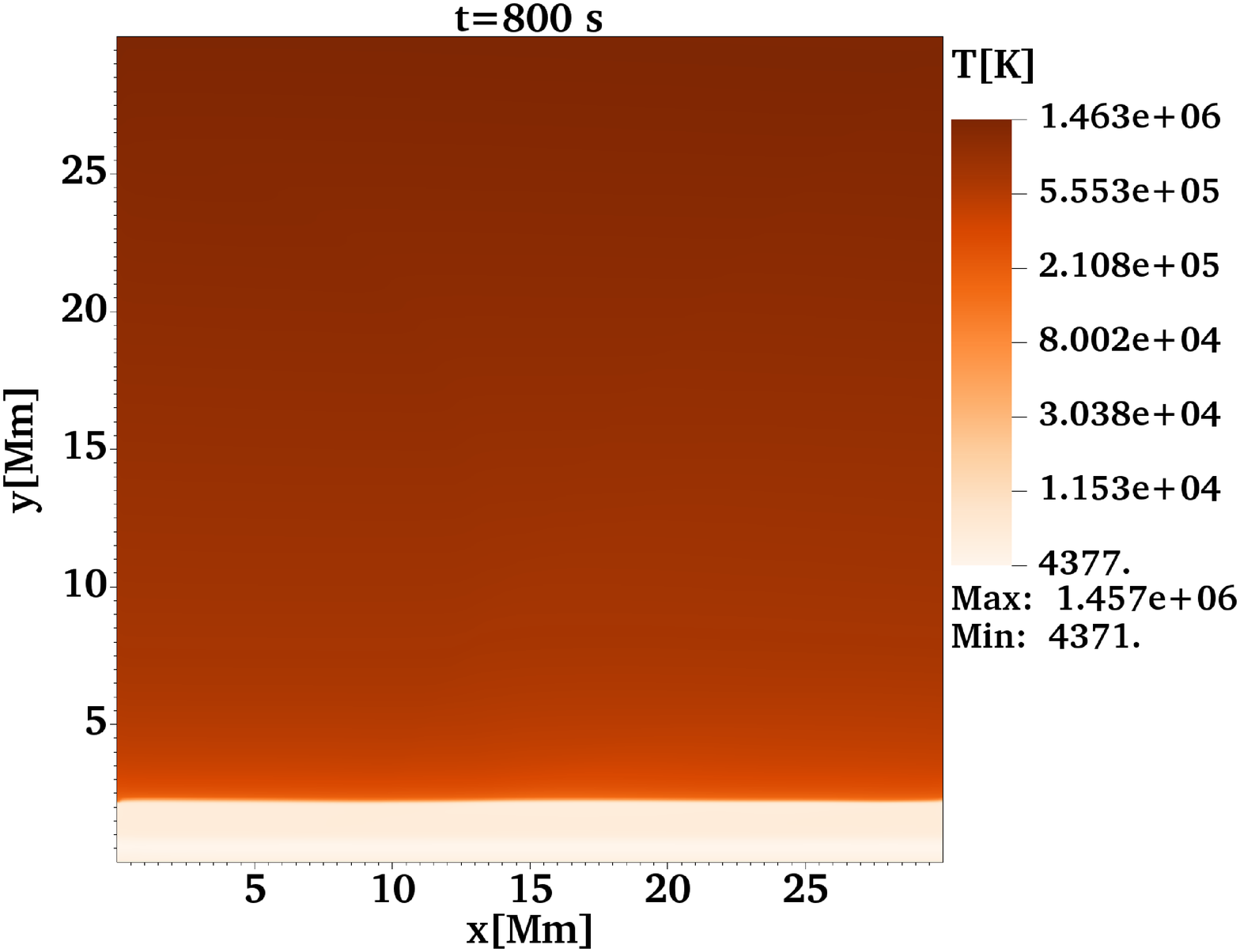}
\includegraphics[width=4.3cm,height=5.5cm]{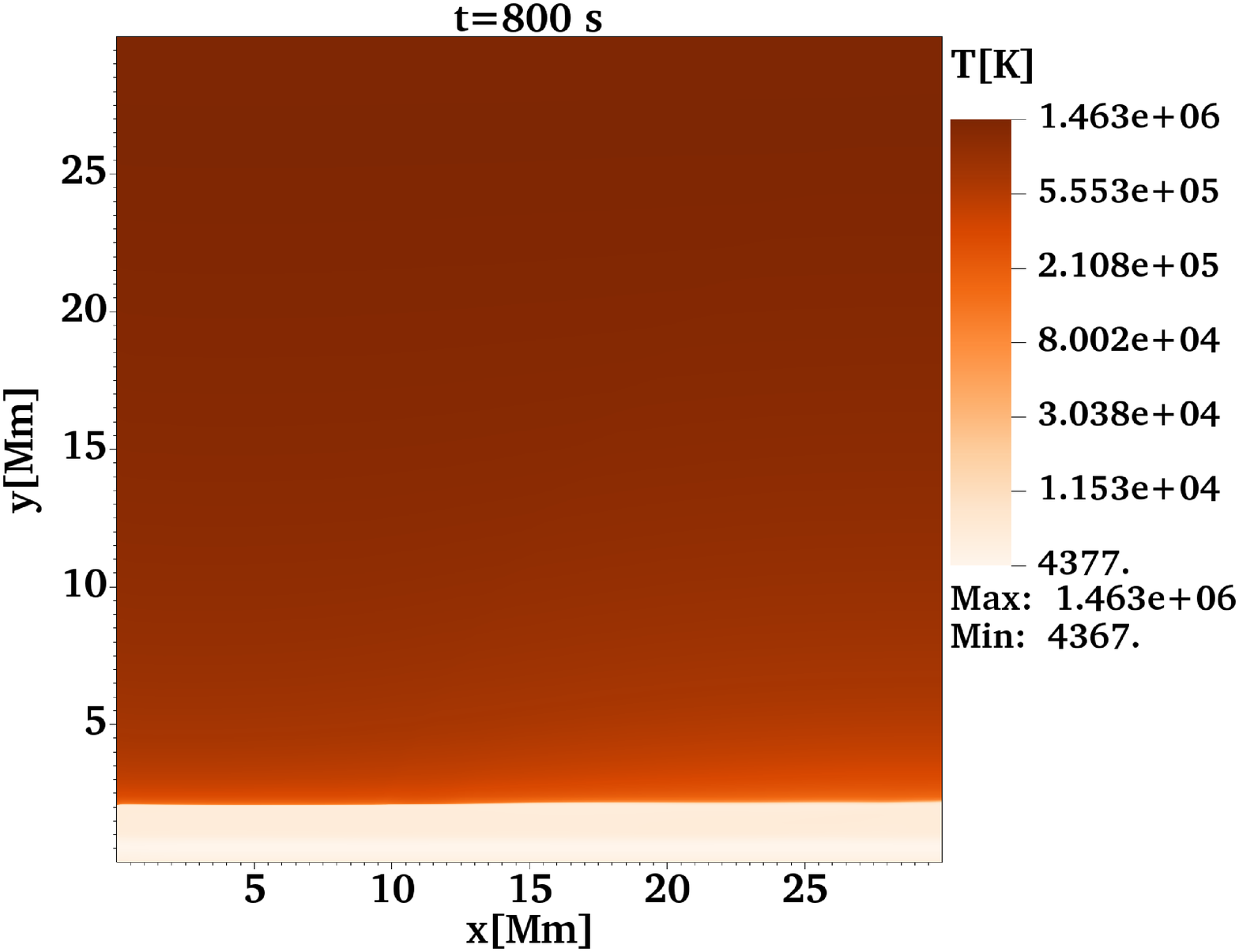}
\includegraphics[width=4.3cm,height=5.5cm]{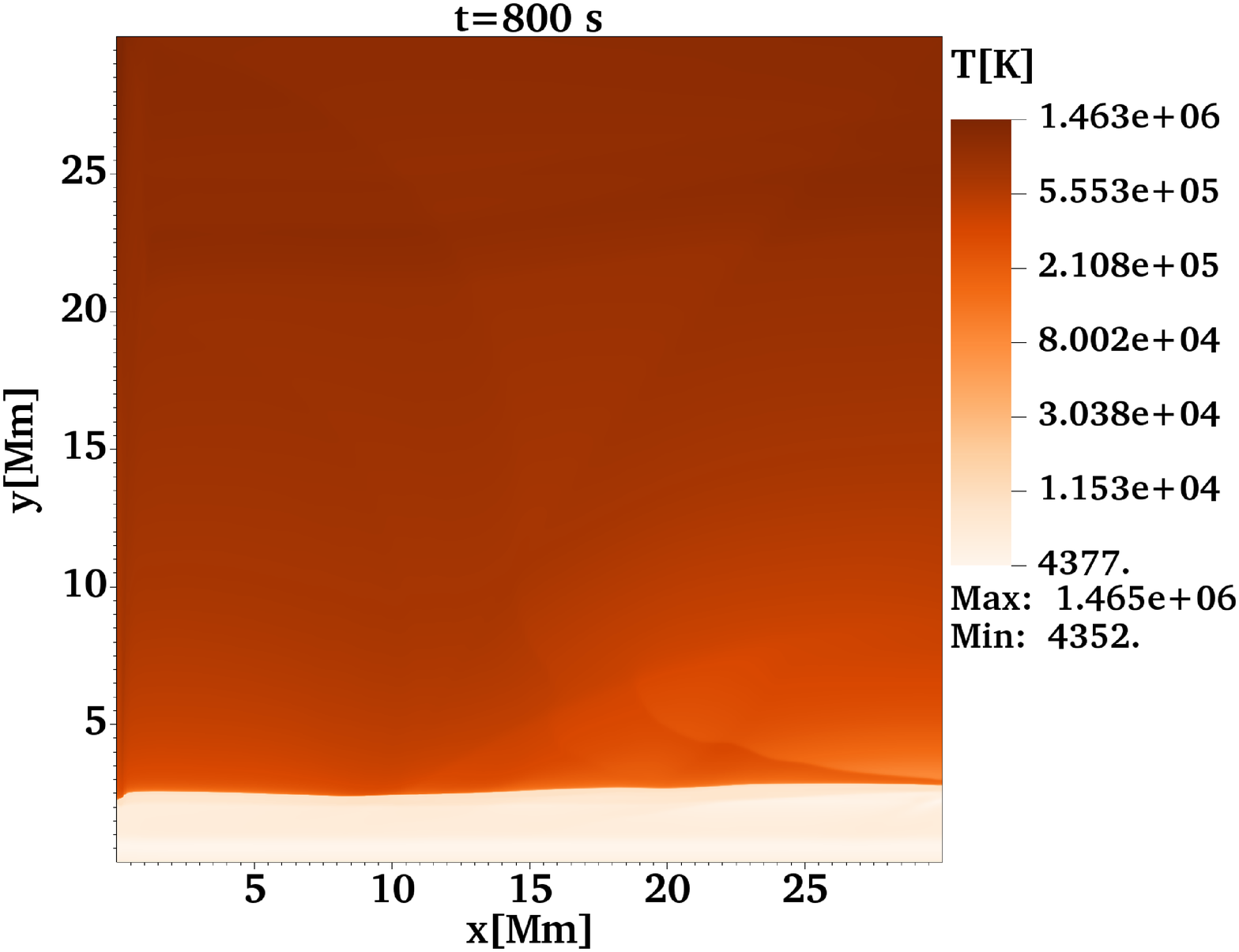}
\includegraphics[width=4.3cm,height=5.5cm]{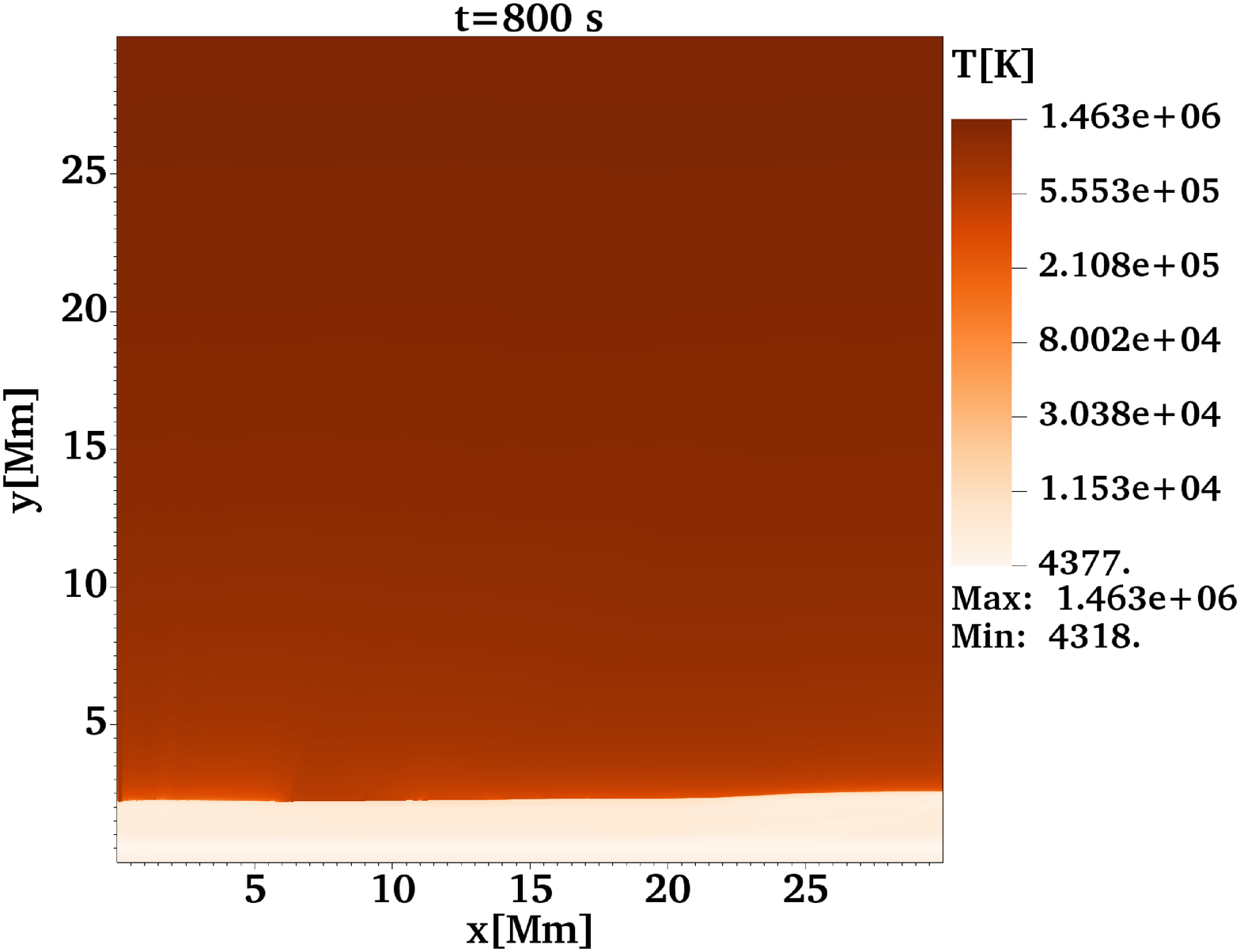}
\caption{Spatial profiles of temperature (in kelvin) at $t=100$ s, $t=400$ s and $t=800$ s in the whole domain for the background test for the adiabatic case (first column), thermal conduction (second column), radiative cooling (third column), and thermal conduction + radiative cooling (fourth column) cases.}
\label{fig:temp_background_test}
\end{figure*}

\begin{figure*}
\centering
\includegraphics[width=6.5cm,height=5.0cm]{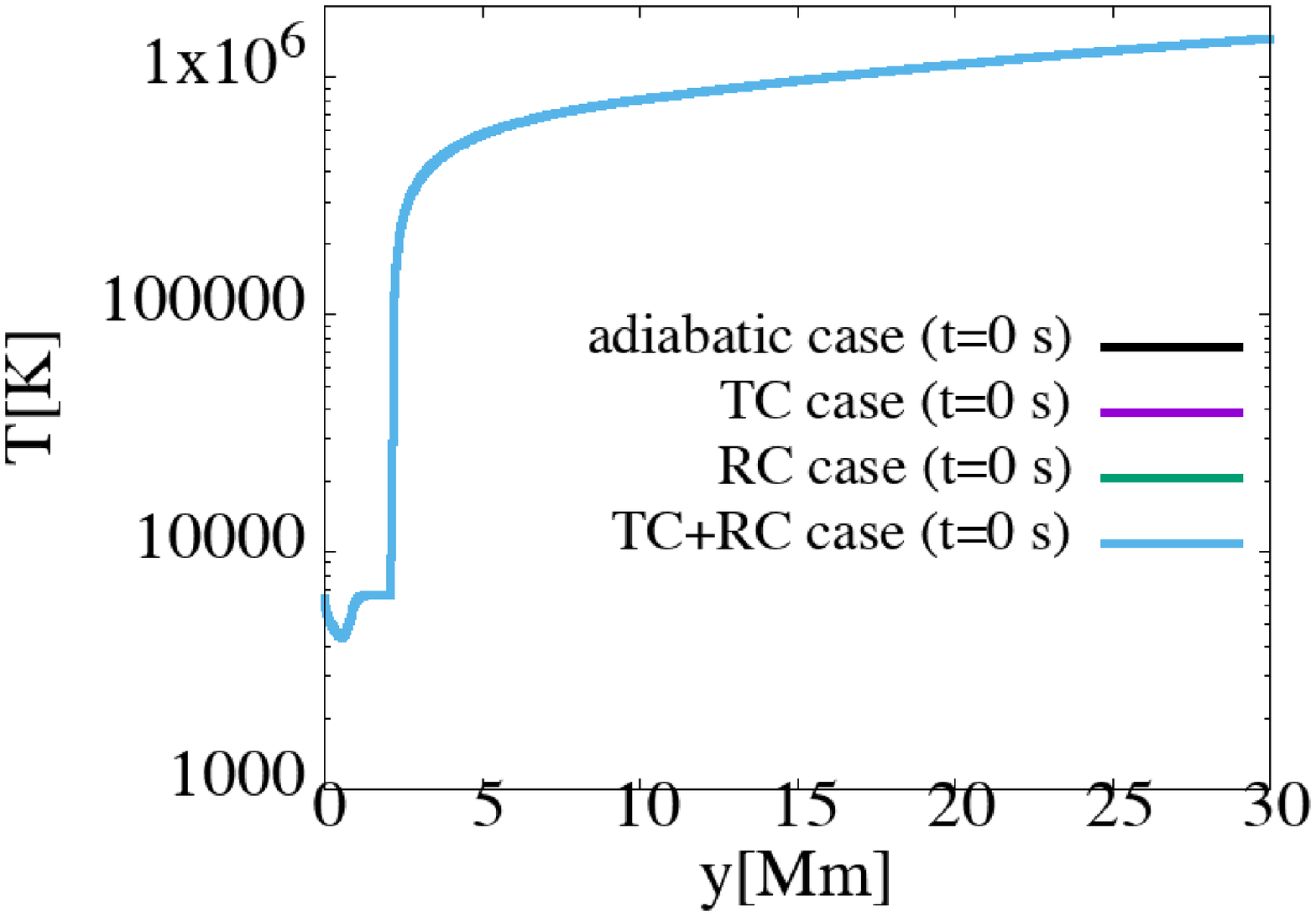}
\includegraphics[width=6.5cm,height=5.0cm]{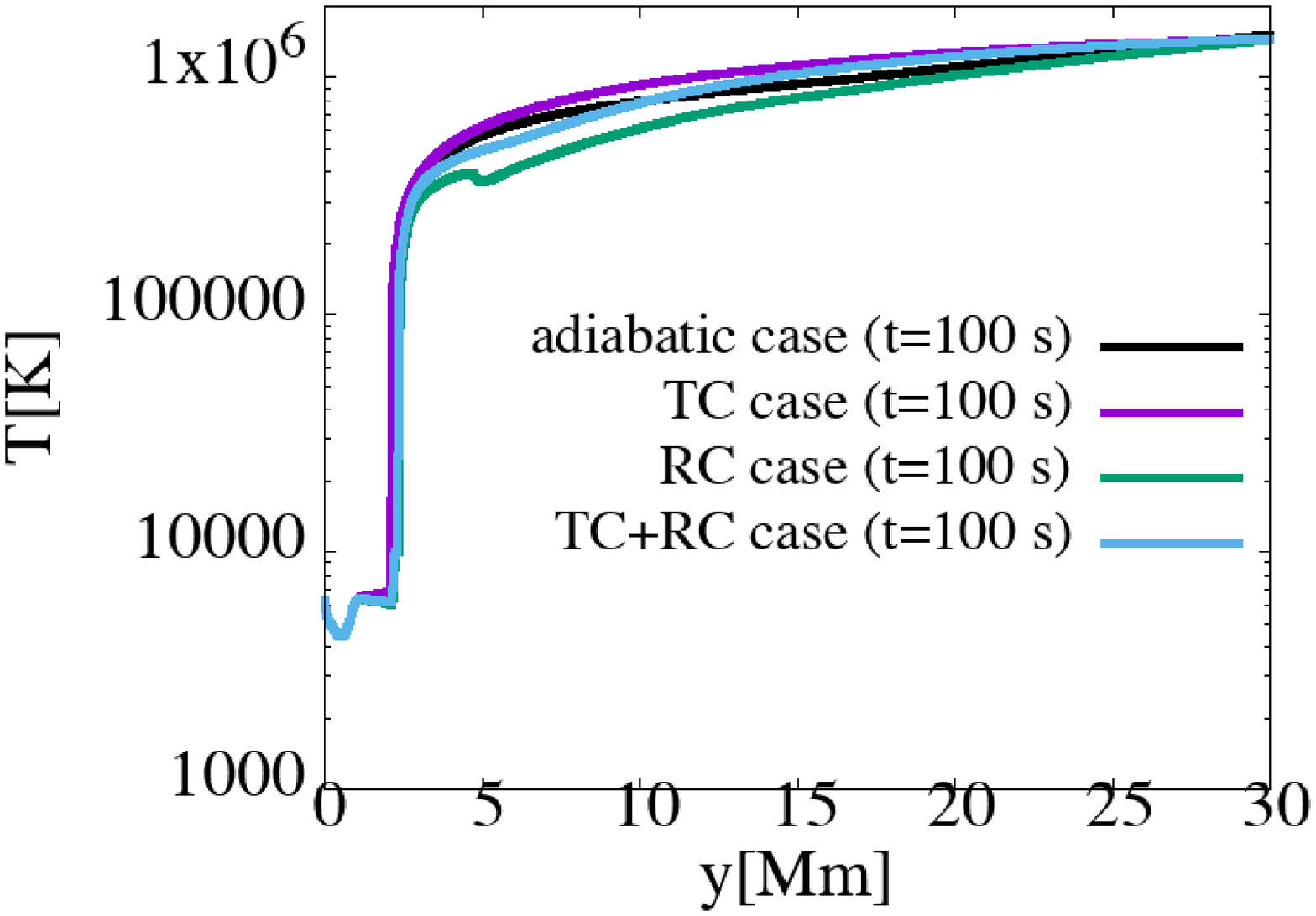}\\
\includegraphics[width=6.5cm,height=5.0cm]{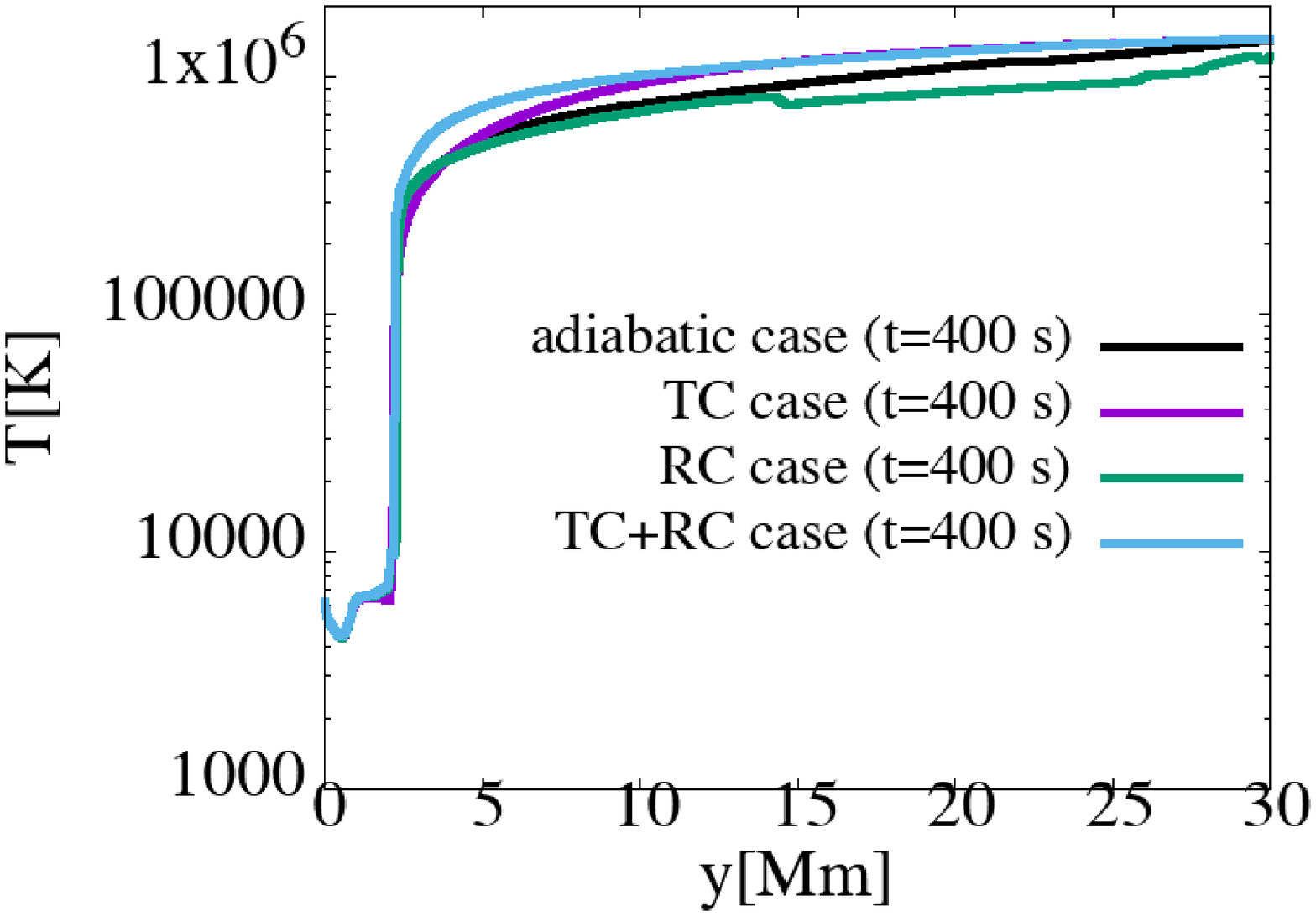}
\includegraphics[width=6.5cm,height=5.0cm]{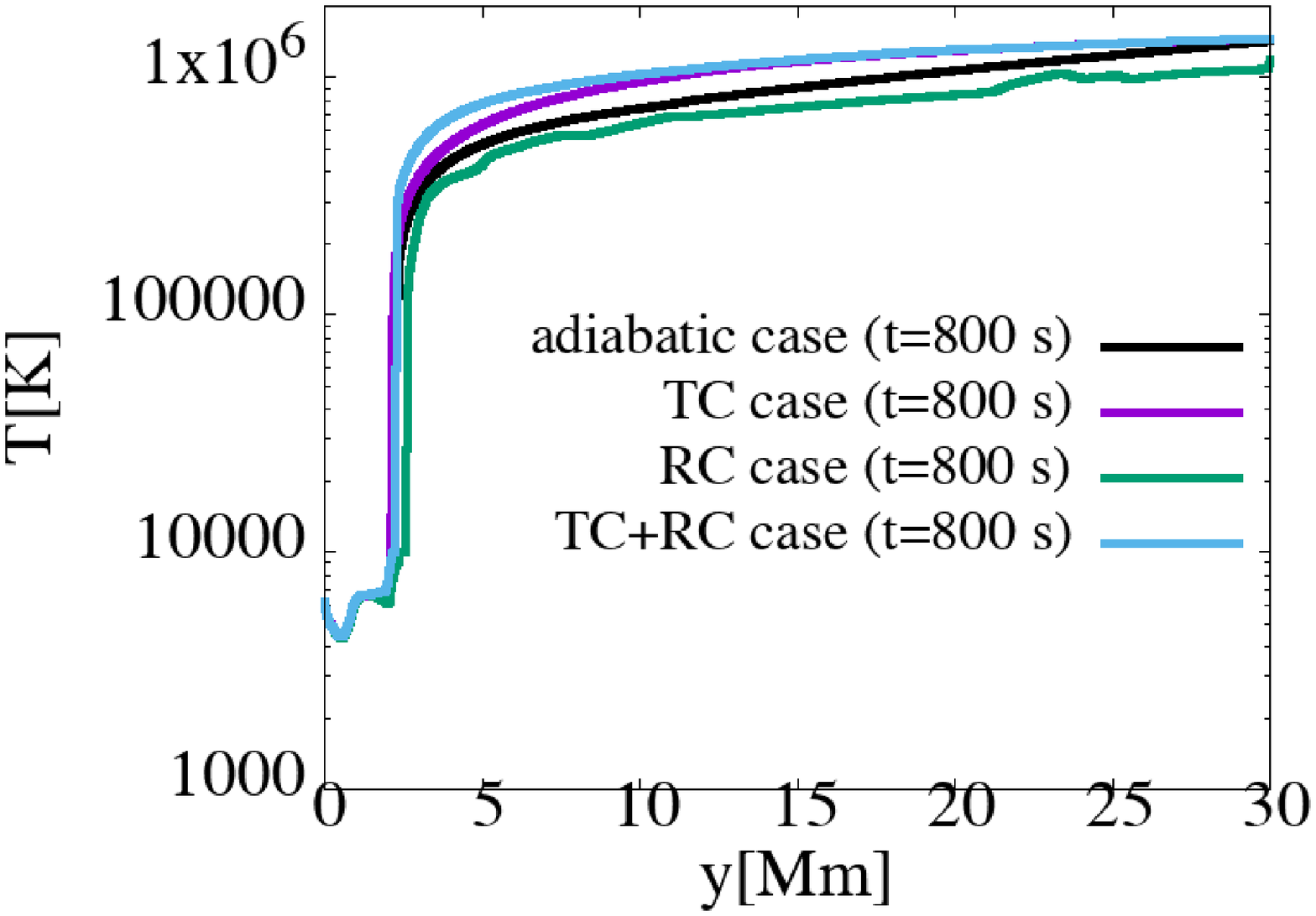}
\caption{Snapshots at four times ($t=0,100,400,800$ s) of the solar background model's plasma temperature in Kelvin as a function of height in Mm corresponding to the four different test cases.}
\label{fig:temp_vs_height_background_test}
\end{figure*}

In the thermal conduction case, we see that the jet reaches a smaller maximum height ($\sim 6$ Mm) than in the adiabatic case as shown in panels (b) of Figure \ref{fig:temp_evolution_comparison}. In this case, the altitude of the jet at $t=800$ s is more similar to the heights $2-6$ Mm observed for type I spicules \citep{Beckers_1968,Suematsu_et_al_1995,Pereira_et_al_2012}. However, in terms of lifetime and velocity, the simulated jet has similar parameters to the adiabatic case. Moreover, the simulation with thermal conduction produces a more efficient interchange between the jet and the solar corona, making the jet hotter ($\sim 5\times10^{5}$ K) than in the adiabatic case. This temperature is in the range of observed spicules surrounded by a hotter shell (1-2)$\times10^{5}$ K \citep{Habbal&Gonzalez_1991}. In the radiative cooling and thermal conduction with radiative cooling cases, we can see that at $t=100$ s, the simulated jets reach a height $\sim 5.5$ Mm in both cases as shown in panels (c) and (d) of Figure \ref{fig:temp_evolution_comparison}. This height is more related to the morphology of the type I spicules than of a macrospicule. However, in the energy imbalance conditions of the solar background atmosphere the radiative cooling quickly dissipates the jet, and the remaining small plasma structures do not resemble either type I spicule or macrospicules. Another essential feature that we can highlight is the parabolic trajectory that follows the jets in the adiabatic and thermal conduction cases, as shown in the time-distance plots of Figures \ref{fig:Distance-time_diagrams_velocity} and \ref{fig:Distance-time_diagrams_velocity}. This parabolic behavior has some similarities with the trajectories of macrospicules observed in the He II 304 {\AA} using high-cadence observations of the AIA, as shown in Figure 2(g) of \cite{Loboda&Bogachev_2019} until 300 s. After that time, the magnetoacoustic shock seems to have a period of $sim 10$ min, i.e., 5 min greater than our simulated jets.

% --------------------------------------------------------------
% ----->        Conclusions and Final comments     <-----
% --------------------------------------------------------------

\section{Conclusions and final comments}
\label{sec:conclusions}

The primary objective of this paper was to test the effects of the background flows caused by thermal conduction and radiative cooling when there is no energy balance in the solar atmosphere model on the behavior of jet with some macrospicules characteristics. Precisely, we estimate the effects of thermal conduction and radiative cooling separately and together on the jets' behavior compared to the adiabatic case, similar to the simulated by \citet{Murawski_et_al_2011}. We excite the jet's formation through a vertical velocity localized pulse, which quickly transforms into shocks in the upper chromosphere. The shock propagates upwards, not lifting the material but propagates the disturbance.

According to the results of the test cases, we found that in the simulation with thermal conduction acting solely, the jet reached a smaller maximum height of about 6.5 Mm than the adiabatic test case. It also makes the corona gets hotter, produced by the most efficient heat driven by the thermal conduction. Another difference is that jets do not fully develop double-structure at the apex; therefore, the bidirectional flows are not present.

In the radiative cooling test case, we found that the jets are shorter ($\approx$ 6 Mm) and slower ($\approx$ 40 km s$^{-1}$) than in the adiabatic and thermal conduction cases. The radiative cooling exerts an effect on the jets, making them dissipated and very short, mainly due to the energy imbalance background atmosphere. It also impacts morphology and the jets' temperature, velocity, and kinetic energy. Especially in the case when thermal conduction and radiative cooling are acting together, we found comparable results to the radiative cooling case itself. In particular, we can identify that radiative cooling dominates over thermal conduction during the jet's evolution. In general terms, we can explain how the radiative cooling naturally induces a cooling of the coronal plasma, which in turn decreases the gravitational scale height in the corona, leads to an enthalpy flux (mass flux) towards the chromosphere. In these terms, the jet propagating up the background flows towards the chromosphere will not reach as high as it would without the background flows being present.

Furthermore, as we show in this paper, the inclusion of effects, such as thermal conduction and radiative cooling, impacts the jets' dynamics as long as we initially consider no external heating that balances such background flows. This scenario may be likely possible if we think that background flows are always present in the solar atmosphere and the corona. Therefore, in this paper, we do not balance the thermal conduction and radiative losses initially. Even though the flows' presence, the corona remains stable, with a temperature of $10^{6}$ K in all the scenarios. Regarding non-equilibrium conditions in the solar atmosphere, we can see, for instance, \cite{Chen_et_al_2014} presented the first model that couples the formation of the corona of an active solar region to a model of the emergence of a sunspot pair. One exciting result of that paper was that the emerging coronal loop never reached an equilibrium state, even when there is initially an energy balance background.
Also, the non-equilibrium conditions are more frequently related to non-equilibrium ionization in the lower solar corona and short-lived releases of energy occurring at low densities \citep{Dudik_et_al_2017}. Furthermore, processes such as thermal instability and non-local thermal equilibrium are present in a significant fraction of coronal loops and on the long-observed coronal rain \citep[see, e.g.,][]{Antolin_2020}. The author mentioned that instability processes are related to a state of the system that undergoes heating and cooling cycles around an equilibrium position due to a feedback mechanism, i.e., the equilibrium may or may not be attainable. Moreover, \cite{Vashalomidze_et_al_2015} found that the energy estimation of coronal loops shows that the radiation is higher than the heat input, which indicates so-called catastrophic cooling. The cooling was accompanied by the formation of coronal rain in the form of falling cold plasma. The latter result makes more plausible this paper's objective and opens possibilities to explore the effects of the energy fluxes on non-equilibrium scenarios. It also allows us to study more appropriate coronal heating functions to balance the energy background flows and be consistent with the coronal heating mechanisms.
 
Finally, in this paper, we do not consider the interaction effects between ions and neutrals in the simulations. However, in \citet{Kuridze_et_al_2016}, the authors showed that ion-neutral collisions are an essential source of dissipation in the dynamics of spicules. For that reason, in a future study, we have to consider the presence of ions and neutral modeled by the two-fluid MHD equations to investigate their possible effect on spicule and macrospicules dynamics, following some ideas of previous investigations \citep[see, e.g.,][]{Kuzma_et_al_2017b, Srivastava_et_al_2018, Wojcik_et_al_2018, Wojcik_et_al_2019}. Therefore, we will include the complete radiative transfer in the model and compare it with synthetic maps of macrospicule observations in a future study.

\section*{Acknowledgements}

We thank the anonymous referee for the thoughtful comments that helped to improve the manuscript. JGA thanks C\'atedras CONACYT (CONACYT Fellow) for partially support this work. The C\'atedras-CONACYT Program, project 1045 sponsor space Weather Service Mexico (SCIESMEX). K.M's work was done within the framework of the projects from the Polish Science Center (NCN) Grant Nos. 2017/25/B/ST9/00506. and 2020/37/B/ST9/00184. The work of TVZ was funded by the Austrian Science Fund (FWF, project P30695-N27). JAGE thanks to CONACYT LN 314845 and CONACYT-AEM 2017-01-292684 grants for supporting this paper. We carry out the simulations in the facilities of ``Centro de Superc\'computo de Clima Espacial (CESCOM)" part of the ``Laboratorio Nacional de Clima Espacial (LANCE)" in Mexico, and of the Big Mamma Cluster at ``Instituto de F\'isica y Matem\'aticas," UMSNH, Morelia, M\'exico. We visualized the simulation data using the VisIt software package \citep{Childs_et_al_2012}. 

%%%%%%%%%%%%%%%%%%%%%%%%%%%%%%%%%%%%%%%%%%%%%%%%%%
\section*{Data Availability}

No new data were generated or analysed in support of this research.



\bibliographystyle{mnras}
\bibliography{Macrospicules_MNRAS_2021} % if your bibtex file is called example.bib

\begin{thebibliography}{}
\makeatletter
\relax
\def\mn@urlcharsother{\let\do\@makeother \do\$\do\&\do\#\do\^\do\_\do\%\do\~}
\def\mn@doi{\begingroup\mn@urlcharsother \@ifnextchar [ {\mn@doi@}
  {\mn@doi@[]}}
\def\mn@doi@[#1]#2{\def\@tempa{#1}\ifx\@tempa\@empty \href
  {http://dx.doi.org/#2} {doi:#2}\else \href {http://dx.doi.org/#2} {#1}\fi
  \endgroup}
\def\mn@eprint#1#2{\mn@eprint@#1:#2::\@nil}
\def\mn@eprint@arXiv#1{\href {http://arxiv.org/abs/#1} {{\tt arXiv:#1}}}
\def\mn@eprint@dblp#1{\href {http://dblp.uni-trier.de/rec/bibtex/#1.xml}
  {dblp:#1}}
\def\mn@eprint@#1:#2:#3:#4\@nil{\def\@tempa {#1}\def\@tempb {#2}\def\@tempc
  {#3}\ifx \@tempc \@empty \let \@tempc \@tempb \let \@tempb \@tempa \fi \ifx
  \@tempb \@empty \def\@tempb {arXiv}\fi \@ifundefined
  {mn@eprint@\@tempb}{\@tempb:\@tempc}{\expandafter \expandafter \csname
  mn@eprint@\@tempb\endcsname \expandafter{\@tempc}}}

\bibitem[\protect\citeauthoryear{Abbett \& Fisher}{Abbett \&
  Fisher}{2012}]{Abbett&Fisher_2012}
Abbett W.~P.,  Fisher G.~H.,  2012, \mn@doi [Solar Physics]
  {10.1007/s11207-011-9817-3}, 277, 176

\bibitem[\protect\citeauthoryear{Alexiades, Amiez  \& Gremaud}{Alexiades
  et~al.}{1996}]{Alexiades_et_al_1996}
Alexiades V.,  Amiez G.,   Gremaud P.-A.,  1996, \mn@doi [Communications in
  Numerical Methods in Engineering]
  {https://doi.org/10.1002/(SICI)1099-0887(199601)12:1<31::AID-CNM950>3.0.CO;2-5},
  12, 31

\bibitem[\protect\citeauthoryear{{Antolin}}{{Antolin}}{2020}]{Antolin_2020}
{Antolin} P.,  2020, \mn@doi [Plasma Physics and Controlled Fusion]
  {10.1088/1361-6587/ab5406}, \href
  {https://ui.adsabs.harvard.edu/abs/2020PPCF...62a4016A} {62, 014016}

\bibitem[\protect\citeauthoryear{{Avrett} \& {Loeser}}{{Avrett} \&
  {Loeser}}{2008}]{AvretLoeser2008}
{Avrett} E.~H.,  {Loeser} R.,  2008, \mn@doi [\apjs] {10.1086/523671}, \href
  {https://ui.adsabs.harvard.edu/abs/2008ApJS..175..229A} {175, 229}

\bibitem[\protect\citeauthoryear{{Beckers}}{{Beckers}}{1968}]{Beckers_1968}
{Beckers} J.~M.,  1968, \mn@doi [\solphys] {10.1007/BF00171614}, \href
  {https://ui.adsabs.harvard.edu/abs/1968SoPh....3..367B} {3, 367}

\bibitem[\protect\citeauthoryear{Bennett \& Erd{\'{e}}lyi}{Bennett \&
  Erd{\'{e}}lyi}{2015}]{Bennet&Erdelyi_2015}
Bennett S.~M.,  Erd{\'{e}}lyi R.,  2015, \mn@doi [The Astrophysical Journal]
  {10.1088/0004-637x/808/2/135}, 808, 135

\bibitem[\protect\citeauthoryear{{Bohlin}, {Vogel}, {Purcell}, {Sheeley},
  {Tousey}  \& {Vanhoosier}}{{Bohlin} et~al.}{1975}]{Bohlin_et_al_1975}
{Bohlin} J.~D.,  {Vogel} S.~N.,  {Purcell} J.~D.,  {Sheeley} N.~R. J.,
  {Tousey} R.,   {Vanhoosier} M.~E.,  1975, \mn@doi [\apjl] {10.1086/181794},
  \href {https://ui.adsabs.harvard.edu/abs/1975ApJ...197L.133B} {197, L133}

\bibitem[\protect\citeauthoryear{{Botha}, {Arber}  \& {Hood}}{{Botha}
  et~al.}{2011}]{Botha_et_al_2011}
{Botha} G. J.~J.,  {Arber} T.~D.,   {Hood} A.~W.,  2011, \mn@doi [A\&A]
  {10.1051/0004-6361/201015534}, 525, A96

\bibitem[\protect\citeauthoryear{Cargill \& Bradshaw}{Cargill \&
  Bradshaw}{2013}]{Cargill&Bradshaw_2013}
Cargill P.~J.,  Bradshaw S.~J.,  2013, \mn@doi [The Astrophysical Journal]
  {10.1088/0004-637x/772/1/40}, 772, 40

\bibitem[\protect\citeauthoryear{{Chen}, {Peter}, {Bingert}  \&
  {Cheung}}{{Chen} et~al.}{2014}]{Chen_et_al_2014}
{Chen} F.,  {Peter} H.,  {Bingert} S.,   {Cheung} M. C.~M.,  2014, \mn@doi
  [A\&A] {10.1051/0004-6361/201322859}, 564, A12

\bibitem[\protect\citeauthoryear{Childs et~al.,}{Childs
  et~al.}{2012}]{Childs_et_al_2012}
Childs H.,  et~al., 2012, {VisIt: An End-User Tool For Visualizing and
  Analyzing Very Large Data}.
Chapman and Hall/CRC

\bibitem[\protect\citeauthoryear{{Dara}, {Koutchmy}  \& {Suematsu}}{{Dara}
  et~al.}{1998}]{Dara_et_al_1998}
{Dara} H.~C.,  {Koutchmy} S.,   {Suematsu} Y.,  1998, in Solar Jets and Coronal
  Plumes. International Meeting on Solar Jets and Coronal Plumes

\bibitem[\protect\citeauthoryear{{De Pontieu}, {Erd{\'e}lyi}  \& {James}}{{De
  Pontieu} et~al.}{2004}]{De_Pontieu_et_al_2004}
{De Pontieu} B.,  {Erd{\'e}lyi} R.,   {James} S.~P.,  2004, \mn@doi [\nat]
  {10.1038/nature02749}, \href
  {https://ui.adsabs.harvard.edu/abs/2004Natur.430..536D} {430, 536}

\bibitem[\protect\citeauthoryear{De~Pontieu et~al.,}{De~Pontieu
  et~al.}{2011}]{De_Pontieu_et_al_2011}
De~Pontieu B.,  et~al., 2011, \mn@doi [Science] {10.1126/science.1197738}, 331,
  55

\bibitem[\protect\citeauthoryear{{Dedner}, {Kemm}, {Kr{\"o}ner}, {Munz},
  {Schnitzer}  \& {Wesenberg}}{{Dedner} et~al.}{2002}]{Dedner_et_al_2002}
{Dedner} A.,  {Kemm} F.,  {Kr{\"o}ner} D.,  {Munz} C.~D.,  {Schnitzer} T.,
  {Wesenberg} M.,  2002, \mn@doi [Journal of Computational Physics]
  {10.1006/jcph.2001.6961}, \href
  {https://ui.adsabs.harvard.edu/abs/2002JCoPh.175..645D} {175, 645}

\bibitem[\protect\citeauthoryear{{Dere, K. P.}}{{Dere, K.
  P.}}{2009}]{Dere_2009}
{Dere, K. P.} 2009, \mn@doi [A\&A] {10.1051/0004-6361/200811329}, 497, 287

\bibitem[\protect\citeauthoryear{{Dere}, {Bartoe}, {Brueckner}, {Cook},
  {Socker}  \& {Ewing}}{{Dere} et~al.}{1989}]{Dere_et_al_1989}
{Dere} K.~P.,  {Bartoe} J. D.~F.,  {Brueckner} G.~E.,  {Cook} J.~W.,  {Socker}
  D.~G.,   {Ewing} J.~W.,  1989, \mn@doi [\solphys] {10.1007/BF00146212}, \href
  {https://ui.adsabs.harvard.edu/abs/1989SoPh..119...55D} {119, 55}

\bibitem[\protect\citeauthoryear{{Dud{\'\i}k} et~al.,}{{Dud{\'\i}k}
  et~al.}{2017}]{Dudik_et_al_2017}
{Dud{\'\i}k} J.,  et~al., 2017, \mn@doi [\solphys] {10.1007/s11207-017-1125-0},
  \href {https://ui.adsabs.harvard.edu/abs/2017SoPh..292..100D} {292, 100}

\bibitem[\protect\citeauthoryear{Fan}{Fan}{2017}]{Fan_2017}
Fan Y.,  2017, \mn@doi [The Astrophysical Journal] {10.3847/1538-4357/aa7a56},
  844, 26

\bibitem[\protect\citeauthoryear{{Georgakilas}, {Koutchmy}  \&
  {Alissandrakis}}{{Georgakilas} et~al.}{1999}]{Georgakilas_et_al_1999}
{Georgakilas} A.~A.,  {Koutchmy} S.,   {Alissandrakis} C.~E.,  1999, \aap,
  \href {https://ui.adsabs.harvard.edu/abs/1999A&A...341..610G} {341, 610}

\bibitem[\protect\citeauthoryear{Gonz{\'{a}}lez-Avil{\'{e}}s, Guzm{\'{a}}n  \&
  Fedun}{Gonz{\'{a}}lez-Avil{\'{e}}s et~al.}{2017}]{Gonzalez-Aviles_et_al_2017}
Gonz{\'{a}}lez-Avil{\'{e}}s J.~J.,  Guzm{\'{a}}n F.~S.,   Fedun V.,  2017,
  \mn@doi [The Astrophysical Journal] {10.3847/1538-4357/836/1/24}, 836, 24

\bibitem[\protect\citeauthoryear{Gonz{\'{a}}lez-Avil{\'{e}}s, Guzm{\'{a}}n,
  Fedun, Verth, Shelyag  \& Regnier}{Gonz{\'{a}}lez-Avil{\'{e}}s
  et~al.}{2018}]{Gonzalez-Aviles_et_al_2018}
Gonz{\'{a}}lez-Avil{\'{e}}s J.~J.,  Guzm{\'{a}}n F.~S.,  Fedun V.,  Verth G.,
  Shelyag S.,   Regnier S.,  2018, \mn@doi [The Astrophysical Journal]
  {10.3847/1538-4357/aab36f}, 856, 176

\bibitem[\protect\citeauthoryear{Gonz{\'{a}}lez-Avil{\'{e}}s, Guzm{\'{a}}n,
  Fedun  \& Verth}{Gonz{\'{a}}lez-Avil{\'{e}}s
  et~al.}{2020}]{Gonzalez-Aviles_et_al_2020}
Gonz{\'{a}}lez-Avil{\'{e}}s J.~J.,  Guzm{\'{a}}n F.~S.,  Fedun V.,   Verth G.,
  2020, \mn@doi [The Astrophysical Journal] {10.3847/1538-4357/ab97b8}, 897,
  153

\bibitem[\protect\citeauthoryear{{Guarrasi, M.}, {Reale, F.}, {Orlando, S.},
  {Mignone, A.}  \& {Klimchuk, J. A.}}{{Guarrasi, M.}
  et~al.}{2014}]{Guarrasi_et_al_2014}
{Guarrasi, M.} {Reale, F.} {Orlando, S.} {Mignone, A.}  {Klimchuk, J. A.} 2014,
  \mn@doi [A\&A] {10.1051/0004-6361/201322848}, 564, A48

\bibitem[\protect\citeauthoryear{{Habbal} \& {Gonzalez}}{{Habbal} \&
  {Gonzalez}}{1991}]{Habbal&Gonzalez_1991}
{Habbal} S.~R.,  {Gonzalez} R.~D.,  1991, \mn@doi [\apjl] {10.1086/186094},
  \href {https://ui.adsabs.harvard.edu/abs/1991ApJ...376L..25H} {376, L25}

\bibitem[\protect\citeauthoryear{{Hansteen}, {De Pontieu}, {Rouppe van der
  Voort}, {van Noort}  \& {Carlsson}}{{Hansteen}
  et~al.}{2006}]{Hansteen_et_al_2006}
{Hansteen} V.~H.,  {De Pontieu} B.,  {Rouppe van der Voort} L.,  {van Noort}
  M.,   {Carlsson} M.,  2006, \mn@doi [\apjl] {10.1086/507452}, \href
  {https://ui.adsabs.harvard.edu/abs/2006ApJ...647L..73H} {647, L73}

\bibitem[\protect\citeauthoryear{{Heggland}, {De Pontieu}  \&
  {Hansteen}}{{Heggland} et~al.}{2007}]{Heggland_et_al_2007}
{Heggland} L.,  {De Pontieu} B.,   {Hansteen} V.~H.,  2007, \mn@doi [\apj]
  {10.1086/518828}, \href
  {https://ui.adsabs.harvard.edu/abs/2007ApJ...666.1277H} {666, 1277}

\bibitem[\protect\citeauthoryear{{Hollweg}}{{Hollweg}}{1982}]{Hollweg_1982}
{Hollweg} J.~V.,  1982, \mn@doi [\apj] {10.1086/159993}, \href
  {https://ui.adsabs.harvard.edu/abs/1982ApJ...257..345H} {257, 345}

\bibitem[\protect\citeauthoryear{{Judge}}{{Judge}}{2006}]{Judge_2006}
{Judge} P.,  2006, in {Leibacher} J.,  {Stein} R.~F.,   {Uitenbroek} H.,  eds,
  Astronomical Society of the Pacific Conference Series Vol. 354, Solar MHD
  Theory and Observations: A High Spatial Resolution Perspective. p.~259

\bibitem[\protect\citeauthoryear{{Kamio}, {Curdt}, {Teriaca}, {Inhester}  \&
  {Solanki}}{{Kamio} et~al.}{2010}]{Kamio_et_al_2010}
{Kamio} S.,  {Curdt} W.,  {Teriaca} L.,  {Inhester} B.,   {Solanki} S.~K.,
  2010, \mn@doi [A\&A] {10.1051/0004-6361/200913269}, 510, L1

\bibitem[\protect\citeauthoryear{{Karovska} \& {Habbal}}{{Karovska} \&
  {Habbal}}{1994}]{Karovska&Habbal_1994}
{Karovska} M.,  {Habbal} S.~R.,  1994, \mn@doi [\apjl] {10.1086/187472}, \href
  {https://ui.adsabs.harvard.edu/abs/1994ApJ...431L..59K} {431, L59}

\bibitem[\protect\citeauthoryear{Kayshap, Srivastava, Murawski  \&
  Tripathi}{Kayshap et~al.}{2013}]{Kayshap_et_al_2013}
Kayshap P.,  Srivastava A.~K.,  Murawski K.,   Tripathi D.,  2013, \mn@doi [The
  Astrophysical Journal] {10.1088/2041-8205/770/1/l3}, 770, L3

\bibitem[\protect\citeauthoryear{Kiss, Gyenge  \& Erd{\'{e}}lyi}{Kiss
  et~al.}{2017}]{Kiss_et_al_2017}
Kiss T.~S.,  Gyenge N.,   Erd{\'{e}}lyi R.,  2017, \mn@doi [The Astrophysical
  Journal] {10.3847/1538-4357/aa5272}, 835, 47

\bibitem[\protect\citeauthoryear{Kuridze, Henriques, Mathioudakis,
  Erd{\'{e}}lyi, Zaqarashvili, Shelyag, Keys  \& Keenan}{Kuridze
  et~al.}{2015}]{Kuridze_et_al_2015}
Kuridze D.,  Henriques V.,  Mathioudakis M.,  Erd{\'{e}}lyi R.,  Zaqarashvili
  T.~V.,  Shelyag S.,  Keys P.~H.,   Keenan F.~P.,  2015, \mn@doi [The
  Astrophysical Journal] {10.1088/0004-637x/802/1/26}, 802, 26

\bibitem[\protect\citeauthoryear{Kuridze, Zaqarashvili, Henriques,
  Mathioudakis, Keenan  \& Hanslmeier}{Kuridze
  et~al.}{2016}]{Kuridze_et_al_2016}
Kuridze D.,  Zaqarashvili T.~V.,  Henriques V.,  Mathioudakis M.,  Keenan
  F.~P.,   Hanslmeier A.,  2016, \mn@doi [The Astrophysical Journal]
  {10.3847/0004-637x/830/2/133}, 830, 133

\bibitem[\protect\citeauthoryear{{Ku{\'{z}}ma, B.}, {Murawski, K.},
  {Zaqarashvili, T. V.}, {Konkol, P.}  \& {Mignone, A.}}{{Ku{\'{z}}ma, B.}
  et~al.}{2017}]{Kuzma_et_al_2017a}
{Ku{\'{z}}ma, B.} {Murawski, K.} {Zaqarashvili, T. V.} {Konkol, P.}  {Mignone,
  A.} 2017, \mn@doi [A\&A] {10.1051/0004-6361/201628747}, 597, A133

\bibitem[\protect\citeauthoryear{Ku{\'{z}}ma, Murawski, Kayshap, W{\'{o}}jcik,
  Srivastava  \& Dwivedi}{Ku{\'{z}}ma et~al.}{2017}]{Kuzma_et_al_2017b}
Ku{\'{z}}ma B.,  Murawski K.,  Kayshap P.,  W{\'{o}}jcik D.,  Srivastava A.~K.,
    Dwivedi B.~N.,  2017, \mn@doi [The Astrophysical Journal]
  {10.3847/1538-4357/aa8ea1}, 849, 78

\bibitem[\protect\citeauthoryear{{Landi}, {Del Zanna}, {Young}, {Dere}  \&
  {Mason}}{{Landi} et~al.}{2012}]{Landi_et_al_2012}
{Landi} E.,  {Del Zanna} G.,  {Young} P.~R.,  {Dere} K.~P.,   {Mason} H.~E.,
  2012, \mn@doi [\apj] {10.1088/0004-637X/744/2/99}, \href
  {https://ui.adsabs.harvard.edu/abs/2012ApJ...744...99L} {744, 99}

\bibitem[\protect\citeauthoryear{{Leenaarts}}{{Leenaarts}}{2020}]{Leenaarts_2020}
{Leenaarts} J.,  2020, \mn@doi [Living Reviews in Solar Physics]
  {10.1007/s41116-020-0024-x}, \href
  {https://ui.adsabs.harvard.edu/abs/2020LRSP...17....3L} {17, 3}

\bibitem[\protect\citeauthoryear{Lipartito, Judge, Reardon  \&
  Cauzzi}{Lipartito et~al.}{2014}]{Lipartito_et_al_2014}
Lipartito I.,  Judge P.~G.,  Reardon K.,   Cauzzi G.,  2014, \mn@doi [The
  Astrophysical Journal] {10.1088/0004-637x/785/2/109}, 785, 109

\bibitem[\protect\citeauthoryear{Loboda \& Bogachev}{Loboda \&
  Bogachev}{2019}]{Loboda&Bogachev_2019}
Loboda I.~P.,  Bogachev S.~A.,  2019, \mn@doi [The Astrophysical Journal]
  {10.3847/1538-4357/aafa7a}, 871, 230

\bibitem[\protect\citeauthoryear{{Low}}{{Low}}{1985}]{Low_1985}
{Low} B.~C.,  1985, \mn@doi [\apj] {10.1086/163211}, \href
  {https://ui.adsabs.harvard.edu/abs/1985ApJ...293...31L} {293, 31}

\bibitem[\protect\citeauthoryear{Mart{\'{\i}}nez-Sykora, Hansteen, Pontieu  \&
  Carlsson}{Mart{\'{\i}}nez-Sykora et~al.}{2009}]{Martinez-Sykora_et_al_2009}
Mart{\'{\i}}nez-Sykora J.,  Hansteen V.,  Pontieu B.~D.,   Carlsson M.,  2009,
  \mn@doi [The Astrophysical Journal] {10.1088/0004-637x/701/2/1569}, 701, 1569

\bibitem[\protect\citeauthoryear{Mignone, Bodo, Massaglia, Matsakos, Tesileanu,
  Zanni  \& Ferrari}{Mignone et~al.}{2007}]{Mignone_et_al_2007}
Mignone A.,  Bodo G.,  Massaglia S.,  Matsakos T.,  Tesileanu O.,  Zanni C.,
  Ferrari A.,  2007, \mn@doi [The Astrophysical Journal Supplement Series]
  {10.1086/513316}, 170, 228

\bibitem[\protect\citeauthoryear{{Miyoshi} \& {Kusano}}{{Miyoshi} \&
  {Kusano}}{2005}]{Miyoshi&Kusano_2005}
{Miyoshi} T.,  {Kusano} K.,  2005, \mn@doi [Journal of Computational Physics]
  {10.1016/j.jcp.2005.02.017}, \href
  {https://ui.adsabs.harvard.edu/abs/2005JCoPh.208..315M} {208, 315}

\bibitem[\protect\citeauthoryear{{Moore}, {Tang}, {Bohlin}  \& {Golub}}{{Moore}
  et~al.}{1977}]{Moore_et_al_1977}
{Moore} R.~L.,  {Tang} F.,  {Bohlin} J.~D.,   {Golub} L.,  1977, \mn@doi [\apj]
  {10.1086/155681}, \href
  {https://ui.adsabs.harvard.edu/abs/1977ApJ...218..286M} {218, 286}

\bibitem[\protect\citeauthoryear{{Murawski} \& {Zaqarashvili}}{{Murawski} \&
  {Zaqarashvili}}{2010}]{Muraswki&Zaqarashvilli_2010}
{Murawski} K.,  {Zaqarashvili} T.~V.,  2010, \mn@doi [\aap]
  {10.1051/0004-6361/201014128}, \href
  {https://ui.adsabs.harvard.edu/abs/2010A&A...519A...8M} {519, A8}

\bibitem[\protect\citeauthoryear{{Murawski}, {Srivastava}  \&
  {Zaqarashvili}}{{Murawski} et~al.}{2011}]{Murawski_et_al_2011}
{Murawski} K.,  {Srivastava} A.~K.,   {Zaqarashvili} T.~V.,  2011, \mn@doi
  [A\&A] {10.1051/0004-6361/201117589}, 535, A58

\bibitem[\protect\citeauthoryear{Navarro, Lora-Clavijo, Murawski  \&
  Poedts}{Navarro et~al.}{2020}]{Navarro_et_al_2020}
Navarro A.,  Lora-Clavijo F.~D.,  Murawski K.,   Poedts S.,  2020, \mn@doi
  [Monthly Notices of the Royal Astronomical Society] {10.1093/mnras/staa3402},
  500, 3329

\bibitem[\protect\citeauthoryear{{Parenti, S.}, {Bromage, B. J. I.}  \&
  {Bromage, G. E.}}{{Parenti, S.} et~al.}{2002}]{Parenti_et_al_2002}
{Parenti, S.} {Bromage, B. J. I.}  {Bromage, G. E.} 2002, \mn@doi [A\&A]
  {10.1051/0004-6361:20011819}, 384, 303

\bibitem[\protect\citeauthoryear{{Pasachoff}, {Jacobson}  \&
  {Sterling}}{{Pasachoff} et~al.}{2009}]{Pasachoff_et_al_2009}
{Pasachoff} J.~M.,  {Jacobson} W.~A.,   {Sterling} A.~C.,  2009, \mn@doi
  [\solphys] {10.1007/s11207-009-9430-x}, \href
  {https://ui.adsabs.harvard.edu/abs/2009SoPh..260...59P} {260, 59}

\bibitem[\protect\citeauthoryear{{Pereira}, {De Pontieu}  \&
  {Carlsson}}{{Pereira} et~al.}{2012}]{Pereira_et_al_2012}
{Pereira} T. M.~D.,  {De Pontieu} B.,   {Carlsson} M.,  2012, \mn@doi [\apj]
  {10.1088/0004-637X/759/1/18}, \href
  {https://ui.adsabs.harvard.edu/abs/2012ApJ...759...18P} {759, 18}

\bibitem[\protect\citeauthoryear{{Petralia, A.}, {Reale, F.}, {Orlando, S.}  \&
  {Klimchuk, J. A.}}{{Petralia, A.} et~al.}{2014}]{Petralia_et_al_2014}
{Petralia, A.} {Reale, F.} {Orlando, S.}  {Klimchuk, J. A.} 2014, \mn@doi
  [A\&A] {10.1051/0004-6361/201323012}, 567, A70

\bibitem[\protect\citeauthoryear{Pontieu, Carlsson, van~der Voort, Rutten,
  Hansteen  \& Watanabe}{Pontieu et~al.}{2012}]{De_Pontieu_et_al_2012}
Pontieu B.~D.,  Carlsson M.,  van~der Voort L. H. M.~R.,  Rutten R.~J.,
  Hansteen V.~H.,   Watanabe H.,  2012, \mn@doi [The Astrophysical Journal]
  {10.1088/2041-8205/752/1/l12}, 752, L12

\bibitem[\protect\citeauthoryear{Priest}{Priest}{2014}]{Priest_2014}
Priest E.,  2014, References.
Cambridge University Press, p. 493–542, \mn@doi{10.1017/CBO9781139020732.017}

\bibitem[\protect\citeauthoryear{{Reale, F.} \& {Landi, E.}}{{Reale, F.} \&
  {Landi, E.}}{2012}]{Reale&Landi_2012}
{Reale, F.} {Landi, E.} 2012, \mn@doi [A\&A] {10.1051/0004-6361/201219280},
  543, A90

\bibitem[\protect\citeauthoryear{{Rempel}}{{Rempel}}{2017}]{Rempel_2017}
{Rempel} M.,  2017, \mn@doi [\apj] {10.3847/1538-4357/834/1/10}, \href
  {https://ui.adsabs.harvard.edu/abs/2017ApJ...834...10R} {834, 10}

\bibitem[\protect\citeauthoryear{Samanta et~al.,}{Samanta
  et~al.}{2019}]{Samanta_et_al_2019}
Samanta T.,  et~al., 2019, \mn@doi [Science] {10.1126/science.aaw2796}, 366,
  890

\bibitem[\protect\citeauthoryear{{Secchi}}{{Secchi}}{1877}]{Secchi_1877}
{Secchi} S.~J.,  1877, {Le Soleil}.
Paris: Gauthier-Villars

\bibitem[\protect\citeauthoryear{{Sekse}, {Rouppe van der Voort}  \& {De
  Pontieu}}{{Sekse} et~al.}{2012}]{Sekse_et_al_2012}
{Sekse} D.~H.,  {Rouppe van der Voort} L.,   {De Pontieu} B.,  2012, \mn@doi
  [\apj] {10.1088/0004-637X/752/2/108}, \href
  {https://ui.adsabs.harvard.edu/abs/2012ApJ...752..108S} {752, 108}

\bibitem[\protect\citeauthoryear{{Shibata}}{{Shibata}}{1982}]{Shibata_1982}
{Shibata} K.,  1982, \mn@doi [\solphys] {10.1007/BF00151974}, \href
  {https://ui.adsabs.harvard.edu/abs/1982SoPh...81....9S} {81, 9}

\bibitem[\protect\citeauthoryear{{Silva}, {Santos}, {B\"uchner}  \&
  {Alves}}{{Silva} et~al.}{2018}]{Silva_et_al_2018}
{Silva} S. S.~A.,  {Santos} J.~C.,  {B\"uchner} J.,   {Alves} M.~V.,  2018,
  \mn@doi [A\&A] {10.1051/0004-6361/201730580}, 615, A32

\bibitem[\protect\citeauthoryear{{Smirnova}, {Konkol}, {Solov'ev}  \&
  {Murawski}}{{Smirnova} et~al.}{2016}]{Smirnova_et_al_2016}
{Smirnova} V.,  {Konkol} P.~M.,  {Solov'ev} A.~A.,   {Murawski} K.,  2016,
  \mn@doi [\solphys] {10.1007/s11207-016-0976-0}, \href
  {https://ui.adsabs.harvard.edu/abs/2016SoPh..291.3207S} {291, 3207}

\bibitem[\protect\citeauthoryear{{Spitzer}}{{Spitzer}}{1962}]{Spitzer_1962}
{Spitzer} L.,  1962, {Physics of Fully Ionized Gases}.
Interscience Publisher, Inc., New York

\bibitem[\protect\citeauthoryear{{Srivastava} et~al.,}{{Srivastava}
  et~al.}{2018}]{Srivastava_et_al_2018}
{Srivastava} A.~K.,  et~al., 2018, \mn@doi [Nature Astronomy]
  {10.1038/s41550-018-0590-1}, \href
  {https://ui.adsabs.harvard.edu/abs/2018NatAs...2..951S} {2, 951}

\bibitem[\protect\citeauthoryear{{Sterling}}{{Sterling}}{2000}]{Sterling_2000}
{Sterling} A.~C.,  2000, \mn@doi [\solphys] {10.1023/A:1005213923962}, \href
  {https://ui.adsabs.harvard.edu/abs/2000SoPh..196...79S} {196, 79}

\bibitem[\protect\citeauthoryear{{Suematsu}, {Shibata}, {Neshikawa}  \&
  {Kitai}}{{Suematsu} et~al.}{1982}]{Suematsu_et_al_1982}
{Suematsu} Y.,  {Shibata} K.,  {Neshikawa} T.,   {Kitai} R.,  1982, \mn@doi
  [\solphys] {10.1007/BF00153464}, \href
  {https://ui.adsabs.harvard.edu/abs/1982SoPh...75...99S} {75, 99}

\bibitem[\protect\citeauthoryear{{Suematsu}, {Wang}  \& {Zirin}}{{Suematsu}
  et~al.}{1995}]{Suematsu_et_al_1995}
{Suematsu} Y.,  {Wang} H.,   {Zirin} H.,  1995, \mn@doi [\apj]
  {10.1086/176151}, \href
  {https://ui.adsabs.harvard.edu/abs/1995ApJ...450..411S} {450, 411}

\bibitem[\protect\citeauthoryear{{Suematsu}, {Ichimoto}, {Katsukawa},
  {Shimizu}, {Okamoto}, {Tsuneta}, {Tarbell}  \& {Shine}}{{Suematsu}
  et~al.}{2008}]{Suematsu_et_al_2008}
{Suematsu} Y.,  {Ichimoto} K.,  {Katsukawa} Y.,  {Shimizu} T.,  {Okamoto} T.,
  {Tsuneta} S.,  {Tarbell} T.,   {Shine} R.~A.,  2008, in {Matthews} S.~A.,
  {Davis} J.~M.,   {Harra} L.~K.,  eds,  Astronomical Society of the Pacific
  Conference Series Vol. 397, First Results From Hinode. p.~27

\bibitem[\protect\citeauthoryear{{Tanaka}}{{Tanaka}}{1974}]{Tanaka_1974}
{Tanaka} K.,  1974, in {Athay} R.~G.,  ed.,  International Astronomical Union
  Series Vol. 56, Chromospheric Fine Structure. p.~239

\bibitem[\protect\citeauthoryear{{Tsiropoula}, {Alissandrakis}  \&
  {Schmieder}}{{Tsiropoula} et~al.}{1994}]{Tsiropoula_et_al_1994}
{Tsiropoula} G.,  {Alissandrakis} C.~E.,   {Schmieder} B.,  1994, \aap, \href
  {https://ui.adsabs.harvard.edu/abs/1994A&A...290..285T} {290, 285}

\bibitem[\protect\citeauthoryear{{Tziotziou, K.}, {Tsiropoula, G.}  \& {Mein,
  P.}}{{Tziotziou, K.} et~al.}{2003}]{Tziotziou_et_al_2003}
{Tziotziou, K.} {Tsiropoula, G.}  {Mein, P.} 2003, \mn@doi [A\&A]
  {10.1051/0004-6361:20030220}, 402, 361

\bibitem[\protect\citeauthoryear{{Tziotziou, K.}, {Tsiropoula, G.}  \& {Mein,
  P.}}{{Tziotziou, K.} et~al.}{2004}]{Tziotziou_et_al_2004}
{Tziotziou, K.} {Tsiropoula, G.}  {Mein, P.} 2004, \mn@doi [A\&A]
  {10.1051/0004-6361:20040173}, 423, 1133

\bibitem[\protect\citeauthoryear{{Vashalomidze}, {Kukhianidze}, {Zaqarashvili},
  {Oliver}, {Shergelashvili}, {Ramishvili}, {Poedts}  \& {De
  Causmaecker}}{{Vashalomidze} et~al.}{2015}]{Vashalomidze_et_al_2015}
{Vashalomidze} Z.,  {Kukhianidze} V.,  {Zaqarashvili} T.~V.,  {Oliver} R.,
  {Shergelashvili} B.,  {Ramishvili} G.,  {Poedts} S.,   {De Causmaecker} P.,
  2015, \mn@doi [A\&A] {10.1051/0004-6361/201424101}, 577, A136

\bibitem[\protect\citeauthoryear{{Wilhelm}}{{Wilhelm}}{2000}]{Wilhelm_2000}
{Wilhelm} K.,  2000, \aap, \href
  {https://ui.adsabs.harvard.edu/abs/2000A&A...360..351W} {360, 351}

\bibitem[\protect\citeauthoryear{{Withbroe} et~al.,}{{Withbroe}
  et~al.}{1976}]{Withbroe_et_al_1976}
{Withbroe} G.~L.,  et~al., 1976, \mn@doi [\apj] {10.1086/154108}, \href
  {https://ui.adsabs.harvard.edu/abs/1976ApJ...203..528W} {203, 528}

\bibitem[\protect\citeauthoryear{W{\'{o}}jcik, Ku{\'{z}}ma, Murawski  \&
  Srivastava}{W{\'{o}}jcik et~al.}{2019}]{Wojcik_et_al_2019}
W{\'{o}}jcik D.,  Ku{\'{z}}ma B.,  Murawski K.,   Srivastava A.~K.,  2019,
  \mn@doi [The Astrophysical Journal] {10.3847/1538-4357/ab26b1}, 884, 127

\bibitem[\protect\citeauthoryear{Wójcik, Murawski  \& Musielak}{Wójcik
  et~al.}{2018}]{Wojcik_et_al_2018}
Wójcik D.,  Murawski K.,   Musielak Z.~E.,  2018, \mn@doi [Monthly Notices of
  the Royal Astronomical Society] {10.1093/mnras/sty2306}, 481, 262

\bibitem[\protect\citeauthoryear{{Zaqarashvili} \&
  {Erd{\'e}lyi}}{{Zaqarashvili} \&
  {Erd{\'e}lyi}}{2009}]{Zaqarashvili&Erdelyi_2009}
{Zaqarashvili} T.~V.,  {Erd{\'e}lyi} R.,  2009, \mn@doi [\ssr]
  {10.1007/s11214-009-9549-y}, \href
  {https://ui.adsabs.harvard.edu/abs/2009SSRv..149..355Z} {149, 355}

\bibitem[\protect\citeauthoryear{{de Pontieu} et~al.,}{{de Pontieu}
  et~al.}{2007}]{De_Pontieu_et_al_2007}
{de Pontieu} B.,  et~al., 2007, \mn@doi [\pasj] {10.1093/pasj/59.sp3.S655},
  \href {https://ui.adsabs.harvard.edu/abs/2007PASJ...59S.655D} {59, S655}

\makeatother
\end{thebibliography}

% Don't change these lines
\bsp	% typesetting comment
\label{lastpage}
\end{document}